\documentclass[11pt]{article}

\usepackage{jheppub}
\pdfoutput=1

\usepackage{epstopdf}
\usepackage{jheppub}
\usepackage{mathrsfs}
\usepackage{psfrag}
\usepackage{color}
\usepackage{hyperref}

\usepackage{slashed}

\usepackage{simplewick}
\usepackage{cancel}

\usepackage{todonotes}
\usepackage{tikz, pict2e}

\usepackage{lmodern}
\usepackage{graphicx}
\usepackage{subcaption}
\usepackage{tkz-euclide}
\usepackage[utf8]{inputenc}

\usetikzlibrary{calc,backgrounds, intersections}
\usetikzlibrary{arrows,shapes,positioning,shadows,trees}
\usetikzlibrary{mindmap}
\usetikzlibrary{decorations.pathreplacing}
\usetikzlibrary{decorations.pathmorphing}
\usetikzlibrary{decorations.markings}
\usetikzlibrary{matrix}

\tikzstyle arrowstyle=[scale=1]
\tikzstyle directed=[postaction={decorate,decoration={markings,
    mark=at position .5 with {\arrow[arrowstyle]{stealth}}}}]
\tikzstyle reverse directed=[postaction={decorate,decoration={markings,
    mark=at position .5 with {\arrowreversed[arrowstyle]{stealth};}}}]

\usepackage{mathtools,bbm,array,amsfonts,graphicx,wrapfig,arydshln,lscape,float,multirow,longtable,rotating,makecell}
\usepackage{url}

\definecolor{redb}{rgb}{0.700, 0.000, 0.300}
\DeclareMathAlphabet\mathbfcal{OMS}{cmsy}{b}{n}

\newcommand{\ra}{\rangle}

\newcommand{\eq}{\begin{equation}}

\newcommand{\eqe}{\end{equation}}
\newcommand{\eqa}{\begin{eqnarray}}
\newcommand{\eqae}{\end{eqnarray}}

\DeclareMathOperator*{\Img}{\mathfrak{Im}}

\newcommand{\xdownarrow}[1]{%
  {\left\downarrow\vbox to #1{}\right.\kern-\nulldelimiterspace}
}

\title{From the flat-space S-matrix to the Wavefunction of the Universe}
\author{Paolo Benincasa,$^{1}$}

\affiliation{$^1$ Niels Bohr International Academy and Discovery Center, University of Copenhagen, Niels Bohr Institute, Blegdamsvej 17, DK-2100, Copenhagen, Denmark}

\emailAdd{pablowellinhouse@anche.no}
\abstract{The physical information encoded in the cosmological late-time wavefunction of the universe is tied to its singularity structure and its behaviour as such singularities are approached. One important singularity is identified by the vanishing of the total energy, where the wavefunction reduces to the physics of scattering in flat space. In this paper, we discuss the behaviour of the perturbative wavefunction as its other singularities are approached and the role played by the flat-space scattering, in the simplified context of the class of toy models admitting a first principle definition in terms of {\it cosmological polytopes}. The problems then translates into the analysis of the structure of its facets, one of which -- the {\it scattering facet} -- beautifully encodes the flat-space S-matrix. We show that all the boundaries of the cosmological polytope encode information about the flat-space physics. In particular, a subset of its facets turns out to have a similar structure as the scattering facet, with the vertices which can be grouped together to form lower dimensional scattering facets. The other facets admit one (and only one) triangulation in terms of products of lower dimensional scattering facets. As a consequence, the whole perturbative wavefunction can be represented as a sum of product of flat-space scattering amplitudes. Finally, we turn the table around and ask whether the knowledge of the flat-space scattering amplitudes suffices to reconstruct the wavefunction of the universe. We show that, at least for our class of toy models, this is indeed the case at tree level if we are also provided with a subset of symmetries that the wavefunction ought to satisfy. Once the tree cosmological polytopes are reconstructed, the loop ones can be obtained as a particular projection of them.}

\begin{document}

\maketitle

%%%%%%%%%%%%%%%%%%%%%%
%%%%%%%%%%%%%%%%%%%%%%
%%%%%%%%%%%%%%%%%%%%%%

\section{Introduction}\label{sec:Intro}

Most of the progress in understanding observables in fundamental physics have been carried out in asymptotically flat and $AdS$ space-times, where the relevant quantum mechanical observables are well defined. In order to be able to make precise quantum mechanical predictions, it is necessary to go to the boundary  of the space-times where it is possible to perform measurements with an infinitely large apparatus and infinitely many times: on the time-like boundary of $AdS$, one can thus define correlation functions, while in asymptotically flat space-times the correct observable is the $S$-matrix. 

However, our universe in neither $AdS$ nor asymptotically flat, but rather cosmological: the universe is in a phase of accelerated expansion and this, by itself, prevents the definition of a quantum mechanical observable, being impossible to have a distinction between an infinitely large measuring apparatus and the system which gets measured. Assuming that  the universe opens up to become infinitely large (and invariant under space translations) at sufficiently late times\footnote{Despite the fact that this is not what is happening in our universe, which, as mentioned earlier, is in a phase of accelerated expansion, this is a reasonable assumption as long as the modes are not as big as the Hubble scale in which case we would have access to a very small amount of data and the spatial averaging looses meaning.}, we can perform measurements at fixed time and compute spatial averages. Thus, in such an approximation, the correct quantum mechanical observables are the late-time spatial correlators, or the wavefunction of the universe generating them.

Not much is known, not even in perturbation theory, about the general properties that these cosmological observables ought to satisfy. In particular, given a certain wavefunction of the universe, which is just a function of data at present time, we still do not know what is the imprint of an underlying consistent cosmological time evolution. In order to address this class of issues, it is of fundamental importance to gain a deeper understanding of the analytic structure of the cosmological observables. Spatial translation invariance implies spatial momentum conservation, while there is no energy conservation due to the fact that these observables are computed at fixed time. Importantly, this fact translates into an interesting feature of the cosmological observables: the sum of the energies of the external states appears as a singularity, which can be reached only after analytic continuation. At this point in energy space, energy conservation is restored as well as the full Lorentz invariance, and the wavefunction reduces to the (high-energy limit of) the flat-space scattering amplitude\footnote{This is similar to what happens in the $AdS$ correlators as shown in \cite{Raju:2012zr, Raju:2012zs}} \cite{Arkani-Hamed:2015bza}. The existence of such a straightforward relation between the wavefunction of the universe and the flat-space scattering amplitude is  already quite remarkable: the vacuum wavefunction {\it knows} about a dynamical problem such as particle scattering. However, the physical interpretation of the other singularities is not quite clear yet. 

Some important first steps in understanding cosmological observables as just function of the {\it boundary} data has been carried out for the wavefunction of the universe in a class of toy models of scalars with polynomial interactions in FRW cosmologies\footnote{The late-time wavefunction of the universe in dS were previously studied in \cite{Anninos:2011kh, Anninos:2014lwa}.} \cite{Arkani-Hamed:2017fdk}, and inflationary correlators \cite{Arkani-Hamed:2018kmz}. While in the second case, it was shown how singularities and symmetries uniquely fix the correlators\footnote{In \cite{Coriano:2013jba, Bzowski:2013sza, Bzowski:2015pba, Bzowski:2017poo, Coriano:2018zdo, Coriano:2018bbe, Isono:2018rrb, Bzowski:2018fql, Coriano:2018bsy} conformal Ward identities in momentum space were used to compute three-point correlators involving currents and stress-tensor.}, in the first one a new representation of the wavefunction was proposed in terms of {\it cosmological polytopes}. These are combinatorial objects with their own first principle definition which encode the singularity structure that we can ascribe to the perturbative wavefunction: their boundaries are lower dimensional polytopes and correspond to the residues of the poles in the wavefunction. There is one special boundary, the {\it scattering facet}, which is related to the total energy pole $E_{\mbox{\tiny tot}}$. Notably, its vertex structure beautifully encodes the cutting rules for flat-space scattering amplitudes, while Lorentz invariance is manifest in its dual.

What about the residues of the other poles of the wavefunction of the universe? Which physical information do they encode? From the cosmological polytope perspective, these questions translate into questions about the structure of the other facets. A direct inspection shows that they encode the factorisation of the wavefunction into a lower point wavefunction and a lower point scattering amplitude: these boundaries are reached as the sum of the energies related to a certain subgraph vanishes so that the local energy conservation is restored which a lower point scattering amplitude correspond to. In this paper we explore more in detail the physical information encoded in these facets. Interestingly, there always exist a number of facets which are isomorphic to the scattering facets and, together with the latter, cover the full cosmological polytope. Even more strikingly, the other facets -- which are no longer simplices -- admit a triangulation such that the canonical form of each simplex is the product of Lorentz-like propagators and, thus, the canonical form of such a facet has a representation as the sum of products of flat-space scattering amplitudes. In other words, {\it all} the physical information of the wavefunction of the universe, which is encoded into the residues if its poles, can be recast in terms of flat-space processes. This is both surprising and remarkable, and immediately brings another question: if all the residues of the poles of the wavefunction of the universe can be understood in terms of flat-space scattering amplitudes, would it be possible to reconstruct the wavefunction of the universe from such flat-space processes? In this paper, we face this question as well, and the answer is indeed positive: if we were given a flat space process {\it and} a subset of symmetries that the wavefunction ought to satisfy, then we could reconstruct the wavefunction of the universe itself. This indeed true for the class of models the cosmological polytope provide a complete description of.

The paper is organised as follows. In Section \ref{sec:CP} we review the salient features of the cosmological polytope and its relation to the wavefunction of the universe. We also point out (Section \ref{subsec:TL}) how a simple projection of the cosmological polytope describing an $L$-loop wavefunction, returns a new cosmological polytope which now describes an $(L+1)$-loop wavefunction. Section \ref{sec:FCP} is devoted to the analysis of all the facets other than the scattering one. Here we show explicitly how there are certain facets are {\it isomorphic} to the scattering facets, and thus their canonical form is just the product of Lorentz-like propagators, as well as how the other facets can be triangulated so to decompose their canonical form into sum of products of Lorentz-like propagators. In Section \ref{sec:SAWU} we turn the table around, and starting from the knowledge of a certain  scattering amplitude and some symmetries of the wavefunction, we can reconstruct the wavefunction. Section \ref{sec:Concl} contains our conclusion and outlook, while the appendices contained a more analytic (rather than combinatorial and geometrical) look at both the map between wavefunctions at different orders in perturbation theory (Appendix \ref{app:TL}), and the relation between the residues of the poles in the wavefunction and the flat-space scattering amplitudes (Appendix \ref{app:WFSm}).

%%%%%%%%%%%%%%%%%%%%%%
%%%%%%%%%%%%%%%%%%%%%%
%%%%%%%%%%%%%%%%%%%%%%

\section{The wavefunction of the universe and the cosmological polytopes}\label{sec:CP}

Let us consider the toy model of scalars in flat space-time with time-dependent polynomial interactions, which contains conformally-coupled scalars with non-conformal polynomial interactions in FRW cosmologies \cite{Arkani-Hamed:2017fdk}
\begin{equation}\label{eq:ccs}
 S[\phi]\:=\:\int d^d x\int_{-\infty}^0d\eta
 \left[
  \frac{1}{2}\left(\partial\phi\right)^2-\sum_{k\ge3}\frac{\lambda_k(\eta)}{k!}\phi^k
 \right],
\end{equation}
with the time-dependent coupling constant which can be conveniently treated in Fourier space: 
\begin{equation*}
 \lambda_k(\eta)\,=\,\int_0^{+\infty} d\varepsilon\,e^{i\varepsilon\eta}\tilde{\lambda}_k(\varepsilon).
\end{equation*}
When the wavefunction $\psi$ is computed perturbatively, each Feynman diagram is a function of sum $x_j\equiv\sum_{k\in v_j}E_k$ of the energies $E_k\equiv|\overrightarrow{p}_k|$ of the external states at each vertex $v_j$, as well as of the internal energies $y_{ij}$ associated to the edges between two vertices $v_i$ and $v_j$\footnote{Feynman diagrams can be conveniently replaced by a reduced graph obtained from it by suppressing the external edges.}. However, considering the couplings in Fourier space, we will focus on the (contribution of the) {\it integrand} $\Psi_{\mathcal{G}}(x,y)$ of the wavefunction from a given graph $\mathcal{G}$, which depends on the coupling energy $\varepsilon$ (it can be absorbed in the definition of $x$) and which at the very end needs to be integrated. It was shown in \cite{Arkani-Hamed:2017fdk} that these graphs are related to a certain cosmological polytope, which is defined from first principle as follows.

Let us consider the space of $n_e$ triangles $\triangle_i$ identified through their midpoints $({\bf x}_i,{\bf y}_{i},{\bf x'}_i)$ or, equivalently, via their vertices $\{{\bf x}_i-{\bf y}_{i}+{\bf x'}_i,\,{\bf x}_i+{\bf y}_{i}-{\bf x'}_i,\,-{\bf x}_i+{\bf y}_{i}+{\bf x'}_i\}$. Such triangles are endowed with the following property: they can be intersected along two of their three edges in their midpoints. This property defines the class of polytopes called {\it cosmological polytopes} as the convex hull of the $3n_e$ vertices of the $n_e$ triangles which are intersected. Such a construction not only mathematically defines a new class of polytopes, but also has the imprint of the space-time causal structure: the two intersectable edges of the triangle represent the two region with a definite causal relation ({\it past} and {\it future}) between events, while the non-intersectable represents the space-like region with no fixed causal relation. These objects live in $\mathbb{P}^{3n_e-r-1}$, where $r$ is the number of independent midpoint identifications. There is a $1-1$ correspondence between the cosmological polytopes and the Feynman graphs. Given a triangle $\triangle_i$, it is possible to associate a two vertex graph, whose vertices correspond to the intersectable edges, while its edge to the non-intersectable one. Then, a cosmological polytope $\mathcal{P}_{\mathcal{G}}$ constructed by intersecting $n_e$ triangles is associated with a graph $\mathcal{G}$ with $n_e$ edges constructed from a collection of $n_e$ two-vertex graphs by identifying some of their vertices:
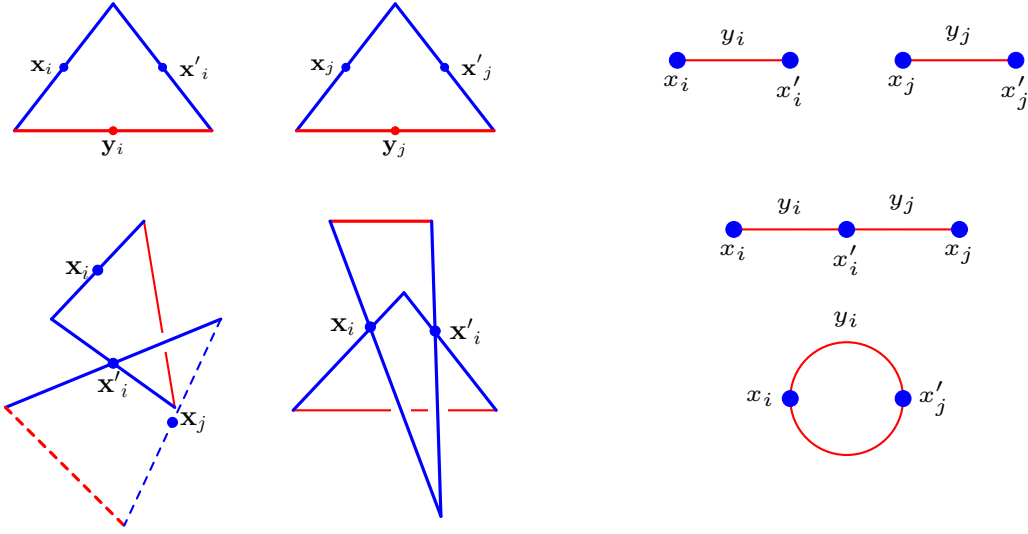
\begin{equation*}
 \begin{tikzpicture}[line join = round, line cap = round, ball/.style = {circle, draw, align=center, anchor=north, inner sep=0}, 
                     axis/.style={very thick, ->, >=stealth'}, pile/.style={thick, ->, >=stealth', shorten <=2pt, shorten>=2pt}, every node/.style={color=black}, scale={1.5}]
  \begin{scope}[shift={(5,2)}, transform shape]
    \coordinate[label=below:{\tiny $x_i$}] (v1) at (0,0);
    \coordinate[label=below:{\tiny $x'_i$}] (v2) at ($(v1)+(1,0)$);
    \coordinate[label=below:{\tiny $x_j$}] (v3) at ($(v2)+(1,0)$);
    \coordinate[label=below:{\tiny $x'_j$}] (v4) at ($(v3)+(1,0)$);
    \coordinate[label=above:{\tiny $y_i$}] (yi) at ($(v1)!0.5!(v2)$);
    \coordinate[label=above:{\tiny $y_j$}] (yj) at ($(v3)!0.5!(v4)$);
    \draw[thick, color=red] (v1) -- (v2);
    \draw[thick, color=red] (v3) -- (v4);
    \draw[fill=blue, color=blue] (v1) circle (2pt);
    \draw[fill=blue, color=blue] (v2) circle (2pt);
    \draw[fill=blue, color=blue] (v3) circle (2pt);
    \draw[fill=blue, color=blue] (v4) circle (2pt);
    \coordinate (t1) at ($(v2)!0.5!(v3)$);
    \coordinate[label=below:{\tiny $x'_i$}] (s2) at ($(t1)+(0,-1.5)$);
    \coordinate[label=below:{\tiny $x_i$}] (s1) at ($(s2)-(1,0)$);
    \coordinate[label=below:{\tiny $x_j$}] (s3) at ($(s2)+(1,0)$);
    \coordinate[label=above:{\tiny $y_i$}] (yyi) at ($(s1)!0.5!(s2)$);
    \coordinate[label=above:{\tiny $y_j$}] (yyj) at ($(s2)!0.5!(s3)$);
    \draw[thick, color=red] (s1) -- (s2) -- (s3);
    \draw[fill=blue, color=blue] (s1) circle (2pt);
    \draw[fill=blue, color=blue] (s2) circle (2pt);
    \draw[fill=blue, color=blue] (s3) circle (2pt);
    \coordinate[label=left:{\tiny $x_i$}] (n1) at ($(s1)!0.5!(s2)+(0,-1.5)$);
    \coordinate[label=right:{\tiny $x'_j$}] (n2) at ($(s2)!0.5!(s3)+(0,-1.5)$);
    \coordinate (nc) at ($(n1)!0.5!(n2)$);
    \coordinate[label=above:{\tiny $y_i$}] (yyyi) at ($(nc)+(0,.5cm)$);
    \draw[thick, color=red] (nc) circle (.5cm);
    \draw[fill=blue, color=blue] (n1) circle (2pt);
    \draw[fill=blue, color=blue] (n2) circle (2pt);
  \end{scope}
  \begin{scope}[shift={(0,2.5)}, scale={.5}]
   \coordinate (A) at (0,0);
   \coordinate (B) at (-1.75,-2.25);
   \coordinate (C) at (+1.75,-2.25);
   \coordinate [label=left:{\footnotesize $\displaystyle {\bf x}_i$}] (m1) at ($(A)!0.5!(B)$);
   \coordinate [label=right:{\footnotesize $\displaystyle \;{\bf x'}_i$}] (m2) at ($(A)!0.5!(C)$);
   \coordinate [label=below:{\footnotesize $\displaystyle {\bf y}_i$}] (m3) at ($(B)!0.5!(C)$);
   \tikzset{point/.style={insert path={ node[scale=2.5*sqrt(\pgflinewidth)]{.} }}} 

   \draw[color=blue,fill=blue] (m1) circle (2pt);
   \draw[color=blue,fill=blue] (m2) circle (2pt);
   \draw[color=red,fill=red] (m3) circle (2pt);

   \draw[-, very thick, color=blue] (B) -- (A);
   \draw[-, very thick, color=blue] (A) -- (C);  
   \draw[-, very thick, color=red] (B) -- (C);    
  \end{scope}
  \begin{scope}[shift={(2.5,2.5)}, scale={.5}]
   \coordinate (A) at (0,0);
   \coordinate (B) at (-1.75,-2.25);
   \coordinate (C) at (+1.75,-2.25);
   \coordinate [label=left:{\footnotesize $\displaystyle {\bf x}_j$}] (m1) at ($(A)!0.5!(B)$);
   \coordinate [label=right:{\footnotesize $\displaystyle \;{\bf x'}_j$}] (m2) at ($(A)!0.5!(C)$);
   \coordinate [label=below:{\footnotesize $\displaystyle {\bf y}_j$}] (m3) at ($(B)!0.5!(C)$);
   \tikzset{point/.style={insert path={ node[scale=2.5*sqrt(\pgflinewidth)]{.} }}} 

   \draw[color=blue,fill=blue] (m1) circle (2pt);
   \draw[color=blue,fill=blue] (m2) circle (2pt);
   \draw[color=red,fill=red] (m3) circle (2pt);

   \draw[-, very thick, color=blue] (B) -- (A);
   \draw[-, very thick, color=blue] (A) -- (C);  
   \draw[-, very thick, color=red] (B) -- (C);    
  \end{scope}
  \begin{scope}[scale={.5}, shift={(0,2)}, transform shape]
   \pgfmathsetmacro{\factor}{1/sqrt(2)};  
   \coordinate  (B2) at (1.5,-3,-1.5*\factor);
   \coordinate  (A1) at (-1.5,-3,-1.5*\factor);
   \coordinate  (B1) at (1.5,-3.75,1.5*\factor);
   \coordinate  (A2) at (-1.5,-3.75,1.5*\factor);  
   \coordinate  (C1) at (0.75,-.65,.75*\factor);
   \coordinate  (C2) at (0.4,-6.05,.75*\factor);
   \coordinate (Int) at (intersection of A2--B2 and B1--C1);
   \coordinate (Int2) at (intersection of A1--B1 and A2--B2);

   \tikzstyle{interrupt}=[
    postaction={
        decorate,
        decoration={markings,
                    mark= at position 0.5 
                          with
                          {
                            \node[rectangle, color=white, fill=white, below=-.1 of Int] {};
                          }}}
   ]
   
   \draw[interrupt,thick,color=red] (B1) -- (C1);
   \draw[-,very thick,color=blue] (A1) -- (B1);
   \draw[-,very thick,color=blue] (A2) -- (B2);
   \draw[-,very thick,color=blue] (A1) -- (C1);
   \draw[-, dashed, very thick, color=red] (A2) -- (C2);
   \draw[-, dashed, thick, color=blue] (B2) -- (C2);

   \coordinate[label=below:{\Large ${\bf x'}_i$}] (x2) at ($(A1)!0.5!(B1)$);
   \draw[fill,color=blue] (x2) circle (2.5pt);   
   \coordinate[label=left:{\Large ${\bf x}_i$}] (x1) at ($(C1)!0.5!(A1)$);
   \draw[fill,color=blue] (x1) circle (2.5pt);
   \coordinate[label=right:{\Large ${\bf x}_j$}] (x3) at ($(B2)!0.5!(C2)$);
   \draw[fill,color=blue] (x3) circle (2.5pt);
  \end{scope}
  \begin{scope}[scale={.6}, shift={(3.75,.75)}, transform shape]
   \pgfmathsetmacro{\factor}{1/sqrt(2)};
   \coordinate  (c1b) at (0.75,0,-.75*\factor);
   \coordinate  (b1b) at (-.75,0,-.75*\factor);
   \coordinate  (a2b) at (0.75,-.65,.75*\factor);
   
   \coordinate  (c2b) at (1.5,-3,-1.5*\factor);
   \coordinate  (b2b) at (-1.5,-3,-1.5*\factor);
   \coordinate  (a1b) at (1.5,-3.75,1.5*\factor); 

   \coordinate (Int1) at (intersection of b2b--c2b and b1b--a1b);
   \coordinate (Int2) at (intersection of b2b--c2b and c1b--a1b);
   \coordinate (Int3) at (intersection of b2b--a2b and b1b--a1b);
   \coordinate (Int4) at (intersection of a2b--c2b and c1b--a1b);
   \tikzstyle{interrupt}=[
    postaction={
        decorate,
        decoration={markings,
                    mark= at position 0.5 
                          with
                          {
                            \node[rectangle, color=white, fill=white] at (Int1) {};
                            \node[rectangle, color=white, fill=white] at (Int2) {};                            
                          }}}
   ]

   \node at (c1b) (c1c) {};
   \node at (b1b) (b1c) {};
   \node at (a2b) (a2c) {};
   \node at (c2b) (c2c) {};
   \node at (b2b) (b2c) {};
   \node at (a1b) (a1c) {};

   \draw[interrupt,thick,color=red] (b2b) -- (c2b);
   \draw[-,very thick,color=red] (b1b) -- (c1b);
   \draw[-,very thick,color=blue] (b1b) -- (a1b);
   \draw[-,very thick,color=blue] (a1b) -- (c1b);   
   \draw[-,very thick,color=blue] (b2b) -- (a2b);
   \draw[-,very thick,color=blue] (a2b) -- (c2b);

   \node[ball,text width=.15cm,fill,color=blue, above=-.06cm of Int3, label=left:{\large ${\bf x}_i$}] (Inta) {};
   \node[ball,text width=.15cm,fill,color=blue, above=-.06cm of Int4, label=right:{\large ${\bf x'}_i$}] (Intb) {};

  \end{scope}
 \end{tikzpicture}
\end{equation*}
The vertices and edges of the graph get labelled by $x_v$ and $y_{e}$ which are respectively related to the midpoints ${\bf x}$ and ${\bf y}$ -- such labels $x_v$ and $y_e$ can be identified with the energies at a vertex $v$ and an edge $e$ respectively. Finally, any cosmological polytope $\mathcal{P}_{\mathcal{G}}$ is endowed with a canonical form, which encodes the contribution $\Psi_{\mathcal{G}}(x_v,y_e)$ of the graph $\mathcal{G}$ to the wavefunction:
\begin{equation}\label{eq:CFWF}
 \Omega\left(\mathcal{Y};\,\mathcal{P}_{\mathcal{G}}\right)\:=\:
  \left(
   \prod_{v,e}dx_vdy_e
  \right)
  \frac{\Psi_{\mathcal{G}}(x_v,y_e)}{\mbox{Vol}\left\{GL(1)\right\}}
\end{equation}
with $\mathcal{Y}\,=\,\sum_{v}x_v{\bf X}_v+\sum_{e}y_e{\bf Y}_e$ being any point of $\mathcal{P}_{\mathcal{G}}$ with ${\bf X}_v$ and ${\bf Y}_e$ being the vectors identifying the independent midpoints. It is characterised by logarithmic singularities on all the faces of $\mathcal{P}_{\mathcal{G}}$, so that once any of these poles is reached, it reduces to a lower dimensional canonical form which identifies a lower dimensional polytope.

%%%%%%%%%%%%%%%%%%%%%%
%%%%%%%%%%%%%%%%%%%%%%

\subsection{From trees to loops}\label{subsec:TL}

Given $n_e$ triangles $\triangle_i$, we can intersect them to generate a number of cosmological polytopes, which corresponds to graph at different order in perturbation theory and with different number of vertices. Interestingly, a cosmological polytope $\mathcal{P}_{\mathcal{G}}\,\in\,\mathbb{P}^{3n_e-r-1}$ generated by imposing $r\,>\,1$ independent identifications of the midpoints and such that $r-n_e\,>\,0$, can be generated from another cosmological polytope obtained by intersecting the very same $n_e$ triangles but imposing $\tilde{r}\,<\,r$ constraints -- thus, the latter lives in higher dimensions -- by allowing for further $r-\tilde{r}$ intersections. This somewhat trivial statement, has an interesting implication. A cosmological polytope living in $\mathbb{P}^{3n_e-r-1}$ is associated to an $L$-loop graph $\mathcal{G}$ with $n_e$ edges and $n_v\,=\,2n_e-r\,=n_e+1-L$ vertices. Thus, the number of loops $L$ of the associated graph $\mathcal{G}$, is determined by the number of independent midpoint identifications to be $L\,=\,1+r-n_e$. Hence, constructing a cosmological polytope in $\mathbb{P}^{3n_e-r-1}$ from another one in higher dimensions $\mathbb{P}^{3n_e-\tilde{r}-1}$ ($\tilde{r}\,<\,r$) translates at graph level into obtaining a higher loop graph starting from a lower point one. In other words, a cosmological polytope encodes the information about a certain class of lower dimensional cosmological polytopes, and, consequently, allows to extract a contribution to the $(L+1)$-loop wavefunction from a contribution to the wavefunction with one loop less.

Formally, this operation is just a projection. Given a cosmological polytope $\mathcal{P}_{\mathcal{G}}$, let ${\bf x}_i$ and ${\bf x}_j$ be the midpoints of two intersectable edges of two distinct triangles $\triangle_i$ and $\triangle_j$ generating $\mathcal{P}_{\mathcal{G}}$. A cosmological polytope $\mathcal{P}'_{\mathcal{G}'}$ living in a one dimension less, can be obtained by projecting $\mathcal{P}$ through a cone with origin in ${\bf x}_i-{\bf x}_j$. 

\begin{figure}
 \centering
 \begin{tikzpicture}[line join = round, line cap = round, ball/.style = {circle, draw, align=center, anchor=north, inner sep=0}]
  \begin{scope}[scale={.45}, transform shape]
   \pgfmathsetmacro{\factor}{1/sqrt(2)}; 
   \coordinate [label=left:{\Large ${\bf x}_2 + {\bf x'} - {\bf y}_2$}] (A2c) at (-1.5,-3.75,1.5*\factor);
   \coordinate [label=below:{\Large ${\bf x}_2 + {\bf y}_2 - {\bf x'}$}] (B2c) at (0,-6.05,.75*\factor);  
   \coordinate [label=left:{\Large ${\bf x}_1 + {\bf x'} - {\bf y}_1$}] (A1c) at (-1.5,-3,-1.5*\factor);
   \coordinate [label=above:{\Large ${\bf x}_1 + {\bf y}_1 - {\bf x'}$}] (B1c) at (0.75,-.65,.75*\factor);
   \coordinate [label=right:{\Large ${\bf y}_1 + {\bf x'} - {\bf x}_1$}] (C1c) at (1.5,-3.75,1.5*\factor);
   \coordinate [label=right:{\Large ${\bf y}_2 + {\bf x'} - {\bf x}_2$}] (C2c) at (1.5,-3,-1.5*\factor);
 
   \coordinate (Int3) at (intersection of A2c--B2c and B1c--C1c);

   \node at (A1c) (A1d) {};
   \node at (B2c) (B2d) {};
   \node at (B1c) (B1d) {};
   \node at (A2c) (A2d) {};
   \node at (C1c) (C1d) {};
   \node at (C2c) (C2d) {};

   \draw[-,dashed,fill=blue!25, opacity=.6] (A1c) -- (C2c) -- (B1c) -- cycle;
   \draw[-,thick,fill=blue!15, opacity=.6] (A1c) -- (A2c) -- (B1c) -- cycle;
   \draw[-,thick,fill=blue!15, opacity=.6] (C1c) -- (C2c) -- (B1c) -- cycle;
   \draw[-,thick,fill=blue!30, opacity=.6] (A2c) -- (C1c) -- (B1c) -- cycle;

   \draw[-,dashed,fill=red!25, opacity=.2] (A1c) -- (C2c) -- (B2c) -- cycle;
   \draw[-,dashed, thick, fill=red!45, opacity=.4] (C2c) -- (C1c) -- (B2c) -- cycle;
   \draw[-,dashed,fill=red!35, opacity=.2] (A1c) -- (A2c) -- (B2c) -- cycle;
   \draw[-,dashed, thick, fill=red!40, opacity=.4] (A2c) -- (C1c) -- (B2c) -- cycle;

   \draw[-, dotted, color=blue] (A1c) -- (B1c);
   \draw[-, dotted, color=blue] (A1c) -- (C1c);
   \draw[-, dotted, color=red] (B1c) -- (C1c);
   \draw[-, dotted, color=blue] (A2c) -- (B2c);
   \draw[-, dotted, color=blue] (A2c) -- (C2c);
   \draw[-, dotted, color=red] (B2c) -- (C2c);
   \draw[fill, color=blue] (intersection of A1c--C1c and A2c--C2c) circle (1.5pt);

   \coordinate (z1) at ($(A1c)!0.5!(B1c)$);
   \coordinate (z2) at ($(B2c)!0.5!(A2c)$);
   \coordinate (z2p) at ($(A1c)!0.5!(C1c)$);
   \coordinate (Ot) at ($(z2)!-5cm!(z2)$);

   \draw [dashed, color=green!70!black] ($(z1)!-6cm!(z2)$) -- ($(z2)!-5cm!(z1)$);   
   \node[ball, text width=.25cm, fill, color=green!70!black, xshift=-1.125cm, yshift=.25cm, label=below:{\Large ${\bf x}_2-{\bf x}_1$}] (O) at (Ot) {};

   \coordinate (a1) at ($(A1c)!-8cm!(O)$);     
   \coordinate (I1) at ($(z1)!-6cm!(z2)$);
   \coordinate (I2) at ($(z2p)!-8cm!(O)$);
   \coordinate (b1) at ($(B1c)!-4.3cm!(O)$);
   \coordinate (c1) at ($(C1c)!-8.2cm!(O)$);
   \coordinate (a2) at ($(A2c)!-6cm!(O)$);
   \coordinate (b2) at ($(B2c)!-18.7cm!(O)$);
   \coordinate (c2) at ($(C2c)!-10.95cm!(O)$);

   \draw [dashed, color=green!70!black] ($(O)!-.0cm!(z2p)$) -- ($(z2p)!-8cm!(O)$);  
   \draw [color=green!70!black] ($(O)!-.0cm!(A1c)$) -- ($(A1c)!-8cm!(O)$);
   \draw [color=green!70!black] ($(O)!-.0cm!(B1c)$) -- ($(B1c)!-4.3cm!(O)$);
   \draw [color=green!70!black] ($(O)!-.0cm!(C1c)$) -- ($(C1c)!-8.2cm!(O)$);

   \draw [color=green!70!black] ($(O)!-.0cm!(A2c)$) -- ($(A2c)!-6cm!(O)$);
   \draw [color=green!70!black] ($(O)!-.0cm!(B2c)$) -- ($(B2c)!-18.7cm!(O)$);
   \draw [color=green!70!black] ($(O)!-.0cm!(C2c)$) -- ($(C2c)!-10.95cm!(O)$);
   
   \draw[-,dashed,fill=green!50,opacity=.6] (c1) -- (b1) -- (b2) -- (c2) -- cycle;
   \draw[draw=none,fill=red!60, opacity=.45, thick] (a1) -- (b2) -- (c2) -- cycle;
   \draw[-,fill=blue!,opacity=.3, thick] (c1) -- (b1) -- (a2) -- cycle; 
   \draw[-,fill=green!50,opacity=.4, thick] (b1) -- (a2) -- (a1) -- (b2) -- cycle;
   \draw[-,fill=green!45!black,opacity=.2, thick] (c1) -- (a2) -- (a1) -- (c2) -- cycle; 
   \draw[-, dotted, color=blue] (a1) -- (b1);
   \draw[-, dotted, color=blue] (a1) -- (c1);
   \draw[-, dotted, color=red] (b1) -- (c1);
   \draw[-, dotted, color=blue] (a2) -- (b2);
   \draw[-, dotted, color=blue] (a2) -- (c2);
   \draw[-, dotted, color=red] (b2) -- (c2);

   \draw[fill=blue, color=blue] (intersection of a1--b1 and a2--b2) circle (1.5pt);
   \draw[fill=blue, color=blue] (intersection of a1--c1 and a2--c2) circle (1.5pt);   
  \end{scope}
  \begin{scope}[shift={(6,-2)}, transform shape]
   \node[ball,text width=.18cm,fill,color=blue,label=below:{\tiny $x_1$}] (z1) at (0,0) {};
   \node[ball,text width=.18cm,fill,color=blue,right=1.5cm of z1.east,label=below:{\tiny $x'$}] (z2) {};
   \node[ball,text width=.18cm,fill,color=blue,right=1.5cm of z2.east,label=below:{\tiny $x_2$}] (z3) {};  
   \draw[-,thick,color=red] (z1.east) edge node [text width=.18cm,color=black,above=-0.05cm,midway] {{\tiny $y_{1}$}} (z2.west);
   \draw[-,thick,color=red] (z2.east) edge node [text width=.18cm,color=black,above=-0.05cm,midway] {{\tiny $y_{2}$}} (z3.west);
  \end{scope}
  \begin{scope}[shift={(7,3)}, transform shape]
   \coordinate [label=left:{\tiny $x$}] (y1) at (0,0);
   \coordinate [label=right:{\tiny $x'$}](y2) at (2,0);  
   \draw[thick,color=red] ($(y1)!0.5!(y2)$) circle (1cm); 
   \draw[fill,color=blue] (y1) circle (2.8pt);      
   \draw[fill,color=blue] (y2) circle (2.8pt);   
  \end{scope}
 \end{tikzpicture}
 \caption{Loops from trees. The projection of the {\it double square pyramid} related to the tree-level three-site graph, which is generated by intersecting two triangles on a midpoint in one of their intersectable edges, through a cone with origin in ${\bf O}\,\equiv\,{\bf x}_2-{\bf x}_1$ maps it into the truncated tetrahedron, which encodes the one-loop two site graph.}
 \label{fig:LT}
\end{figure}
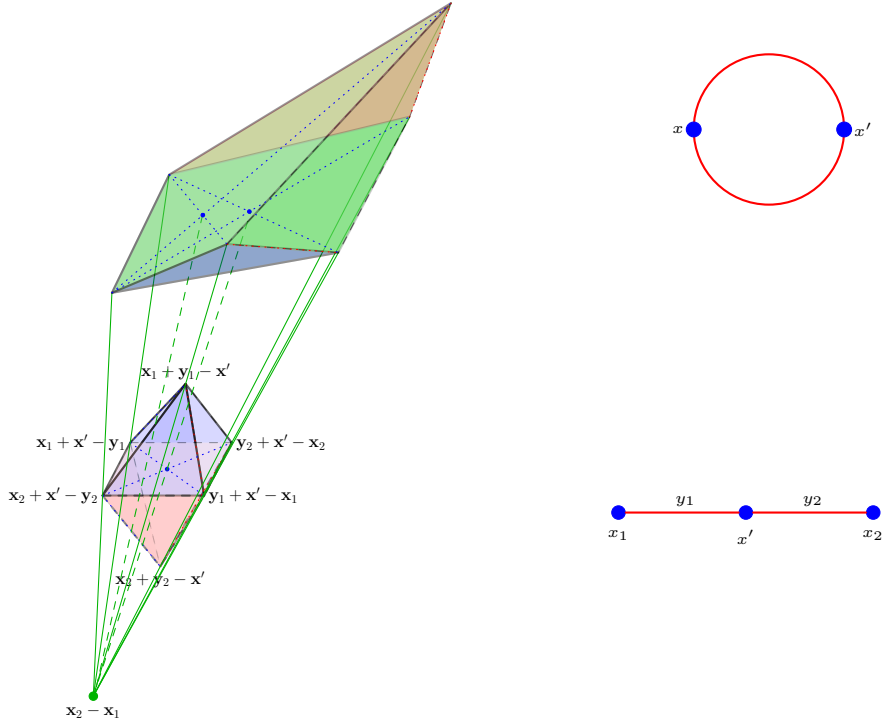

A {\it nearly} visualisable example is obtained by consider $n_e\,=\,2$ triangles (see Figure \ref{fig:LT}). 
A single identification of two midpoints belonging to different triangles, generates the {\it double squared pyramid} in $\mathbb{P}^4$ which encodes the contribution to the wavefunction given by tree-level three-site line graph. Such a cosmological polytope is identified by six vertices $\{{\bf x}_j-{\bf y}_j+{\bf x'},\,{\bf x}_j+{\bf y}_j-{\bf x'},\,-{\bf x}_j+{\bf y}_j+{\bf x'}\}$ ($j\,=\,1,\,2$). Let us now consider the point ${\bf O}\,\equiv\,{\bf x}_2-{\bf x}_1$ and project $\mathcal{P}_{\mathcal{G}}$ through the cone with origin in ${\bf O}$. The result is now a polytope $\mathcal{P}'_{\mathcal{G}'}\,\in\,\mathbb{P}^3$ identified by, again, six vertices $\{{\bf x}-{\bf y}_j+{\bf x'},\,{\bf x}+{\bf y}_j-{\bf x'},\,-{\bf x}+{\bf y}_j+{\bf x'}\}$ ($j\,=\,1,\,2$), where ${\bf x}\,\equiv\,{\bf x}_1\,\equiv\,{\bf x}_2$: $\mathcal{P}'_{\mathcal{G}'}$ is nothing but the cosmological polytope encoding the one-loop two-site graph. It is clear that, when $\mathcal{P}_{\mathcal{G}}$ is projected through the cone with origin in ${\bf O}$, the two free midpoints of the generating triangles get identified and this is how the cosmological polytope related to the one-loop two-site graph is generated. It is really striking how the cosmological polytopes related to graphs at a certain order in perturbation theory, really encode polytopes related to lower-order graphs!

As described, this projection is applied choosing the origin of the cone at ${\bf O}\,\equiv\,{\bf x}_i-{\bf x}_j$, with ${\bf x}_i$ and ${\bf x}_j$ being midpoints of edges belonging to different triangles. However, we can also choose the origin ${\bf O}$ of the projective cone taking ${\bf x}_i$ and ${\bf x}_j$ to be the midpoints of the two intersectable edges of the same triangle. It is straightforward to see that the resulting cosmological polytope is related to tadpole-like graphs\footnote{Cosmological polytopes generated this way can be thought of as generated by intersecting triangles and a segment -- the segment is the result of projecting a given triangle, with the non-intersectable edge which reduces to a point.}.

%%%%%%%%%%%%%%%%%%%%%%
%%%%%%%%%%%%%%%%%%%%%%
%%%%%%%%%%%%%%%%%%%%%%

\section{Facets of the cosmological polytopes}\label{sec:FCP}

Much of the physical information of the wavefunction of the universe is encoded in the boundaries of the cosmological polytopes. There is one particular facet, the {\it scattering facet}, which has been studied in detail \cite{Arkani-Hamed:2018ahb}: it encodes all the properties of the flat-space scattering amplitude, and, on it, Lorentz invariance and unitarity emerge beautifully. What about the other facets? In principle, a first look at the vertex structure suggests that, at any boundary other than the scattering facet, the wavefunction of the universe factorises into a product of a lower-point wavefunction and a lower-point amplitude.

The facets can be easily identified as the collection $\mathcal{V}_{\mbox{\tiny $\mathcal{F}$}}$ of vertices ${\bf V}_a^I$ ($a\,=\,1,\,\ldots,\,3n_e$) of $\mathcal{P}_{\mathcal{G}}$ such that, for a certain hyperplane $\mathcal{W}_I\,=\,\tilde{x}_v{\bf \tilde{X}_{vI}}+\tilde{y}_e{\bf \tilde{Y}}_{eI}$, $\mathcal{W}_I{\bf V}_a^I\,=\,0$ while $\mathcal{W}_I{\bf V}_a^I\,>\,0$ for any ${\bf V}_a^I$ which does not belong to $\mathcal{V}_{\mbox{\tiny $\mathcal{F}$}}$, compatibly with the constraints on the midpoints of the generating triangles. Graphically, it is convenient to keep track of the vertices which {\it do not} belong to a given hyperplane $\mathcal{W}$ by associating a marking to the relevant graph:
\begin{equation*}
 \begin{tikzpicture}[ball/.style = {circle, draw, align=center, anchor=north, inner sep=0}, cross/.style={cross out, draw, minimum size=2*(#1-\pgflinewidth), inner sep=0pt, outer sep=0pt}, scale={1.125}, transform shape]
  \begin{scope}%[scale={.45}]
   \node[ball,text width=.18cm,fill,color=black,label=below:{\footnotesize $v\phantom{'}$}] at (0,0) (v1) {};
   \node[ball,text width=.18cm,fill,color=black,label=below:{\footnotesize $v'$},right=1.5cm of v1.east] (v2) {};  
   \draw[-,thick,color=black] (v1.east) edge node [text width=.18cm,below=.1cm,midway] {\footnotesize $e$} (v2.west);
   \node[very thick, cross=4pt, rotate=0, color=blue, right=.7cm of v1.east]{};
   \coordinate (x) at ($(v1)!0.5!(v2)$);
   \node[right=1.5cm of v2, scale=.9] (lb1) {$\mathcal{W}\cdot({\bf x}_v+{\bf x}_{v'}-{\bf y}_e)>\,0$};  
  \end{scope}
  \begin{scope}[shift={(0,-1)}]%, scale={.45}]
   \node[ball,text width=.18cm,fill,color=black,label=below:{\footnotesize $v\phantom{'}$}] at (0,0) (v1) {};
   \node[ball,text width=.18cm,fill,color=black,label=below:{\footnotesize $v'$},right=1.5cm of v1.east] (v2) {};  
   \draw[-,thick,color=black] (v1.east) edge node [text width=.18cm,below=.1cm,midway] {\footnotesize $e$} (v2.west);
   \node[very thick, cross=4pt, rotate=0, color=blue, left=.1cm of v2.west]{};
   \coordinate (x) at ($(v1)!0.5!(v2)$);
   \node[right=1.5cm of v2, scale=.9] (lb1) {$\mathcal{W}\cdot({\bf x}_{v'}+{\bf y}_e-{\bf x}_v)>\,0$};  
  \end{scope}
  \begin{scope}[shift={(0,-2)}]%, scale={.45}]
   \node[ball,text width=.18cm,fill,color=black,label=below:{\footnotesize $v\phantom{'}$}] at (0,0) (v1) {};
   \node[ball,text width=.18cm,fill,color=black,label=below:{\footnotesize $v'$},right=1.5cm of v1.east] (v2) {};  
   \draw[-,thick,color=black] (v1.east) edge node [text width=.18cm,below=.1cm,midway] {\footnotesize $e$} (v2.west);
   \node[very thick, cross=4pt, rotate=0, color=blue, right=.1cm of v1.east]{};
   \coordinate (x) at ($(v1)!0.5!(v2)$);
   \node[right=1.5cm of v2, scale=.9] (lb1) {$\mathcal{W}\cdot({\bf x}_v+{\bf y}_e-{\bf x}_{v'})>\,0$};  
  \end{scope}
 \end{tikzpicture}
\end{equation*}
Given a graph $\mathcal{G}$ and a certain subgraph $\mathfrak{g}$, it is easy to check that the hyperplane identifying a facet is given by $\mathcal{W}\,=\,\sum_{v\in\mathfrak{g}}\tilde{x}_v{\bf \tilde{X}}_v+\sum_{e\in\mathcal{E}_{\mathfrak{g}}^{\mbox{\tiny ext}}}\tilde{y}_e{\bf \tilde{Y}}_e$, $\mathcal{E}_{\mathfrak{g}}^{\mbox{\tiny ext}}$ being the set of edges external to $\mathfrak{g}$. If $\mathcal{G}\,=\,\mathfrak{g}$, then all the edges are marked in their middle and the hyperplane identifying this facet is given by $\mathcal{W}\,=\,\sum_{v}\tilde{x}_v{\bf \tilde{X}}_v$, {\it i.e.} it is the scattering facet. For more general facets, we have
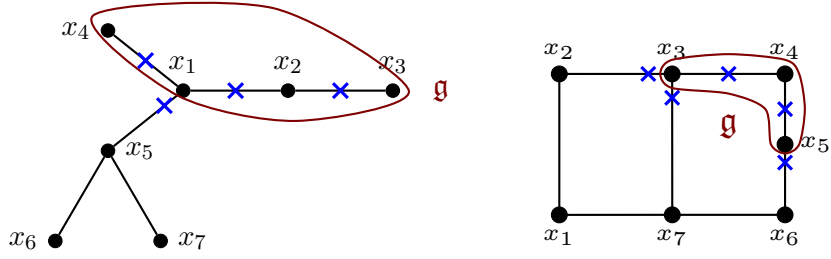
\begin{equation*}
  \begin{tikzpicture}[ball/.style = {circle, draw, align=center, anchor=north, inner sep=0}, cross/.style={cross out, draw, minimum size=2*(#1-\pgflinewidth), inner sep=0pt, outer sep=0pt}, scale=1, transform shape]
   \begin{scope}
    \node[ball,text width=.18cm,fill,color=black,label=above:{$x_1$}] at (0,0) (x1) {};    
    \node[ball,text width=.18cm,fill,color=black,right=1.2cm of x1.east, label=above:{$x_2$}] (x2) {};    
    \node[ball,text width=.18cm,fill,color=black,right=1.2cm of x2.east, label=above:{$x_3$}] (x3) {};
    \node[ball,text width=.18cm,fill,color=black, label=left:{$x_4$}] at (-1,.8) (x4) {};    
    \node[ball,text width=.18cm,fill,color=black, label=right:{$x_5$}] at (-1,-.8) (x5) {};    
    \node[ball,text width=.18cm,fill,color=black, label=left:{$x_6$}] at (-1.7,-2) (x6) {};    
    \node[ball,text width=.18cm,fill,color=black, label=right:{$x_7$}] at (-.3,-2) (x7) {};

    \node[above=.35cm of x5.north] (ref2) {};
    \coordinate (Int2) at (intersection of x5--x1 and ref2--x2);  

    \coordinate (t1) at (x3.east);
    \coordinate (t2) at (x4.west);
    \coordinate (t3) at (x1.south west);
    \coordinate (t4) at (x2.south);

    \draw[-,thick,color=black] (x1) -- (x2) -- (x3); 
    \draw[-,thick,color=black] (x1) -- (x4);
    \draw[-,thick,color=black] (x5) -- (x1);
    \draw[-,thick,color=black] (x5) -- (x7);   
    \draw[-,thick,color=black] (x5) -- (x6); 
    \draw[red!50!black, thick] plot [smooth cycle] coordinates {(3,-.1) (1.2,1) (-1.2,.9) (t3) (1.5,-.5)};
    \node[color=red!50!black,right=.3cm of x3.east] {\large $\mathfrak{g}$}; 

    \coordinate (m1) at ($(x1)!0.5!(x4)$);
    \coordinate (m2) at ($(x1)!0.5!(x2)$);
    \coordinate (m3) at ($(x2)!0.5!(x3)$);
    \coordinate (e1) at ($(x1)!0.25!(x5)$);
    \node[very thick, cross=4pt, rotate=0, color=blue] at (m1) {};
    \node[very thick, cross=4pt, rotate=0, color=blue] at (m2) {};
    \node[very thick, cross=4pt, rotate=0, color=blue] at (m3) {};  
    \node[very thick, cross=4pt, rotate=0, color=blue] at (e1) {};  
   \end{scope}
   \begin{scope}[shift={(5,-1.75)}, scale={1.5}, transform shape]
   \coordinate[label=below:{\tiny $x_1$}] (v1) at (0,0);
   \coordinate[label=above:{\tiny $x_2$}] (v2) at ($(v1)+(0,1.25)$);
   \coordinate[label=above:{\tiny $x_3$}] (v3) at ($(v2)+(1,0)$);
   \coordinate[label=above:{\tiny $x_4$}] (v4) at ($(v3)+(1,0)$);
   \coordinate[label=right:{\tiny $x_5$}] (v5) at ($(v4)-(0,.625)$);   
   \coordinate[label=below:{\tiny $x_6$}] (v6) at ($(v5)-(0,.625)$);
   \coordinate[label=below:{\tiny $x_7$}] (v7) at ($(v6)-(1,0)$);
   \draw[thick] (v1) -- (v2) -- (v3) -- (v4) -- (v5) -- (v6) -- (v7) -- cycle;
   \draw[thick] (v3) -- (v7);
   \draw[fill=black] (v1) circle (2pt);
   \draw[fill=black] (v2) circle (2pt);
   \draw[fill=black] (v3) circle (2pt);
   \draw[fill=black] (v4) circle (2pt);
   \draw[fill=black] (v5) circle (2pt);
   \draw[fill=black] (v6) circle (2pt);
   \draw[fill=black] (v7) circle (2pt);   
   \coordinate (v12) at ($(v1)!0.5!(v2)$);   
   \coordinate (v23) at ($(v2)!0.5!(v3)$);
   \coordinate (v34) at ($(v3)!0.5!(v4)$);
   \coordinate (v45) at ($(v4)!0.5!(v5)$);   
   \coordinate (v56) at ($(v5)!0.5!(v6)$);   
   \coordinate (v67) at ($(v6)!0.5!(v7)$);
   \coordinate (v71) at ($(v7)!0.5!(v1)$);   
   \coordinate (v37) at ($(v3)!0.5!(v7)$);   
   \node[very thick, cross=4pt, rotate=0, color=blue, scale=.625] at (v34) {};
   \node[very thick, cross=4pt, rotate=0, color=blue, scale=.625] at (v45) {};
   \node[very thick, cross=4pt, rotate=0, color=blue, scale=.625, left=.15cm of v3] (v3l) {};
   \node[very thick, cross=4pt, rotate=0, color=blue, scale=.625, below=.15cm of v3] (v3b) {};   
   \node[very thick, cross=4pt, rotate=0, color=blue, scale=.625, below=.1cm of v5] (v5b){};
   \coordinate (a) at ($(v3l)!0.5!(v3)$);
   \coordinate (b) at ($(v3)+(0,.125)$);
   \coordinate (c) at ($(v34)+(0,.175)$);
   \coordinate (d) at ($(v4)+(0,.125)$);
   \coordinate (e) at ($(v4)+(.125,0)$);
   \coordinate (f) at ($(v45)+(.175,0)$);
   \coordinate (g) at ($(v5)+(.125,0)$);
   \coordinate (h) at ($(v5b)!0.5!(v5)$);
   \coordinate (i) at ($(v5)-(.125,0)$);
   \coordinate (j) at ($(v45)-(.175,0)$);
   \coordinate (k) at ($(v34)-(0,.175)$);
   \coordinate (l) at ($(v3)-(0,.125)$);
   \draw [thick, red!50!black] plot [smooth cycle] coordinates {(a) (b) (c) (d) (e) (f) (g) (h) (i) (j) (k) (l)};
   \node[below=.05cm of k, color=red!50!black] {\footnotesize $\displaystyle\mathfrak{g}$};   
  \end{scope}
 \end{tikzpicture}  
\end{equation*}
We can equivalently indicate the vertices {\it belonging} to the facet by marking them with a circle
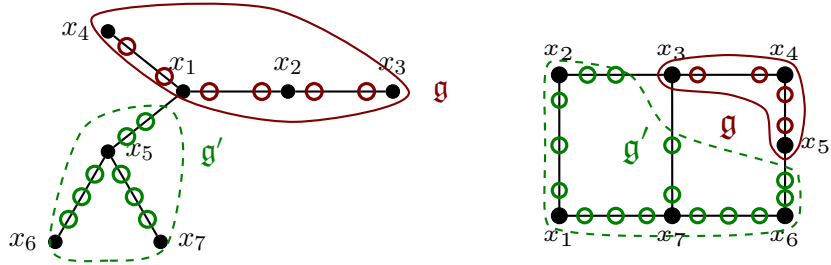
\begin{equation*}
  \begin{tikzpicture}[ball/.style = {circle, draw, align=center, anchor=north, inner sep=0}, cross/.style={cross out, draw, minimum size=2*(#1-\pgflinewidth), inner sep=0pt, outer sep=0pt}, scale=1, transform shape]
   \begin{scope}
    \node[ball,text width=.18cm,fill,color=black,label=above:{$x_1$}] at (0,0) (x1) {};    
    \node[ball,text width=.18cm,fill,color=black,right=1.2cm of x1.east, label=above:{$x_2$}] (x2) {};    
    \node[ball,text width=.18cm,fill,color=black,right=1.2cm of x2.east, label=above:{$x_3$}] (x3) {};
    \node[ball,text width=.18cm,fill,color=black, label=left:{$x_4$}] at (-1,.8) (x4) {};    
    \node[ball,text width=.18cm,fill,color=black, label=right:{$x_5$}] at (-1,-.8) (x5) {};    
    \node[ball,text width=.18cm,fill,color=black, label=left:{$x_6$}] at (-1.7,-2) (x6) {};    
    \node[ball,text width=.18cm,fill,color=black, label=right:{$x_7$}] at (-.3,-2) (x7) {};

    \node[above=.35cm of x5.north] (ref2) {};
    \coordinate (Int2) at (intersection of x5--x1 and ref2--x2);  

    \coordinate (t1) at (x3.east);
    \coordinate (t2) at (x4.west);
    \coordinate (t3) at (x1.south west);
    \coordinate (t4) at (x2.south);

    \draw[-,thick,color=black] (x1) -- (x2) -- (x3); 
    \draw[-,thick,color=black] (x1) -- (x4);
    \draw[-,thick,color=black] (x5) -- (x1);
    \draw[-,thick,color=black] (x5) -- (x7);   
    \draw[-,thick,color=black] (x5) -- (x6); 
    \draw[red!50!black, thick] plot [smooth cycle] coordinates {(3,-.1) (1.2,1) (-1.2,.9) (t3) (1.5,-.5)};
    \node[color=red!50!black,right=.3cm of x3.east] {\large $\mathfrak{g}$}; 

    \coordinate (m1a) at ($(x1)!0.25!(x4)$);
    \coordinate (m1b) at ($(x1)!0.75!(x4)$);    
    \coordinate (m2a) at ($(x1)!0.25!(x2)$);
    \coordinate (m2b) at ($(x1)!0.75!(x2)$);
    \coordinate (m3a) at ($(x2)!0.25!(x3)$);
    \coordinate (m3b) at ($(x2)!0.75!(x3)$);
    \draw[very thick, color=red!50!black] (m1a) circle (3pt);
    \draw[very thick, color=red!50!black] (m1b) circle (3pt);
    \draw[very thick, color=red!50!black] (m2a) circle (3pt);
    \draw[very thick, color=red!50!black] (m2b) circle (3pt);
    \draw[very thick, color=red!50!black] (m3a) circle (3pt);
    \draw[very thick, color=red!50!black] (m3b) circle (3pt);

    \coordinate (m5a) at ($(x5)!0.25!(x6)$);
    \coordinate (m5b) at ($(x5)!0.25!(x7)$);    
    \coordinate (m6a) at ($(x6)!0.25!(x5)$);
    \coordinate (m6b) at ($(x6)!0.50!(x5)$);
    \coordinate (m7a) at ($(x7)!0.25!(x5)$);
    \coordinate (m7b) at ($(x7)!0.50!(x5)$);
    \draw[very thick, color=green!50!black] (m5a) circle (3pt);
    \draw[very thick, color=green!50!black] (m5b) circle (3pt);
    \draw[very thick, color=green!50!black] (m6a) circle (3pt);
    \draw[very thick, color=green!50!black] (m6b) circle (3pt);
    \draw[very thick, color=green!50!black] (m7a) circle (3pt);
    \draw[very thick, color=green!50!black] (m7b) circle (3pt);
    \coordinate (a) at ($(x7.south east)$);
    \coordinate (b) at ($(x6)!0.5!(x7)-(0,.25)$);
    \coordinate (c) at ($(x6.south west)$);
    \coordinate (d) at ($(x5.north west)+(-.25,.25)$);
    \coordinate (e) at ($(x1)!0.25!(x5)$);
    \coordinate (f) at ($(x5)+(1,0)$);
    \draw [dashed,thick, green!50!black] plot [smooth cycle] coordinates {(a) (b) (c) (d) (e) (f)}; 
    \node[color=green!50!black,right=1cm of x5] {\large $\mathfrak{g}'$};  
   
    \coordinate (m15c) at ($(x1)!0.50!(x5)$);
    \coordinate (m15b) at ($(x1)!0.75!(x5)$);
    \draw[very thick, color=green!50!black] (m15c) circle (3pt);
    \draw[very thick, color=green!50!black] (m15b) circle (3pt);    
   \end{scope}
   \begin{scope}[shift={(5,-1.75)}, scale={1.5}, transform shape]
    \coordinate[label=below:{\tiny $x_1$}] (v1) at (0,0);
    \coordinate[label=above:{\tiny $x_2$}] (v2) at ($(v1)+(0,1.25)$);
    \coordinate[label=above:{\tiny $x_3$}] (v3) at ($(v2)+(1,0)$);
    \coordinate[label=above:{\tiny $x_4$}] (v4) at ($(v3)+(1,0)$);
    \coordinate[label=right:{\tiny $x_5$}] (v5) at ($(v4)-(0,.625)$);   
    \coordinate[label=below:{\tiny $x_6$}] (v6) at ($(v5)-(0,.625)$);
    \coordinate[label=below:{\tiny $x_7$}] (v7) at ($(v6)-(1,0)$);
    \draw[thick] (v1) -- (v2) -- (v3) -- (v4) -- (v5) -- (v6) -- (v7) -- cycle;
    \draw[thick] (v3) -- (v7);
    \draw[fill=black] (v1) circle (2pt);
    \draw[fill=black] (v2) circle (2pt);
    \draw[fill=black] (v3) circle (2pt);
    \draw[fill=black] (v4) circle (2pt);
    \draw[fill=black] (v5) circle (2pt);
    \draw[fill=black] (v6) circle (2pt);
    \draw[fill=black] (v7) circle (2pt);   
    \coordinate (v12) at ($(v1)!0.5!(v2)$);   
    \coordinate (v23) at ($(v2)!0.5!(v3)$);
    \coordinate (v34) at ($(v3)!0.5!(v4)$);
    \coordinate (v45) at ($(v4)!0.5!(v5)$);   
    \coordinate (v56) at ($(v5)!0.5!(v6)$);   
    \coordinate (v67) at ($(v6)!0.5!(v7)$);
    \coordinate (v71) at ($(v7)!0.5!(v1)$);   
    \coordinate (v37) at ($(v3)!0.5!(v7)$);
    \node[ball,text width=.20cm,very thick,color=red!50!black,right=.15cm of v3, scale=.625] {};   
    \node[ball,text width=.20cm,very thick,color=red!50!black,left=.15cm of v4, scale=.625] {};
    \node[ball,text width=.20cm,very thick,color=red!50!black,below=.1cm of v4, scale=.625] {};
    \node[ball,text width=.20cm,very thick,color=red!50!black,above=.1cm of v5, scale=.625] {};
    \draw[very thick,color=green!50!black] ($(v2)!0.25!(v3)$) circle (1.875pt);
    \draw[very thick,color=green!50!black] ($(v3)!0.85!(v7)$) circle (1.875pt);
    \draw[very thick,color=green!50!black] ($(v5)!0.75!(v6)$) circle (1.875pt);
    \draw[very thick,color=green!50!black] (v23) circle (1.875pt);
    \draw[very thick,color=green!50!black] (v37) circle (1.875pt);
    \draw[very thick,color=green!50!black] (v56) circle (1.875pt);
    \draw[very thick,color=green!50!black] (v12) circle (1.875pt);
    \draw[very thick,color=green!50!black] (v71) circle (1.875pt);
    \draw[very thick,color=green!50!black] (v67) circle (1.875pt);
    \node[ball,text width=.20cm,very thick,color=green!50!black,below=.15cm of v2, scale=.625] {};
    \node[ball,text width=.20cm,very thick,color=green!50!black,above=.15cm of v1, scale=.625] {};
    \node[ball,text width=.20cm,very thick,color=green!50!black,right=.15cm of v1, scale=.625] {};
    \node[ball,text width=.20cm,very thick,color=green!50!black,left=.15cm of v7, scale=.625] {};
    \node[ball,text width=.20cm,very thick,color=green!50!black,right=.15cm of v7, scale=.625] {};
    \node[ball,text width=.20cm,very thick,color=green!50!black,left=.15cm of v6, scale=.625] {}; 
    \coordinate (a) at ($(v3)-(.125,0)$);
    \coordinate (b) at ($(v3)+(0,.125)$);
    \coordinate (c) at ($(v34)+(0,.175)$);
    \coordinate (d) at ($(v4)+(0,.125)$);
    \coordinate (e) at ($(v4)+(.125,0)$);
    \coordinate (f) at ($(v45)+(.175,0)$);
    \coordinate (g) at ($(v5)+(.125,0)$);
    \coordinate (h) at ($(v5)-(0,.125)$);
    \coordinate (i) at ($(v5)-(.125,0)$);
    \coordinate (j) at ($(v45)-(.175,0)$);
    \coordinate (k) at ($(v34)-(0,.175)$);
    \coordinate (l) at ($(v3)-(0,.125)$);
    \draw [thick, red!50!black] plot [smooth cycle] coordinates {(a) (b) (c) (d) (e) (f) (g) (h) (i) (j) (k) (l)};
    \node[below=.05cm of k, color=red!50!black] {\footnotesize $\displaystyle\mathfrak{g}$};
    \coordinate (n) at ($(v2)+(0,.125)$);
    \coordinate (o) at ($(v23)+(3.5pt,0)$);
    \coordinate (p) at ($(v37)+(0,3.5pt)$);
    \coordinate (q) at ($(v56)+(0,3.5pt)$);
    \coordinate (r) at ($(v56)+(3.5pt,0)$);
    \coordinate (tb) at ($(v6)+(.125,0)$);
    \coordinate (tc) at ($(v6)-(0,.125)$);
    \coordinate (td) at ($(v67)-(0,.175)$);
    \coordinate (u) at ($(v71)-(0,.175)$);
    \coordinate (w) at ($(v1)-(0,.125)$);
    \coordinate (x) at ($(v1)-(.125,0)$);
    \coordinate (y) at ($(v12)-(.125,0)$);
    \coordinate (z) at ($(v2)-(.125,0)$);
    \draw [dashed, thick, green!50!black] plot [smooth cycle] coordinates {(n) (o) (p) (q) (r) (tb) (tc) (td) (u) (w) (x) (y) (z)};   
    \node[left=.05cm of v37, color=green!50!black] {\footnotesize $\displaystyle\mathfrak{g}'$};     
   \end{scope}
 \end{tikzpicture}  
\end{equation*}
The subgraph $\mathfrak{g}$ corresponds to $\sum_{v\in\mathfrak{g}}x_v+\sum_{e\in\mathcal{E}_{\mathfrak{g}}^{\mbox{\tiny ext}}}y_e\,\longrightarrow\,0$, with the vertices
$
 \begin{tikzpicture}[ball/.style = {circle, draw, align=center, anchor=north, inner sep=0}]
   \node[ball,text width=.18cm,thick,color=red!50!black] at (0,0) {};     
 \end{tikzpicture}
$
identifying a lower-dimensional scattering facet. The vertices
$
 \begin{tikzpicture}[ball/.style = {circle, draw, align=center, anchor=north, inner sep=0}]
   \node[ball,text width=.18cm,thick,color=green!50!black] at (0,0) {};     
 \end{tikzpicture}
$
are related to {\it a facet} of a smaller wavefunction. Hence, the canonical form for this facet is given by
\begin{equation}\label{eq:CFGF}
 \Omega\:=\:\mathcal{A}[\mathfrak{g}]\times\hat{\Psi}_{\mathfrak{g}'}. 
\end{equation}
Similarly, we can consider any type of subgraph, and the vertex structure of the related facet will always be a product of a scattering facet times $\tilde{n}_e$ facets of smaller wavefunctions, $\tilde{n}_e$ being the number of edges connecting $\mathfrak{g}$ to disconnected subgraphs:
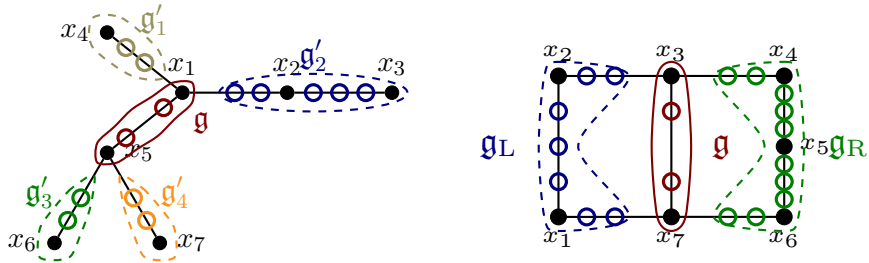
\begin{equation*}
  \begin{tikzpicture}[ball/.style = {circle, draw, align=center, anchor=north, inner sep=0}, cross/.style={cross out, draw, minimum size=2*(#1-\pgflinewidth), inner sep=0pt, outer sep=0pt}, scale=1, transform shape]
   \begin{scope}
    \node[ball,text width=.18cm,fill,color=black,label=above:{$x_1$}] at (0,0) (x1) {};    
    \node[ball,text width=.18cm,fill,color=black,right=1.2cm of x1.east, label=above:{$x_2$}] (x2) {};    
    \node[ball,text width=.18cm,fill,color=black,right=1.2cm of x2.east, label=above:{$x_3$}] (x3) {};
    \node[ball,text width=.18cm,fill,color=black, label=left:{$x_4$}] at (-1,.8) (x4) {};    
    \node[ball,text width=.18cm,fill,color=black, label=right:{$x_5$}] at (-1,-.8) (x5) {};    
    \node[ball,text width=.18cm,fill,color=black, label=left:{$x_6$}] at (-1.7,-2) (x6) {};    
    \node[ball,text width=.18cm,fill,color=black, label=right:{$x_7$}] at (-.3,-2) (x7) {};

    \node[above=.35cm of x5.north] (ref2) {};
    \coordinate (Int2) at (intersection of x5--x1 and ref2--x2);  

    \coordinate (t1) at (x3.east);
    \coordinate (t2) at (x4.west);
    \coordinate (t3) at (x1.south west);
    \coordinate (t4) at (x2.south);

    \draw[-,thick,color=black] (x1) -- (x2) -- (x3); 
    \draw[-,thick,color=black] (x1) -- (x4);
    \draw[-,thick,color=black] (x5) -- (x1);
    \draw[-,thick,color=black] (x5) -- (x7);   
    \draw[-,thick,color=black] (x5) -- (x6); 

    \coordinate (m1a) at ($(x1)!0.50!(x4)$);
    \coordinate (m1b) at ($(x1)!0.75!(x4)$);    
    \coordinate (m2a) at ($(x1)!0.50!(x2)$);
    \coordinate (m2b) at ($(x1)!0.75!(x2)$);
    \coordinate (m3a) at ($(x2)!0.25!(x3)$);
    \coordinate (m3b) at ($(x2)!0.75!(x3)$);
    \coordinate (m3c) at ($(x2)!0.50!(x3)$);
    
    \draw[very thick, color=yellow!50!black] (m1a) circle (3pt);
    \draw[very thick, color=yellow!50!black] (m1b) circle (3pt);
    \coordinate (a) at ($(x4.north west)+(-.125,.125)$);
    \coordinate (b) at ($(x4.north east)+(.125,.125)$);
    \coordinate (c) at ($(x4)!0.50!(x1)+(.175,.175)$);
    \coordinate (d) at ($(x4)!0.75!(x1)$);
    \coordinate (e) at ($(x4)!0.50!(x1)-(.175, .175)$);
    \coordinate (f) at ($(x4.south west)-(.125, .125)$);
    \draw [dashed,thick, yellow!50!black] plot [smooth cycle] coordinates {(a) (b) (c) (d) (e) (f)}; 
    \node[color=yellow!50!black] at ($(x4)!0.25!(x1)+(.375,.375)$) {\large $\mathfrak{g}_1'$};

    \draw[very thick, color=blue!50!black] (m2a) circle (3pt);
    \draw[very thick, color=blue!50!black] (m2b) circle (3pt);
    \draw[very thick, color=blue!50!black] (m3a) circle (3pt);
    \draw[very thick, color=blue!50!black] (m3b) circle (3pt);
    \draw[very thick, color=blue!50!black] (m3c) circle (3pt);
    \draw[dashed,thick, color=blue!50!black] ($(m2a)!0.50!(x3)$) ellipse (1.25cm and .25cm);
    \node[color=blue!50!black] at ($(m2a)!0.50!(x3)+(0,.5)$) {\large $\mathfrak{g}_2'$};    

    \coordinate (m5a) at ($(x5)!0.25!(x6)$);
    \coordinate (m5b) at ($(x5)!0.25!(x7)$);    
    \coordinate (m6a) at ($(x6)!0.25!(x5)$);
    \coordinate (m6b) at ($(x6)!0.50!(x5)$);
    \coordinate (m7a) at ($(x7)!0.25!(x5)$);
    \coordinate (m7b) at ($(x7)!0.50!(x5)$);
    \draw[very thick, color=green!50!black] (m6a) circle (3pt);
    \draw[very thick, color=green!50!black] (m6b) circle (3pt);
    \coordinate (ag) at ($(m6a)+(-.175,+.175)$);
    \coordinate (bg) at ($(x6.north west)+(-.125,.125)$);
    \coordinate (cg) at ($(x6.south west)-(.125,.125)$);
    \coordinate (dg) at ($(x6.south east)+(.125,-.125)$);
    \coordinate (eg) at ($(m6a)+(.175,-.175)$);
    \coordinate (fg) at ($(m6b)+(-.125,.125)$);
    \coordinate (gg) at ($(m6b)+(.125,-.125)$);
    \draw [dashed,thick, green!50!black] plot [smooth cycle] coordinates {(m5a) (fg) (ag) (bg) (cg) (dg) (eg) (gg)}; 
    \node[color=green!50!black] at ($(m6a)+(-.375,.375)$) {\large $\mathfrak{g}_3'$};

    \draw[very thick, color=yellow!50!red] (m7a) circle (3pt);
    \draw[very thick, color=yellow!50!red] (m7b) circle (3pt);
    \coordinate (ao) at ($(m7a)+(.175,+.175)$);
    \coordinate (bo) at ($(x7.north east)+(.125,.125)$);
    \coordinate (co) at ($(x7.south east)+(.125,-.125)$);
    \coordinate (do) at ($(x7.south west)+(-.125,-.125)$);
    \coordinate (eo) at ($(m7a)+(-.175,-.175)$);
    \coordinate (fo) at ($(m7b)+(.125,.125)$);
    \coordinate (go) at ($(m7b)+(-.125,-.125)$);
    \draw [dashed,thick, yellow!50!red] plot [smooth cycle] coordinates {(m5b) (fo) (ao) (bo) (co) (do) (eo) (go)}; 
    \node[color=yellow!50!red] at ($(m7a)+(.375,.375)$) {\large $\mathfrak{g}_4'$};
  
    \coordinate (m15c) at ($(x1)!0.25!(x5)$);
    \coordinate (m15b) at ($(x1)!0.75!(x5)$);
    \draw[very thick, color=red!50!black] (m15c) circle (3pt);
    \draw[very thick, color=red!50!black] (m15b) circle (3pt);    
    \coordinate (ar) at ($(x1.north east)+(+.0625,+.0625)$);
    \coordinate (kr) at ($(x1.north west)+(-.0625,+.0625)$);
    \coordinate (lr) at ($(x1.south east)+(+.0625,-.0625)$);
    \coordinate (br) at ($(m15c)+(-.15,+.15)$);
    \coordinate (cr) at ($(x1)!0.50!(x5)+(-.175,+.175)$);
    \coordinate (dr) at ($(m15b)+(-.15,+.15)$);
    \coordinate (er) at ($(x5.north west)+(-.0625,+.0625)$);
    \coordinate (fr) at ($(x5.south west)+(-.0625,-.0625)$);
    \coordinate (gr) at ($(x5.south east)+(+.0625,-.0625)$);
    \coordinate (hr) at ($(m15b)+(.15,-.15)$);
    \coordinate (ir) at ($(x1)!0.50!(x5)+(+.175,-.175)$);
    \coordinate (jr) at ($(m15c)+(.15,-.15)$);    
    \draw [thick, red!50!black] plot [smooth cycle] coordinates {(ar) (kr) (br) (cr) (dr) (er) (fr) (gr) (hr) (ir) (jr) (lr)}; 
    \node[color=red!50!black] at ($(x1)!0.50!(x5)+(.75,0)$) {\large $\mathfrak{g}$};
   \end{scope}
   \begin{scope}[shift={(5,-1.75)}, scale={1.5}, transform shape]
   \coordinate[label=below:{\tiny $x_1$}] (v1) at (0,0);
   \coordinate[label=above:{\tiny $x_2$}] (v2) at ($(v1)+(0,1.25)$);
   \coordinate[label=above:{\tiny $x_3$}] (v3) at ($(v2)+(1,0)$);
   \coordinate[label=above:{\tiny $x_4$}] (v4) at ($(v3)+(1,0)$);
   \coordinate[label=right:{\tiny $x_5$}] (v5) at ($(v4)-(0,.625)$);   
   \coordinate[label=below:{\tiny $x_6$}] (v6) at ($(v5)-(0,.625)$);
   \coordinate[label=below:{\tiny $x_7$}] (v7) at ($(v6)-(1,0)$);
   \draw[thick] (v1) -- (v2) -- (v3) -- (v4) -- (v5) -- (v6) -- (v7) -- cycle;
   \draw[thick] (v3) -- (v7);
   \draw[fill=black] (v1) circle (2pt);
   \draw[fill=black] (v2) circle (2pt);
   \draw[fill=black] (v3) circle (2pt);
   \draw[fill=black] (v4) circle (2pt);
   \draw[fill=black] (v5) circle (2pt);
   \draw[fill=black] (v6) circle (2pt);
   \draw[fill=black] (v7) circle (2pt);   
   \coordinate (v12) at ($(v1)!0.5!(v2)$);   
   \coordinate (v23) at ($(v2)!0.5!(v3)$);
   \coordinate (v34) at ($(v3)!0.5!(v4)$);
   \coordinate (v45) at ($(v4)!0.5!(v5)$);   
   \coordinate (v56) at ($(v5)!0.5!(v6)$);   
   \coordinate (v67) at ($(v6)!0.5!(v7)$);
   \coordinate (v71) at ($(v7)!0.5!(v1)$);   
   \coordinate (v37) at ($(v3)!0.5!(v7)$);
   \draw[very thick,color=red!50!black] ($(v3)!0.25!(v7)$) circle (2pt);
   \draw[very thick,color=green!50!black] ($(v3)!0.75!(v4)$) circle (2pt);
   \draw[very thick,color=green!50!black] (v34) circle (2pt);
   \draw[very thick,color=green!50!black] ($(v4)!0.25!(v5)$) circle (2pt);
   \draw[very thick,color=green!50!black] ($(v45)$) circle (2pt);
   \draw[very thick,color=green!50!black] ($(v4)!0.75!(v5)$) circle (2pt);
   \draw[very thick,color=blue!50!black] ($(v2)!0.25!(v3)$) circle (2pt);
   \draw[very thick,color=blue!50!black] (v23) circle (2pt);
   \draw[very thick,color=red!50!black] ($(v3)!0.75!(v7)$) circle (2pt);
   \draw[very thick,color=green!50!black] ($(v5)!0.75!(v6)$) circle (2pt);
   \draw[very thick,color=blue!50!black] ($(v1)!0.75!(v2)$) circle (2pt);
   \draw[very thick,color=blue!50!black] (v12) circle (2pt);
   \draw[very thick,color=blue!50!black] ($(v1)!0.25!(v2)$) circle (2pt);
   \draw[very thick,color=blue!50!black] ($(v1)!0.25!(v7)$) circle (2pt);
   \draw[very thick,color=blue!50!black] (v71) circle (2pt);
   \draw[very thick,color=green!50!black] (v56) circle (2pt);   
   \draw[very thick,color=green!50!black] ($(v5)!0.25!(v6)$) circle (2pt);
   \draw[very thick,color=green!50!black] ($(v6)!0.25!(v7)$) circle (2pt); 
   \draw[very thick,color=green!50!black] (v67) circle (2pt);     
   \coordinate (a) at ($(v3)-(.125,0)$);
   \coordinate (b) at ($(v3)+(0,.125)$);
   \coordinate (c) at ($(v3)+(.125,0)$);
   \coordinate (d) at ($(v37)+(.175,0)$);
   \coordinate (e) at ($(v7)+(.125,0)$);
   \coordinate (f) at ($(v7)-(0,.125)$);
   \coordinate (g) at ($(v7)-(.125,0)$);
   \coordinate (h) at ($(v37)-(.175,0)$);
   \draw [thick, red!50!black] plot [smooth cycle] coordinates {(a) (b) (c) (d) (e) (f) (g) (h)};
   \node[right=.05cm of d, color=red!50!black] {\footnotesize $\displaystyle\mathfrak{g}$};
   \coordinate (al) at ($(v2)-(.125,0)$);
   \coordinate (bl) at ($(v2)+(0,.125)$);
   \coordinate (cl) at ($(v23)+(.125,0)$);
   \coordinate (dl) at ($(v12)+(.175,0)$);
   \coordinate (el) at ($(v71)+(.125,0)$);
   \coordinate (fl) at ($(v1)-(0,.125)$);
   \coordinate (gl) at ($(v1)-(.125,0)$);
   \coordinate (hl) at ($(v12)-(.175,0)$);
   \draw [dashed, thick, blue!50!black] plot [smooth cycle] coordinates {(al) (bl) (cl) (dl) (el) (fl) (gl) (hl)};
   \node[left=.05cm of hl, color=blue!50!black] {\footnotesize $\displaystyle\mathfrak{g}_{\mbox{\tiny L}}$};
   \coordinate (ar) at ($(v34)-(.125,0)$);
   \coordinate (br) at ($(v4)+(0,.125)$);
   \coordinate (cr) at ($(v4)+(.125,0)$);
   \coordinate (dr) at ($(v5)+(.175,0)$);
   \coordinate (er) at ($(v6)+(.125,0)$);
   \coordinate (fr) at ($(v6)-(0,.125)$);
   \coordinate (gr) at ($(v67)-(.125,0)$);
   \coordinate (hr) at ($(v5)-(.175,0)$);
   \draw [dashed, thick, green!50!black] plot [smooth cycle] coordinates {(ar) (br) (cr) (dr) (er) (fr) (gr) (hr)};
   \node[right=.08cm of dr, color=green!50!black] {\footnotesize $\displaystyle\mathfrak{g}_{\mbox{\tiny R}}$};
  \end{scope}
 \end{tikzpicture}  
\end{equation*}
with the canonical form of such facets becoming
\begin{equation}\label{eq:CFGF2}
 \Omega\:=\:\mathcal{A}[\mathfrak{g}]\times\prod_{j=1}^{\tilde{n}_e}\hat{\Psi}_{\mathfrak{g}'_j}. 
\end{equation}
However, this is not the end of the story. First of all, it has to be noticed that there is a number of facets which are {\it isomorphic} to the scattering facet: each of them corresponds to a subgraph whose external edges connect it to subgraphs made out by a single vertex only. They have $2n_e\,=\,n_e+n_v-1+L$ vertices, where $n_e$, $n_v$ and $L$ indicated the number of edges, vertices and loops of the graph associated to the cosmological polytope. As the scattering facet, they have $L$ vertices more than a simplex -- they are a simplex just for $L\,=\,0$ (tree level). We are going to discuss the facets other than the scattering facet for the cosmological polytopes at tree and loop level separately.

%%%%%%%%%%%%%%%%%%%%%%%
%%%%%%%%%%%%%%%%%%%%%%%

\subsection{Facets of the tree cosmological polytopes}\label{subsec:TCP}

Let us begin with considering the structure of those facets in the $L\,=\,0$ case with $2n_e$ vertices. Being each of them a simplex, any of their facets is a simplex as well.

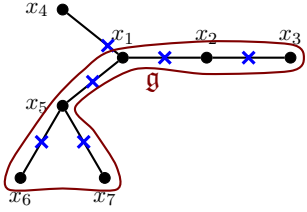
\begin{wrapfigure}{l}{4.5cm}
 \begin{tikzpicture}[ball/.style = {circle, draw, align=center, anchor=north, inner sep=0}, cross/.style={cross out, draw, minimum size=2*(#1-\pgflinewidth), inner sep=0pt, outer sep=0pt}, scale=.8, transform shape]
  \begin{scope}
   \node[ball,text width=.18cm,fill,color=black,label=above:{$x_1$}] at (0,0) (x1) {};    
   \node[ball,text width=.18cm,fill,color=black,right=1.2cm of x1.east, label=above:{$x_2$}] (x2) {};    
   \node[ball,text width=.18cm,fill,color=black,right=1.2cm of x2.east, label=above:{$x_3$}] (x3) {};
   \node[ball,text width=.18cm,fill,color=black, label=left:{$x_4$}] at (-1,.8) (x4) {};    
   \node[ball,text width=.18cm,fill,color=black, label=left:{$x_5$}] at (-1,-.8) (x5) {};    
   \node[ball,text width=.18cm,fill,color=black, label=below:{$x_6$}] at (-1.7,-2) (x6) {};    
   \node[ball,text width=.18cm,fill,color=black, label=below:{$x_7$}] at (-.3,-2) (x7) {};

   \node[above=.35cm of x5.north] (ref2) {};
   \coordinate (Int2) at (intersection of x5--x1 and ref2--x2);  

   \coordinate (t1) at (x3.east);
   \coordinate (t2) at (x4.west);
   \coordinate (t3) at (x1.south west);
   \coordinate (t4) at (x2.south);

   \draw[-,thick,color=black] (x1) -- (x2) -- (x3); 
   \draw[-,thick,color=black] (x1) -- (x4);
   \draw[-,thick,color=black] (x5) -- (x1);
   \draw[-,thick,color=black] (x5) -- (x7);   
   \draw[-,thick,color=black] (x5) -- (x6); 
%  \node[ball,text width=.18cm,thick,color=blue,left=-.09cm of Int2] {};

   \coordinate (x12) at ($(x1)!0.50!(x2)$);
   \coordinate (x23) at ($(x2)!0.50!(x3)$);
   \coordinate (x14) at ($(x1)!0.25!(x4)$);
   \coordinate (x15) at ($(x1)!0.50!(x5)$);
   \coordinate (x56) at ($(x5)!0.50!(x6)$);
   \coordinate (x57) at ($(x5)!0.50!(x7)$);
   \node[very thick, cross=4pt, rotate=0, color=blue] at (x12) {};
   \node[very thick, cross=4pt, rotate=0, color=blue] at (x23) {};
   \node[very thick, cross=4pt, rotate=0, color=blue] at (x14) {};
   \node[very thick, cross=4pt, rotate=0, color=blue] at (x15) {};
   \node[very thick, cross=4pt, rotate=0, color=blue] at (x56) {};
   \node[very thick, cross=4pt, rotate=0, color=blue] at (x57) {};

   \coordinate (a) at ($(x7.south east)+(.125,-.125)$);
   \coordinate (b) at ($(x7)!0.5!(x6)+(0,-.175)$);
   \coordinate (c) at ($(x6.south west)+(-.125,-.125)$);
   \coordinate (d) at ($(x6.north west)+(-.125,+.125)$);
   \coordinate (e) at ($(x6)!0.50!(x5)+(-.175,+.175)$);
   \coordinate (f) at ($(x5)+(-.125,+.125)$);
   \coordinate (g) at ($(x5)!0.50!(x1)+(-.175,+.175)$);
   \coordinate (h) at ($(x1.north west)+(-.0625,+.0625)$);
   \coordinate (i) at ($(x2.north)+(0,+.175)$);
   \coordinate (j) at ($(x3.north)+(0,+.125)$);
   \coordinate (k) at ($(x3.east)+(+.125,0)$);
   \coordinate (l) at ($(x3.south)+(0,-.125)$);
   \coordinate (m) at ($(x2.south)+(0,-.175)$);
   \coordinate (n) at ($(x1.south east)+(+.125,-.125)$);
   \coordinate (o) at ($(x5.east)+(+.125,0)$);
   \coordinate (p) at ($(x5)!0.5!(x7)+(+.175,+.175)$);
   \coordinate (q) at ($(x7.north east)+(+.125,+.125)$);
  
   \draw[thick, red!50!black] plot [smooth cycle] coordinates {(a) (b) (c) (d) (e) (f) (g) (h) (i) (j) (k) (l) (m) (n) (o) (p) (q)};
   \node[color=red!50!black, scale=1.25] at ($(x5)!0.50!(x1)+(1,0)$) {$\mathfrak{g}$};
  \end{scope}
 \end{tikzpicture}  
\end{wrapfigure}
For the sake of concreteness, let us consider the facet identified by the subgraph $\mathfrak{g}$ as in the figure here on the left. In general, the canonical form associated to a polytope living in $\mathbb{P}^{N-1}$ can be represented as a contour integral \cite{Arkani-Hamed:2017tmz}
\begin{equation}\label{eq:CFgF}
 \Omega\:\sim\:\int\prod_{j=1}^{\nu}\frac{dc_j}{c_j-i\varepsilon}\delta^{\mbox{\tiny $(N)$}}\left(\mathcal{Y}-\sum_{j=1}^{\nu}c_j{\bf V}^{\mbox{\tiny $(j)$}}\right),
\end{equation}
where $\nu$ is the number of vertices ${\bf V}$ of the polytope, and $\sim$ indicates the suppression of numerical factors which are irrelevant for the present discussion. The facet related to codimension-$1$ subgraphs such as $\mathfrak{g}$, admits such a representation, with $\nu\,=\,2n_e\,=\,n_e+n_v-1$ and $N\,=\,n_e+n_e-1$, and the integral \eqref{eq:CFgF} gets localised -- as already mentioned, these facets {\it are} simplices. For each edge of the graph, the facet related to $\mathfrak{g}$ has two vertices: for all edges but one, they correspond to open circle markings on either of the ends, while for the remaining one -- which is external with respect to $\mathfrak{g}$ -- they are related to one of its ends and to its middle. 

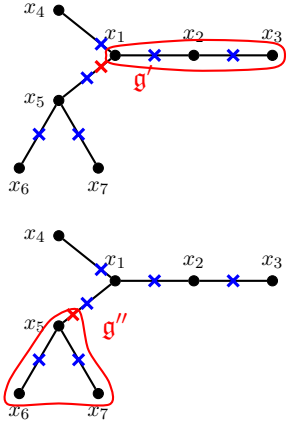
\begin{wrapfigure}{l}{4.5cm}
 \begin{tikzpicture}[ball/.style = {circle, draw, align=center, anchor=north, inner sep=0}, cross/.style={cross out, draw, minimum size=2*(#1-\pgflinewidth), inner sep=0pt, outer sep=0pt}, scale=.75, transform shape]
  \begin{scope}
   \node[ball,text width=.18cm,fill,color=black,label=above:{$x_1$}] at (0,0) (x1) {};    
   \node[ball,text width=.18cm,fill,color=black,right=1.2cm of x1.east, label=above:{$x_2$}] (x2) {};    
   \node[ball,text width=.18cm,fill,color=black,right=1.2cm of x2.east, label=above:{$x_3$}] (x3) {};
   \node[ball,text width=.18cm,fill,color=black, label=left:{$x_4$}] at (-1,.8) (x4) {};    
   \node[ball,text width=.18cm,fill,color=black, label=left:{$x_5$}] at (-1,-.8) (x5) {};    
   \node[ball,text width=.18cm,fill,color=black, label=below:{$x_6$}] at (-1.7,-2) (x6) {};    
   \node[ball,text width=.18cm,fill,color=black, label=below:{$x_7$}] at (-.3,-2) (x7) {};

   \node[above=.35cm of x5.north] (ref2) {};
   \coordinate (Int2) at (intersection of x5--x1 and ref2--x2);  

   \coordinate (t1) at (x3.east);
   \coordinate (t2) at (x4.west);
   \coordinate (t3) at (x1.south west);
   \coordinate (t4) at (x2.south);

   \draw[-,thick,color=black] (x1) -- (x2) -- (x3); 
   \draw[-,thick,color=black] (x1) -- (x4);
   \draw[-,thick,color=black] (x5) -- (x1);
   \draw[-,thick,color=black] (x5) -- (x7);   
   \draw[-,thick,color=black] (x5) -- (x6); 

   \coordinate (x12) at ($(x1)!0.50!(x2)$);
   \coordinate (x23) at ($(x2)!0.50!(x3)$);
   \coordinate (x14) at ($(x1)!0.25!(x4)$);
   \coordinate (x15) at ($(x1)!0.50!(x5)$);
   \coordinate (x56) at ($(x5)!0.50!(x6)$);
   \coordinate (x57) at ($(x5)!0.50!(x7)$);
   \coordinate (x15d) at ($(x1)!0.25!(x5)$);
   \node[very thick, cross=4pt, rotate=0, color=blue] at (x12) {};
   \node[very thick, cross=4pt, rotate=0, color=blue] at (x23) {};
   \node[very thick, cross=4pt, rotate=0, color=blue] at (x14) {};
   \node[very thick, cross=4pt, rotate=0, color=blue] at (x15) {};
   \node[very thick, cross=4pt, rotate=0, color=blue] at (x56) {};
   \node[very thick, cross=4pt, rotate=0, color=blue] at (x57) {};
   \node[very thick, cross=4pt, rotate=0, color=red] at (x15d) {};   

   \coordinate (g) at ($(x1.west)+(-.0625,0)$);
   \coordinate (h) at ($(x1.north)+(0,+.0625)$);
   \coordinate (i) at ($(x2.north)+(0,+.175)$);
   \coordinate (j) at ($(x3.north)+(0,+.125)$);
   \coordinate (k) at ($(x3.east)+(+.125,0)$);
   \coordinate (l) at ($(x3.south)+(0,-.125)$);
   \coordinate (m) at ($(x2.south)+(0,-.175)$);
   \coordinate (n) at ($(x1.south)+(0,-.0625)$);
  
   \draw[thick, red] plot [smooth cycle] coordinates {(g) (h) (i) (j) (k) (l) (m) (n)};
   \node[color=red, scale=1.25] at ($(x5)!0.50!(x1)+(1,0)$) {$\mathfrak{g}'$};

  \end{scope}
  \begin{scope}[shift={(0,-4)}, transform shape]
   \node[ball,text width=.18cm,fill,color=black,label=above:{$x_1$}] at (0,0) (x1) {};    
   \node[ball,text width=.18cm,fill,color=black,right=1.2cm of x1.east, label=above:{$x_2$}] (x2) {};    
   \node[ball,text width=.18cm,fill,color=black,right=1.2cm of x2.east, label=above:{$x_3$}] (x3) {};
   \node[ball,text width=.18cm,fill,color=black, label=left:{$x_4$}] at (-1,.8) (x4) {};    
   \node[ball,text width=.18cm,fill,color=black, label=left:{$x_5$}] at (-1,-.8) (x5) {};    
   \node[ball,text width=.18cm,fill,color=black, label=below:{$x_6$}] at (-1.7,-2) (x6) {};    
   \node[ball,text width=.18cm,fill,color=black, label=below:{$x_7$}] at (-.3,-2) (x7) {};

   \node[above=.35cm of x5.north] (ref2) {};
   \coordinate (Int2) at (intersection of x5--x1 and ref2--x2);  

   \coordinate (t1) at (x3.east);
   \coordinate (t2) at (x4.west);
   \coordinate (t3) at (x1.south west);
   \coordinate (t4) at (x2.south);

   \draw[-,thick,color=black] (x1) -- (x2) -- (x3); 
   \draw[-,thick,color=black] (x1) -- (x4);
   \draw[-,thick,color=black] (x5) -- (x1);
   \draw[-,thick,color=black] (x5) -- (x7);   
   \draw[-,thick,color=black] (x5) -- (x6); 

   \coordinate (x12) at ($(x1)!0.50!(x2)$);
   \coordinate (x23) at ($(x2)!0.50!(x3)$);
   \coordinate (x14) at ($(x1)!0.25!(x4)$);
   \coordinate (x15) at ($(x1)!0.50!(x5)$);
   \coordinate (x56) at ($(x5)!0.50!(x6)$);
   \coordinate (x57) at ($(x5)!0.50!(x7)$);
   \coordinate (x15u) at ($(x1)!0.75!(x5)$);   
   \node[very thick, cross=4pt, rotate=0, color=blue] at (x12) {};
   \node[very thick, cross=4pt, rotate=0, color=blue] at (x23) {};
   \node[very thick, cross=4pt, rotate=0, color=blue] at (x14) {};
   \node[very thick, cross=4pt, rotate=0, color=blue] at (x15) {};
   \node[very thick, cross=4pt, rotate=0, color=blue] at (x56) {};
   \node[very thick, cross=4pt, rotate=0, color=blue] at (x57) {};
   \node[very thick, cross=4pt, rotate=0, color=red] at (x15u) {};      

   \coordinate (a) at ($(x7.south east)+(.125,-.125)$);
   \coordinate (b) at ($(x7)!0.5!(x6)+(0,-.175)$);
   \coordinate (c) at ($(x6.south west)+(-.125,-.125)$);
   \coordinate (d) at ($(x6.north west)+(-.125,+.125)$);
   \coordinate (e) at ($(x6)!0.50!(x5)+(-.175,+.175)$);
   \coordinate (f) at ($(x5)+(-.125,+.125)$);
   \coordinate (g) at ($(x5)!0.35!(x1)$);
%   \coordinate (o) at ($(x5.east)+(+.125,0)$);
   \coordinate (p) at ($(x5)!0.5!(x7)+(+.175,+.175)$);
   \coordinate (q) at ($(x7.north east)+(+.125,+.125)$);
  
   \draw[thick, red] plot [smooth cycle] coordinates {(a) (b) (c) (d) (e) (f) (g) (p) (q)};
   \node[color=red, scale=1.25] at ($(x5)+(1,0)$) {$\mathfrak{g}''$};
  \end{scope}
 \end{tikzpicture}  
\end{wrapfigure}
The solutions for each $c_j$ is given by the hyperplane which {\it does not} contain the vertex ${\bf V}^{\mbox{\tiny $(j)$}}$ associated to it. Graphically, one can mark such a vertex with a red cross. For all the edges where these vertex can be related to either ends, then the two hyperplanes $\omega$ fixing the relevant $c_j$'s are given by the facet identified by the two markings are related to subgraphs $\mathfrak{g}'$ and $\mathfrak{g}''$:
\begin{equation}\label{eq:Pe1a}
 \begin{split}
  &\omega_{\mathfrak{g}'}\cdot\mathcal{Y}\:=\:y_e+y_{\bar{e}}+\sum_{v\in\mathfrak{g}'}x_v,\\
  &\omega_{\mathfrak{g}''}\cdot\mathcal{Y}\:=\:y_e+\sum_{v\in\mathfrak{g}''}x_v 
 \end{split}
\end{equation}
$e$ and $\bar{e}$ being the edge marked by the red cross and the one external to $\mathfrak{g}$. Notice that both \eqref{eq:Pe1a} live on the facet identified by $y_{\bar{e}}+\sum_{v\in\mathfrak{g}}x_v\,=\,0$, where they can be written in the Lorentz invariant form $y_e^2-\left(\sum_{v\in\mathfrak{g}''}x_v\right)^2$!

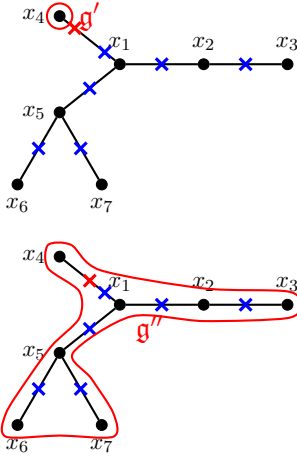
\begin{wrapfigure}{l}{4.5cm}
 \begin{tikzpicture}[ball/.style = {circle, draw, align=center, anchor=north, inner sep=0}, cross/.style={cross out, draw, minimum size=2*(#1-\pgflinewidth), inner sep=0pt, outer sep=0pt}, scale=.8, transform shape]
  \begin{scope}
   \node[ball,text width=.18cm,fill,color=black,label=above:{$x_1$}] at (0,0) (x1) {};    
   \node[ball,text width=.18cm,fill,color=black,right=1.2cm of x1.east, label=above:{$x_2$}] (x2) {};    
   \node[ball,text width=.18cm,fill,color=black,right=1.2cm of x2.east, label=above:{$x_3$}] (x3) {};
   \node[ball,text width=.18cm,fill,color=black, label=left:{$x_4$}] at (-1,.8) (x4) {};    
   \node[ball,text width=.18cm,fill,color=black, label=left:{$x_5$}] at (-1,-.8) (x5) {};    
   \node[ball,text width=.18cm,fill,color=black, label=below:{$x_6$}] at (-1.7,-2) (x6) {};    
   \node[ball,text width=.18cm,fill,color=black, label=below:{$x_7$}] at (-.3,-2) (x7) {};

   \node[above=.35cm of x5.north] (ref2) {};
   \coordinate (Int2) at (intersection of x5--x1 and ref2--x2);  

   \coordinate (t1) at (x3.east);
   \coordinate (t2) at (x4.west);
   \coordinate (t3) at (x1.south west);
   \coordinate (t4) at (x2.south);

   \draw[-,thick,color=black] (x1) -- (x2) -- (x3); 
   \draw[-,thick,color=black] (x1) -- (x4);
   \draw[-,thick,color=black] (x5) -- (x1);
   \draw[-,thick,color=black] (x5) -- (x7);   
   \draw[-,thick,color=black] (x5) -- (x6); 

   \coordinate (x12) at ($(x1)!0.50!(x2)$);
   \coordinate (x23) at ($(x2)!0.50!(x3)$);
   \coordinate (x14) at ($(x1)!0.25!(x4)$);
   \coordinate (x14u) at ($(x1)!0.75!(x4)$);
   \coordinate (x15) at ($(x1)!0.50!(x5)$);
   \coordinate (x56) at ($(x5)!0.50!(x6)$);
   \coordinate (x57) at ($(x5)!0.50!(x7)$);
   \node[very thick, cross=4pt, rotate=0, color=blue] at (x12) {};
   \node[very thick, cross=4pt, rotate=0, color=blue] at (x23) {};
   \node[very thick, cross=4pt, rotate=0, color=blue] at (x14) {};
   \node[very thick, cross=4pt, rotate=0, color=blue] at (x15) {};
   \node[very thick, cross=4pt, rotate=0, color=blue] at (x56) {};
   \node[very thick, cross=4pt, rotate=0, color=blue] at (x57) {};
   \node[very thick, cross=4pt, rotate=0, color=red] at (x14u) {};   

   \draw[thick, red] (x4) circle (6pt);
   \node[color=red, scale=1.25] at ($(x4)+(.5,0)$) {$\mathfrak{g}'$};
  \end{scope}
  \begin{scope}[shift={(0,-4)}, transform shape]
   \node[ball,text width=.18cm,fill,color=black,label=above:{$x_1$}] at (0,0) (x1) {};    
   \node[ball,text width=.18cm,fill,color=black,right=1.2cm of x1.east, label=above:{$x_2$}] (x2) {};    
   \node[ball,text width=.18cm,fill,color=black,right=1.2cm of x2.east, label=above:{$x_3$}] (x3) {};
   \node[ball,text width=.18cm,fill,color=black, label=left:{$x_4$}] at (-1,.8) (x4) {};    
   \node[ball,text width=.18cm,fill,color=black, label=left:{$x_5$}] at (-1,-.8) (x5) {};    
   \node[ball,text width=.18cm,fill,color=black, label=below:{$x_6$}] at (-1.7,-2) (x6) {};    
   \node[ball,text width=.18cm,fill,color=black, label=below:{$x_7$}] at (-.3,-2) (x7) {};

   \node[above=.35cm of x5.north] (ref2) {};
   \coordinate (Int2) at (intersection of x5--x1 and ref2--x2);  

   \coordinate (t1) at (x3.east);
   \coordinate (t2) at (x4.west);
   \coordinate (t3) at (x1.south west);
   \coordinate (t4) at (x2.south);

   \draw[-,thick,color=black] (x1) -- (x2) -- (x3); 
   \draw[-,thick,color=black] (x1) -- (x4);
   \draw[-,thick,color=black] (x5) -- (x1);
   \draw[-,thick,color=black] (x5) -- (x7);   
   \draw[-,thick,color=black] (x5) -- (x6); 

   \coordinate (x12) at ($(x1)!0.50!(x2)$);
   \coordinate (x23) at ($(x2)!0.50!(x3)$);
   \coordinate (x14) at ($(x1)!0.25!(x4)$);
   \coordinate (x15) at ($(x1)!0.50!(x5)$);
   \coordinate (x56) at ($(x5)!0.50!(x6)$);
   \coordinate (x57) at ($(x5)!0.50!(x7)$);
   \coordinate (x14c) at ($(x1)!0.50!(x4)$);   
   \node[very thick, cross=4pt, rotate=0, color=blue] at (x12) {};
   \node[very thick, cross=4pt, rotate=0, color=blue] at (x23) {};
   \node[very thick, cross=4pt, rotate=0, color=blue] at (x14) {};
   \node[very thick, cross=4pt, rotate=0, color=blue] at (x15) {};
   \node[very thick, cross=4pt, rotate=0, color=blue] at (x56) {};
   \node[very thick, cross=4pt, rotate=0, color=blue] at (x57) {};
   \node[very thick, cross=4pt, rotate=0, color=red] at (x14c) {};   

   \coordinate (a) at ($(x7.south east)+(.125,-.125)$);
   \coordinate (b) at ($(x7)!0.5!(x6)+(0,-.175)$);
   \coordinate (c) at ($(x6.south west)+(-.125,-.125)$);
   \coordinate (d) at ($(x6.north west)+(-.125,+.125)$);
   \coordinate (e) at ($(x6)!0.50!(x5)+(-.175,+.175)$);
   \coordinate (f) at ($(x5)+(-.125,+.125)$);
   \coordinate (g) at ($(x5)!0.50!(x1)+(-.175,+.175)$);
   \coordinate (ga) at ($(x1)!0.50!(x4)+(-.175,-.175)$);
   \coordinate (gb) at ($(x4.south west)+(-.125,-.125)$);
   \coordinate (gc) at ($(x4.north west)+(-.125,+.125)$);
   \coordinate (gd) at ($(x4.north east)+(+.125,+.125)$);
   \coordinate (ge) at ($(x4)!0.50!(x1)+(+.175,+.175)$);
   \coordinate (h) at ($(x1.north west)+(-.125,+.125)$);
   \coordinate (i) at ($(x2.north)+(0,+.175)$);
   \coordinate (j) at ($(x3.north)+(0,+.125)$);
   \coordinate (k) at ($(x3.east)+(+.125,0)$);
   \coordinate (l) at ($(x3.south)+(0,-.125)$);
   \coordinate (m) at ($(x2.south)+(0,-.175)$);
   \coordinate (n) at ($(x1.south east)+(+.125,-.125)$);
   \coordinate (o) at ($(x5.east)+(+.125,0)$);
   \coordinate (p) at ($(x5)!0.5!(x7)+(+.175,+.175)$);
   \coordinate (q) at ($(x7.north east)+(+.125,+.125)$);
  
   \draw[thick, red] plot [smooth cycle] coordinates {(a) (b) (c) (d) (e) (f) (g) (ga) (gb) (gc) (gd) (ge) (i) (j) (k) (l) (m) (n) (o) (p) (q)};
   \node[color=red, scale=1.25] at ($(x5)!0.50!(x1)+(1,0)$) {$\mathfrak{g}''$};
  \end{scope}
 \end{tikzpicture}  
\end{wrapfigure}

Let us now discuss the solutions for the $c_j$'s related to the edge $\bar{e}$ which is external to $\mathfrak{g}$. One of the relevant $c_j$ is identified by marking the edge in its end closer to the outer vertex $x_{\bar{v}}$ ($x_{\bar{v}}\,=\,x_4$ in the picture) and it is determined by the hyperplane associated to the subgraph $\mathfrak{g}'$ enclosing just such a vertex. The hyperplane fixing the other $c_j$ is instead related to the graph $\mathfrak{g}''\,=\,\mathcal{G}$. Hence
\begin{equation}\label{eq:Pe1b}
 \begin{split}
  &\omega_{\mathfrak{g}'}\cdot\mathcal{Y}\:=\:y_{\bar{e}}+x_{\bar{v}},\\
  &\omega_{\mathfrak{g}''}\cdot\mathcal{Y}\:=\:\sum_{v\in\mathfrak{g}''}x_v 
 \end{split}
\end{equation}
which on the facet $y_{\bar{e}}+\sum_{v\in\mathfrak{g}}x_v\,=\,0$ together form -- up to a sign -- the Lorentz invariant propagator $y_{\bar{e}}^2-x_{\bar{v}}^2$! Putting all these contributions together, the canonical form associated to these facets can be written as a (minus) flat-space scattering amplitude! Explicitly:
\begin{equation}\label{eq:CFgF2}
 \Omega\:\sim\:-\frac{1}{y_{\bar{e}}^2-x_{\bar{v}}^2}\prod_{e\in\mathcal{E}\setminus\{\bar{e}\}}\frac{1}{y_e^2-\left(\sum_{v\in\mathfrak{g}''}x_v\right)^2}.
\end{equation}
In other words, the canonical form of {\it all} these facets encode the Lorentz invariant flat-space amplitude associated to the graph $\mathcal{G}$!

Let us now consider the facet associated to the subgraph $\mathfrak{g}$ characterised by the maximum number of external edges connecting $\mathfrak{g}$ to subgraphs made out of single vertices only

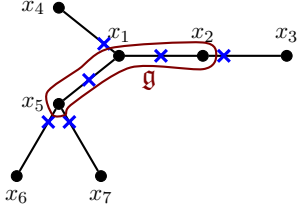
\begin{wrapfigure}{l}{4.5cm}
 \begin{tikzpicture}[ball/.style = {circle, draw, align=center, anchor=north, inner sep=0}, cross/.style={cross out, draw, minimum size=2*(#1-\pgflinewidth), inner sep=0pt, outer sep=0pt}, scale=.8, transform shape]
  \begin{scope}
   \node[ball,text width=.18cm,fill,color=black,label=above:{$x_1$}] at (0,0) (x1) {};    
   \node[ball,text width=.18cm,fill,color=black,right=1.2cm of x1.east, label=above:{$x_2$}] (x2) {};    
   \node[ball,text width=.18cm,fill,color=black,right=1.2cm of x2.east, label=above:{$x_3$}] (x3) {};
   \node[ball,text width=.18cm,fill,color=black, label=left:{$x_4$}] at (-1,.8) (x4) {};    
   \node[ball,text width=.18cm,fill,color=black, label=left:{$x_5$}] at (-1,-.8) (x5) {};    
   \node[ball,text width=.18cm,fill,color=black, label=below:{$x_6$}] at (-1.7,-2) (x6) {};    
   \node[ball,text width=.18cm,fill,color=black, label=below:{$x_7$}] at (-.3,-2) (x7) {};

   \node[above=.35cm of x5.north] (ref2) {};
   \coordinate (Int2) at (intersection of x5--x1 and ref2--x2);  

   \coordinate (t1) at (x3.east);
   \coordinate (t2) at (x4.west);
   \coordinate (t3) at (x1.south west);
   \coordinate (t4) at (x2.south);

   \draw[-,thick,color=black] (x1) -- (x2) -- (x3); 
   \draw[-,thick,color=black] (x1) -- (x4);
   \draw[-,thick,color=black] (x5) -- (x1);
   \draw[-,thick,color=black] (x5) -- (x7);   
   \draw[-,thick,color=black] (x5) -- (x6); 

   \coordinate (x12) at ($(x1)!0.50!(x2)$);
   \coordinate (x23) at ($(x2)!0.25!(x3)$);
   \coordinate (x14) at ($(x1)!0.25!(x4)$);
   \coordinate (x15) at ($(x1)!0.50!(x5)$);
   \coordinate (x56) at ($(x5)!0.25!(x6)$);
   \coordinate (x57) at ($(x5)!0.25!(x7)$);
   \node[very thick, cross=4pt, rotate=0, color=blue] at (x12) {};
   \node[very thick, cross=4pt, rotate=0, color=blue] at (x23) {};
   \node[very thick, cross=4pt, rotate=0, color=blue] at (x14) {};
   \node[very thick, cross=4pt, rotate=0, color=blue] at (x15) {};
   \node[very thick, cross=4pt, rotate=0, color=blue] at (x56) {};
   \node[very thick, cross=4pt, rotate=0, color=blue] at (x57) {};

   \coordinate (b) at ($(x7)!0.85!(x5)$);
   \coordinate (e) at ($(x6)!0.85!(x5)$);
   \coordinate (f) at ($(x5.north west)+(-.125,+.125)$);
   \coordinate (g) at ($(x5)!0.50!(x1)+(-.175,+.175)$);
   \coordinate (h) at ($(x1.north west)+(-.0625,+.0625)$);
   \coordinate (i) at ($(x2.north)+(0,+.125)$);
   \coordinate (ia) at ($(x2.east)+(+.125,0)$);
   \coordinate (m) at ($(x2.south)+(0,-.125)$);
   \coordinate (n) at ($(x1.south east)+(+.125,-.125)$);
   \coordinate (o) at ($(x5.east)+(+.125,0)$);
  
   \draw[thick, red!50!black] plot [smooth cycle] coordinates {(b) (e) (f) (g) (h) (i) (ia) (m) (n) (o)};
   \node[color=red!50!black, scale=1.25] at ($(x5)!0.50!(x1)+(1,0)$) {$\mathfrak{g}$};
  \end{scope}
 \end{tikzpicture}  
\end{wrapfigure}

Following exactly the procedure described above, the $c_j$'s related to the vertices which are marked to either ends of a given edge are given by:
\begin{equation}\label{eq:Pe2a}
 \begin{split}
  &\omega_{\mathfrak{g}'}\cdot\mathcal{Y}\:\sim\:y_e+\sum_{\bar{e}\in\mathcal{E}^{\mbox{\tiny ext}}_1}y_{\bar{e}}+\sum_{v\in\mathfrak{g}'}x_v,\\
  &\omega_{\mathfrak{g}''}\cdot\mathcal{Y}\:\sim\:y_e+\sum_{\bar{e}\in\mathcal{E}^{\mbox{\tiny ext}}_2}y_{\bar{e}}+\sum_{v\in\mathfrak{g}''}x_v,  
 \end{split}
\end{equation}
$\mathcal{E}^{\mbox{\tiny ext}}_1$ and $\mathcal{E}^{\mbox{\tiny ext}}_2$ being the set of edges which are simultaneously external to the pairs $(\mathfrak{g},\mathfrak{g}')$ and $(\mathfrak{g},\mathfrak{g}'')$ respectively. Notice that the set of edges which are external to $\mathfrak{g}$ is given by $\mathcal{E}^{\mbox{\tiny ext}}\:=\:\mathcal{E}^{\mbox{\tiny ext}}_1\,\cup\,\mathcal{E}^{\mbox{\tiny ext}}_2$ as well as the set of vertices in $\mathfrak{g}$ is the union of the sets of vertices of $\mathfrak{g}'$ and $\mathfrak{g}''$. Thus, the condition identifying the facet, allows to recast these two linear contributions as a single Lorentz invariant propagator $y_e^2-\left(\sum_{\bar{e}\in\mathcal{E}^{\mbox{\tiny ext}}_1}y_{\bar{e}}+\sum_{v\in\mathfrak{g}'}x_v\right)^2$. When instead we consider the vertices related to edges external to $\mathfrak{g}$, then
\begin{equation}\label{eq:Pe2n}
 \begin{split}
  &\omega_{\mathfrak{g}'}\cdot\mathcal{Y}\:\sim\:y_{\bar{e}}+x_{\bar{v}},\\
  &\omega_{\mathfrak{g}''}\cdot\mathcal{Y}\:\sim\:\sum_{\hat{e}\in\mathcal{E}^{\mbox{\tiny ext}}\setminus\{\bar{e}\}}y_{\hat{e}}+\sum_{v\in\mathfrak{g}''}x_v,  
 \end{split}
\end{equation}
which, because of the condition identifying the facet, can be group together in the form $-(y_{\bar{e}}^2-x_{\bar{v}}^2)$. Hence, the canonical form of the facet is given by
\begin{equation}\label{eq:CFg}
 \begin{split}
  \Omega\:&\sim\:\prod_{\bar{e}\in\mathcal{E}^{\mbox{\tiny ext}}}\frac{-1}{y_{\bar{e}}^2-x_{\bar{v}}^2}\prod_{e\in\mathfrak{g}}\frac{1}{y_e^2-\left(\sum_{\bar{e}\in\mathcal{E}^{\mbox{\tiny ext}}_1}y_{\bar{e}}+\sum_{v\in\mathfrak{g}'}x_v\right)^2}\:=\\
  &=\:(-1)^{\mbox{\tiny dim$\{\mathcal{E}^{\mbox{\tiny ext}}\}$}}\left(\prod_{\bar{e}\in\mathcal{E}^{\mbox{\tiny ext}}}\mathcal{A}_{\bar{e}}\right)\times\mathcal{A}[\mathfrak{g}]
   \:\equiv\:(-1)^{\mbox{\tiny dim$\{\mathcal{E}^{\mbox{\tiny ext}}\}$}}\mathcal{A}[\mathcal{G}],
 \end{split}
\end{equation}
where $\mathcal{A}_{\bar{e}}$ is the flat-space amplitude related the edge $\bar{e}$ external to $\mathfrak{g}$, while $\mathcal{A}[\mathfrak{g}]$ and $\mathcal{A}[\mathcal{G}]$ are the flat-space amplitude related to the subgraphs $\mathfrak{g}$ and $\mathcal{G}$ respectively.

It is really striking how these subset of facets turn out to encode Lorentz invariance as the scattering facet! This precisely characterises the statement that these facets are {\it isomorphic} to the scattering facet. Interestingly, all the faces of the facets we have just analysed, which differ by a vertex related to the same edge $e$ of the graph, produces contributions $\omega\cdot\mathcal{Y}$ which differ by the sign of the energy $y_e$ related to the edge $e$.

Finally, let us discuss the facets related to higher codimension subgraphs. In this case the facets are no longer simplices, and the facets with the highest number of vertices correspond to the subgraph encircling one of the most outer vertices ($\nu\,=\,3n_e-1$) -- the facets corresponding to subgraphs made out of a single internal vertex have $\nu\,=\,3n_e-n_{\tilde{e}}$ vertices ($n_{\tilde{e}}$ is the number of edges departing from such a subgraph). Without loss of generality, let us consider the highest codimension subgraph.

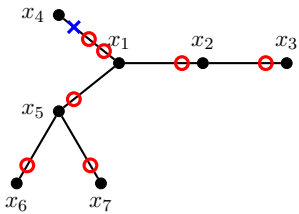
\begin{wrapfigure}{l}{4.5cm}
 \begin{tikzpicture}[ball/.style = {circle, draw, align=center, anchor=north, inner sep=0}, cross/.style={cross out, draw, minimum size=2*(#1-\pgflinewidth), inner sep=0pt, outer sep=0pt}, scale=.8, transform shape]
  \begin{scope}
   \node[ball,text width=.18cm,fill,color=black,label=above:{$x_1$}] at (0,0) (x1) {};    
   \node[ball,text width=.18cm,fill,color=black,right=1.2cm of x1.east, label=above:{$x_2$}] (x2) {};    
   \node[ball,text width=.18cm,fill,color=black,right=1.2cm of x2.east, label=above:{$x_3$}] (x3) {};
   \node[ball,text width=.18cm,fill,color=black, label=left:{$x_4$}] at (-1,.8) (x4) {};    
   \node[ball,text width=.18cm,fill,color=black, label=left:{$x_5$}] at (-1,-.8) (x5) {};    
   \node[ball,text width=.18cm,fill,color=black, label=below:{$x_6$}] at (-1.7,-2) (x6) {};    
   \node[ball,text width=.18cm,fill,color=black, label=below:{$x_7$}] at (-.3,-2) (x7) {};

   \node[above=.35cm of x5.north] (ref2) {};
   \coordinate (Int2) at (intersection of x5--x1 and ref2--x2);  

   \coordinate (t1) at (x3.east);
   \coordinate (t2) at (x4.west);
   \coordinate (t3) at (x1.south west);
   \coordinate (t4) at (x2.south);

   \draw[-,thick,color=black] (x1) -- (x2) -- (x3); 
   \draw[-,thick,color=black] (x1) -- (x4);
   \draw[-,thick,color=black] (x5) -- (x1);
   \draw[-,thick,color=black] (x5) -- (x7);   
   \draw[-,thick,color=black] (x5) -- (x6); 

   \coordinate (x12) at ($(x1)!0.50!(x2)$);
   \coordinate (x23) at ($(x2)!0.50!(x3)$);
   \coordinate (x14) at ($(x1)!0.25!(x4)$);
   \coordinate (x14u) at ($(x1)!0.75!(x4)$);
   \coordinate (x15) at ($(x1)!0.50!(x5)$);
   \coordinate (x56) at ($(x5)!0.50!(x6)$);
   \coordinate (x57) at ($(x5)!0.50!(x7)$);
   \node[very thick, cross=4pt, rotate=0, color=blue] at (x14u) {};   
%   \node[very thick, cross=4pt, rotate=0, color=red] at (x12) {};   
%   \node[very thick, cross=4pt, rotate=0, color=red] at (x23) {};   
%   \node[very thick, cross=4pt, rotate=0, color=red] at (x15) {};   
%   \node[very thick, cross=4pt, rotate=0, color=red] at (x56) {};   
%   \node[very thick, cross=4pt, rotate=0, color=red] at (x57) {};   
   \coordinate (x12r) at ($(x1)!0.75!(x2)$) {};
   \coordinate (x14c) at ($(x1)!0.50!(x4)$) {};
   \coordinate (x23r) at ($(x2)!0.75!(x3)$) {};   
   \coordinate (x15b) at ($(x1)!0.75!(x5)$) {};   
   \coordinate (x56b) at ($(x5)!0.75!(x6)$) {};
   \coordinate (x57b) at ($(x5)!0.75!(x7)$) {};  

   \draw[very thick, color=red] (x12r) circle (3pt);
   \draw[very thick, color=red] (x23r) circle (3pt);
   \draw[very thick, color=red] (x14c) circle (3pt);
   \draw[very thick, color=red] (x14) circle (3pt);   
   \draw[very thick, color=red] (x15b) circle (3pt);
   \draw[very thick, color=red] (x56b) circle (3pt);
   \draw[very thick, color=red] (x57b) circle (3pt);
  \end{scope}
 \end{tikzpicture}
\end{wrapfigure}

In the contour integral representation, the delta-functions do not fully localise the integration -- this facet is not a simplex -- and the remaining integrations can be performed by closing the integration contour in either the upper-half (UHP) or the lower-half plane (LHP) of each $c_j$ chosen as left free. Indeed, there is certain freedom in choosing such $c_j$'s. However, irrespectively of such a choice, all the possible combinations of the various contours, return different triangulations of the facet. Among all the possible triangulations, there is a special class of particular interest for us: it is characterised by having {\it all} its simplices with two vertices {\it for each} edge. Importantly, the contour integral representation allows to identify those vertices which are common to all the simplices: they are given by the vertices related to the external end of the outer edges (the related $c_j$'s do not depend on the ones chosen as independent and, thus, they are fixed once for all) and by those ones which identify a subgraph and whose $c_j$'s depend on a subset of the independent ones. This latter class of vertices turn out to be related to the end of each internal edge further away from the excluded vertex identifying the facet. Let us graphically indicate all those vertices with a red open circle
$
 \begin{tikzpicture}[ball/.style = {circle, draw, align=center, anchor=north, inner sep=0}]
   \node[ball,text width=.18cm,thick,color=red] at (0,0) {};     
 \end{tikzpicture}
$
. Then, the simplices providing this triangulation can be identified by all those possible markings with two vertices for each edge (they are $2^{n_e-1}$), such that the vertices 
$
 \begin{tikzpicture}[ball/.style = {circle, draw, align=center, anchor=north, inner sep=0}]
   \node[ball,text width=.18cm,thick,color=red] at (0,0) {};     
 \end{tikzpicture}
$
are always included. It is important to stress that for the tree-level graphs a triangulation with these features, is {\it unique}. 

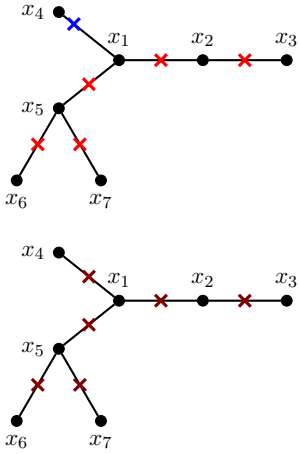
\begin{wrapfigure}{l}{4.5cm}
 \begin{tikzpicture}[ball/.style = {circle, draw, align=center, anchor=north, inner sep=0}, cross/.style={cross out, draw, minimum size=2*(#1-\pgflinewidth), inner sep=0pt, outer sep=0pt}, scale=.8, transform shape]
  \begin{scope}
   \node[ball,text width=.18cm,fill,color=black,label=above:{$x_1$}] at (0,0) (x1) {};    
   \node[ball,text width=.18cm,fill,color=black,right=1.2cm of x1.east, label=above:{$x_2$}] (x2) {};    
   \node[ball,text width=.18cm,fill,color=black,right=1.2cm of x2.east, label=above:{$x_3$}] (x3) {};
   \node[ball,text width=.18cm,fill,color=black, label=left:{$x_4$}] at (-1,.8) (x4) {};    
   \node[ball,text width=.18cm,fill,color=black, label=left:{$x_5$}] at (-1,-.8) (x5) {};    
   \node[ball,text width=.18cm,fill,color=black, label=below:{$x_6$}] at (-1.7,-2) (x6) {};    
   \node[ball,text width=.18cm,fill,color=black, label=below:{$x_7$}] at (-.3,-2) (x7) {};

   \node[above=.35cm of x5.north] (ref2) {};
   \coordinate (Int2) at (intersection of x5--x1 and ref2--x2);  

   \coordinate (t1) at (x3.east);
   \coordinate (t2) at (x4.west);
   \coordinate (t3) at (x1.south west);
   \coordinate (t4) at (x2.south);

   \draw[-,thick,color=black] (x1) -- (x2) -- (x3); 
   \draw[-,thick,color=black] (x1) -- (x4);
   \draw[-,thick,color=black] (x5) -- (x1);
   \draw[-,thick,color=black] (x5) -- (x7);   
   \draw[-,thick,color=black] (x5) -- (x6); 

   \coordinate (x12) at ($(x1)!0.50!(x2)$);
   \coordinate (x23) at ($(x2)!0.50!(x3)$);
   \coordinate (x14) at ($(x1)!0.25!(x4)$);
   \coordinate (x14u) at ($(x1)!0.75!(x4)$);
   \coordinate (x15) at ($(x1)!0.50!(x5)$);
   \coordinate (x56) at ($(x5)!0.50!(x6)$);
   \coordinate (x57) at ($(x5)!0.50!(x7)$);
   \node[very thick, cross=4pt, rotate=0, color=blue] at (x14u) {};   
   \node[very thick, cross=4pt, rotate=0, color=red] at (x12) {};   
   \node[very thick, cross=4pt, rotate=0, color=red] at (x23) {};   
   \node[very thick, cross=4pt, rotate=0, color=red] at (x15) {};   
   \node[very thick, cross=4pt, rotate=0, color=red] at (x56) {};   
   \node[very thick, cross=4pt, rotate=0, color=red] at (x57) {};   
   \coordinate (x12r) at ($(x1)!0.75!(x2)$) {};
   \coordinate (x14c) at ($(x1)!0.50!(x4)$) {};
   \coordinate (x23r) at ($(x2)!0.75!(x3)$) {};   
   \coordinate (x15b) at ($(x1)!0.75!(x5)$) {};   
   \coordinate (x56b) at ($(x5)!0.75!(x6)$) {};
   \coordinate (x57b) at ($(x5)!0.75!(x7)$) {};  

  \end{scope}
  \begin{scope}[shift={(0,-4)}, transform shape]
   \node[ball,text width=.18cm,fill,color=black,label=above:{$x_1$}] at (0,0) (x1) {};    
   \node[ball,text width=.18cm,fill,color=black,right=1.2cm of x1.east, label=above:{$x_2$}] (x2) {};    
   \node[ball,text width=.18cm,fill,color=black,right=1.2cm of x2.east, label=above:{$x_3$}] (x3) {};
   \node[ball,text width=.18cm,fill,color=black, label=left:{$x_4$}] at (-1,.8) (x4) {};    
   \node[ball,text width=.18cm,fill,color=black, label=left:{$x_5$}] at (-1,-.8) (x5) {};    
   \node[ball,text width=.18cm,fill,color=black, label=below:{$x_6$}] at (-1.7,-2) (x6) {};    
   \node[ball,text width=.18cm,fill,color=black, label=below:{$x_7$}] at (-.3,-2) (x7) {};

   \node[above=.35cm of x5.north] (ref2) {};
   \coordinate (Int2) at (intersection of x5--x1 and ref2--x2);  

   \coordinate (t1) at (x3.east);
   \coordinate (t2) at (x4.west);
   \coordinate (t3) at (x1.south west);
   \coordinate (t4) at (x2.south);

   \draw[-,thick,color=black] (x1) -- (x2) -- (x3); 
   \draw[-,thick,color=black] (x1) -- (x4);
   \draw[-,thick,color=black] (x5) -- (x1);
   \draw[-,thick,color=black] (x5) -- (x7);   
   \draw[-,thick,color=black] (x5) -- (x6); 

   \coordinate (x12) at ($(x1)!0.50!(x2)$);
   \coordinate (x23) at ($(x2)!0.50!(x3)$);
   \coordinate (x14) at ($(x1)!0.25!(x4)$);
   \coordinate (x14u) at ($(x1)!0.75!(x4)$);
   \coordinate (x14c) at ($(x1)!0.50!(x4)$) {};
   \coordinate (x15) at ($(x1)!0.50!(x5)$);
   \coordinate (x56) at ($(x5)!0.50!(x6)$);
   \coordinate (x57) at ($(x5)!0.50!(x7)$);
   \node[very thick, cross=4pt, rotate=0, color=red!50!black] at (x14c) {};   
   \node[very thick, cross=4pt, rotate=0, color=red!50!black] at (x12) {};   
   \node[very thick, cross=4pt, rotate=0, color=red!50!black] at (x23) {};   
   \node[very thick, cross=4pt, rotate=0, color=red!50!black] at (x15) {};   
   \node[very thick, cross=4pt, rotate=0, color=red!50!black] at (x56) {};   
   \node[very thick, cross=4pt, rotate=0, color=red!50!black] at (x57) {};   
  \end{scope}
 \end{tikzpicture}
\end{wrapfigure}

For each of the simplices of this triangulation, we can perform the very same analysis of their faces carried out for the other facets. Given one of these simplices, it is again straightforward to see that the contributions coming from two faces which differ of a vertex related to the same edge $e$, will differ for just a sign of the energy $y_e$ of $e$: they thus group together to form a Lorentz-invariant propagator and this class of facets can be expressed as a sum of products of Lorentz invariant propagators! However, there is a further feature. {\it All} the simplices of the triangulation we are considering differ from simplices which describe other facets of the cosmological polytope by a vertex on the same edge of the graph: the two canonical forms are then the same up to a sign!

Summarising, all the facets of the cosmological polytope $\mathcal{P}_{\mbox{\tiny $\mathcal{G}$}}$ {\it know} about flat-space scattering processes: while the facets which have $2n_e$ vertices are all isomorphic to the scattering facet and, thus, their canonical form returns, up to a sign, the contribution $\mathcal{A}_{\mbox{\tiny $\mathcal{G}$}}$  to the scattering amplitude, the ones with an higher number of vertices have one (and only one) triangulation whose terms have a canonical form which is a product of Lorentz invariant propagators (which are organised as products of lower point amplitudes) and thus the canonical form of the full facet can be expressed as sum of products of flat-space scattering amplitudes.

%%%%%%%%%%%%%%%%%%%%%%
%%%%%%%%%%%%%%%%%%%%%%

\subsection{Facets of the loop cosmological polytopes}\label{subsec:LCP}

We can proceed for the facets of the loop cosmological polytope along the same lines, together with the argument used in \cite{Arkani-Hamed:2018ahb} to prove the emergence of Lorentz invariance on the scattering amplitude facet. 

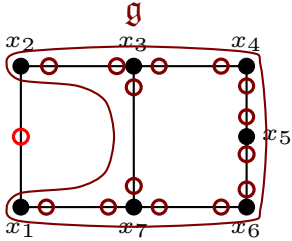
\begin{wrapfigure}{l}{4.5cm}
 \begin{tikzpicture}[ball/.style = {circle, draw, align=center, anchor=north, inner sep=0}, cross/.style={cross out, draw, minimum size=2*(#1-\pgflinewidth), inner sep=0pt, outer sep=0pt}]
  \begin{scope}[shift={(5,-1.75)}, scale={1.5}, transform shape]
    \coordinate[label=below:{\tiny $x_1$}] (v1) at (0,0);
    \coordinate[label=above:{\tiny $x_2$}] (v2) at ($(v1)+(0,1.25)$);
    \coordinate[label=above:{\tiny $x_3$}] (v3) at ($(v2)+(1,0)$);
    \coordinate[label=above:{\tiny $x_4$}] (v4) at ($(v3)+(1,0)$);
    \coordinate[label=right:{\tiny $x_5$}] (v5) at ($(v4)-(0,.625)$);   
    \coordinate[label=below:{\tiny $x_6$}] (v6) at ($(v5)-(0,.625)$);
    \coordinate[label=below:{\tiny $x_7$}] (v7) at ($(v6)-(1,0)$);
    \draw[thick] (v1) -- (v2) -- (v3) -- (v4) -- (v5) -- (v6) -- (v7) -- cycle;
    \draw[thick] (v3) -- (v7);
    \draw[fill=black] (v1) circle (2pt);
    \draw[fill=black] (v2) circle (2pt);
    \draw[fill=black] (v3) circle (2pt);
    \draw[fill=black] (v4) circle (2pt);
    \draw[fill=black] (v5) circle (2pt);
    \draw[fill=black] (v6) circle (2pt);
    \draw[fill=black] (v7) circle (2pt);   
    \coordinate (v12) at ($(v1)!0.5!(v2)$);   
    \coordinate (v23) at ($(v2)!0.5!(v3)$);
    \coordinate (v34) at ($(v3)!0.5!(v4)$);
    \coordinate (v45) at ($(v4)!0.5!(v5)$);   
    \coordinate (v56) at ($(v5)!0.5!(v6)$);   
    \coordinate (v67) at ($(v6)!0.5!(v7)$);
    \coordinate (v71) at ($(v7)!0.5!(v1)$);   
    \coordinate (v37) at ($(v3)!0.5!(v7)$);
    \node[ball,text width=.20cm,very thick,color=red!50!black,right=.15cm of v3, scale=.625] {};   
    \node[ball,text width=.20cm,very thick,color=red!50!black,left=.15cm of v4, scale=.625] {};
    \node[ball,text width=.20cm,very thick,color=red!50!black,below=.1cm of v4, scale=.625] {};
    \node[ball,text width=.20cm,very thick,color=red!50!black,above=.1cm of v5, scale=.625] {};
    \draw[very thick,color=red!50!black] ($(v2)!0.25!(v3)$) circle (1.875pt);
    \draw[very thick,color=red!50!black] ($(v3)!0.85!(v7)$) circle (1.875pt);
    \draw[very thick,color=red!50!black] ($(v5)!0.75!(v6)$) circle (1.875pt);
    \draw[very thick,color=red!50!black] ($(v5)!0.25!(v6)$) circle (1.875pt);    
    \draw[very thick,color=red!50!black] ($(v2)!0.85!(v3)$) circle (1.875pt);    
    \draw[very thick,color=red!50!black] ($(v3)!0.15!(v7)$) circle (1.875pt);    
    \draw[very thick,color=red] (v12) circle (1.875pt);
    \node[ball,text width=.20cm,very thick,color=red!50!black,right=.15cm of v1, scale=.625] {};
    \node[ball,text width=.20cm,very thick,color=red!50!black,left=.15cm of v7, scale=.625] {};
    \node[ball,text width=.20cm,very thick,color=red!50!black,right=.15cm of v7, scale=.625] {};
    \node[ball,text width=.20cm,very thick,color=red!50!black,left=.15cm of v6, scale=.625] {}; 
    \coordinate (a) at ($(v2)+(0,.125)$);
    \coordinate (b) at ($(v3)+(0,.175)$);
    \coordinate (c) at ($(v4)+(0,.125)$);
    \coordinate (d) at ($(v4)+(.125,0)$);
    \coordinate (e) at ($(v5)+(.175,0)$);
    \coordinate (f) at ($(v6)+(.125,0)$);
    \coordinate (g) at ($(v6)-(0,.125)$);
    \coordinate (h) at ($(v7)-(0,.175)$);
    \coordinate (i) at ($(v1)-(0,.125)$);
    \coordinate (j) at ($(v1)-(.125,0)$);
    \coordinate (k) at ($(v1)+(0,.125)$);
    \coordinate (l) at ($(v71)+(0,.175)$);
    \coordinate (ln) at ($(v71)!0.5!(v37)$);
    \coordinate (n) at ($(v37)-(.175,0)$);
    \coordinate (no) at ($(v37)!0.5!(v23)$);
    \coordinate (o) at ($(v23)-(0,.175)$);
    \coordinate (p) at ($(v2)-(0,.125)$);
    \coordinate (q) at ($(v2)-(.125,0)$);
    \draw [thick, red!50!black] plot [smooth cycle] coordinates {(a) (b) (c) (d) (e) (f) (g) (h) (i) (j) (k) (l) (ln) (n) (no) (o) (p) (q)};
    \node[above=.05cm of b, color=red!50!black] {\footnotesize $\displaystyle\mathfrak{g}$};
   \end{scope}
 \end{tikzpicture}  
\end{wrapfigure}

Let us start  with those facets identified by a subgraph $\mathfrak{g}$ containing all the vertices of the graph $\mathcal{G}$ and crossing one of the edges twice. Such facets live in $\mathbb{P}^{n_e+n_v-2}$ and have $\nu\,=\,2n_e-1\,\equiv\,n_e+n_v-2+L$ vertices. In the contour integral representation \eqref{eq:CFgF}, the delta functions localise all the $c_j$'s but $L-1$. As usual, we have the freedom to choose which of the $c_j$'s are kept free, identifying those hyperplanes which the vertices attached to the free $c_j$'s {\it do not} belong to. Then, a given $c_j$ is determined in terms of the hyperplane $\omega$ which {\it also} do not contain the vertex associated to it and it is a linear function of the free variables. Interestingly, the variable related to the  vertex in the middle of the edge $e$ crossed twice by the line that identifies the subgraph $\mathfrak{g}$ (in the graph above is indicated with the red open circle) is the only one which is determined by the hyperplane which do not contain any of the three vertices associated to the edge $e$ and turns out {\it not} to depend on {\it any} of the free variables. Furthermore, the vertex structure included into $\mathfrak{g}$ identifies an $(L-1)$-loop scattering facet, and thus, the canonical form is given by
\begin{equation}\label{eq:CFL1}
 \Omega\:\sim\:\frac{1}{y_e}\times\mathcal{A}[\mathfrak{g}],
\end{equation}
$e$ being the edge marked by the red open circle -- this can be seen explicitly by realising that the solutions of the delta functions for the other variables are determined in terms of subgraphs identified by the excluded vertices and that, after having localised the delta-function, the denominators can be grouped together to form a product of Lorentz invariant propagators, with the free $L-1$ $ c_j$ which can be interpreted as $l_0$ integrations. 

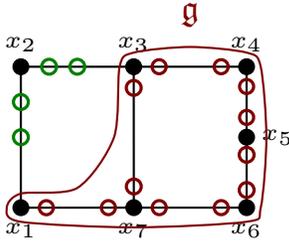
\begin{wrapfigure}{l}{4.5cm}
 \begin{tikzpicture}[ball/.style = {circle, draw, align=center, anchor=north, inner sep=0}, cross/.style={cross out, draw, minimum size=2*(#1-\pgflinewidth), inner sep=0pt, outer sep=0pt}]
  \begin{scope}[shift={(5,-1.75)}, scale={1.5}, transform shape]
    \coordinate[label=below:{\tiny $x_1$}] (v1) at (0,0);
    \coordinate[label=above:{\tiny $x_2$}] (v2) at ($(v1)+(0,1.25)$);
    \coordinate[label=above:{\tiny $x_3$}] (v3) at ($(v2)+(1,0)$);
    \coordinate[label=above:{\tiny $x_4$}] (v4) at ($(v3)+(1,0)$);
    \coordinate[label=right:{\tiny $x_5$}] (v5) at ($(v4)-(0,.625)$);   
    \coordinate[label=below:{\tiny $x_6$}] (v6) at ($(v5)-(0,.625)$);
    \coordinate[label=below:{\tiny $x_7$}] (v7) at ($(v6)-(1,0)$);
    \draw[thick] (v1) -- (v2) -- (v3) -- (v4) -- (v5) -- (v6) -- (v7) -- cycle;
    \draw[thick] (v3) -- (v7);
    \draw[fill=black] (v1) circle (2pt);
    \draw[fill=black] (v2) circle (2pt);
    \draw[fill=black] (v3) circle (2pt);
    \draw[fill=black] (v4) circle (2pt);
    \draw[fill=black] (v5) circle (2pt);
    \draw[fill=black] (v6) circle (2pt);
    \draw[fill=black] (v7) circle (2pt);   
    \coordinate (v12) at ($(v1)!0.5!(v2)$);   
    \coordinate (v23) at ($(v2)!0.5!(v3)$);
    \coordinate (v34) at ($(v3)!0.5!(v4)$);
    \coordinate (v45) at ($(v4)!0.5!(v5)$);   
    \coordinate (v56) at ($(v5)!0.5!(v6)$);   
    \coordinate (v67) at ($(v6)!0.5!(v7)$);
    \coordinate (v71) at ($(v7)!0.5!(v1)$);   
    \coordinate (v37) at ($(v3)!0.5!(v7)$);
    \node[ball,text width=.20cm,very thick,color=red!50!black,right=.15cm of v3, scale=.625] {};   
    \node[ball,text width=.20cm,very thick,color=red!50!black,left=.15cm of v4, scale=.625] {};
    \node[ball,text width=.20cm,very thick,color=red!50!black,below=.1cm of v4, scale=.625] {};
    \node[ball,text width=.20cm,very thick,color=red!50!black,above=.1cm of v5, scale=.625] {};
    \draw[very thick,color=green!50!black] ($(v1)!0.75!(v2)$) circle (1.875pt);    
    \draw[very thick,color=green!50!black] ($(v2)!0.25!(v3)$) circle (1.875pt);
    \draw[very thick,color=red!50!black] ($(v3)!0.85!(v7)$) circle (1.875pt);
    \draw[very thick,color=red!50!black] ($(v5)!0.75!(v6)$) circle (1.875pt);
    \draw[very thick,color=red!50!black] ($(v5)!0.25!(v6)$) circle (1.875pt);    
    \draw[very thick,color=green!50!black] ($(v2)!0.50!(v3)$) circle (1.875pt);    
    \draw[very thick,color=red!50!black] ($(v3)!0.15!(v7)$) circle (1.875pt);    
    \draw[very thick,color=green!50!black] (v12) circle (1.875pt);
    \node[ball,text width=.20cm,very thick,color=red!50!black,right=.15cm of v1, scale=.625] {};
    \node[ball,text width=.20cm,very thick,color=red!50!black,left=.15cm of v7, scale=.625] {};
    \node[ball,text width=.20cm,very thick,color=red!50!black,right=.15cm of v7, scale=.625] {};
    \node[ball,text width=.20cm,very thick,color=red!50!black,left=.15cm of v6, scale=.625] {}; 
    \coordinate (a) at ($(v3)+(0,.125)$);
    \coordinate (b) at ($(v34)+(0,.175)$);
    \coordinate (c) at ($(v4)+(0,.125)$);
    \coordinate (d) at ($(v4)+(.125,0)$);
    \coordinate (e) at ($(v5)+(.175,0)$);
    \coordinate (f) at ($(v6)+(.125,0)$);
    \coordinate (g) at ($(v6)-(0,.125)$);
    \coordinate (h) at ($(v7)-(0,.175)$);
    \coordinate (i) at ($(v1)-(0,.125)$);
    \coordinate (j) at ($(v1)-(.125,0)$);
    \coordinate (k) at ($(v1)+(0,.125)$);
    \coordinate (l) at ($(v71)+(0,.175)$);
    \coordinate (n) at ($(v37)-(.175,0)$);
    \coordinate (q) at ($(v3)-(.125,0)$);
    \draw [thick, red!50!black] plot [smooth cycle] coordinates {(a) (b) (c) (d) (e) (f) (g) (h) (i) (j) (k) (l) (n) (q)};
    \node[above=.05cm of b, color=red!50!black] {\footnotesize $\displaystyle\mathfrak{g}$};
   \end{scope}
 \end{tikzpicture}  
\end{wrapfigure}

Another class of facets of particular interests are the ones identified by a subgraph $\mathfrak{g}$ with all but one sites of the graph $\mathcal{G}$ and with two edges being external. The facet related to this graph has exactly $2n_e\,=\,n_e+n_v-1+L$ vertices, as the scattering amplitude facet. A subset of the vertices of such facets, that in the picture on the left have been with open circles
$
 \begin{tikzpicture}[ball/.style = {circle, draw, align=center, anchor=north, inner sep=0}, cross/.style={cross out, draw, minimum size=2*(#1-\pgflinewidth), inner sep=0pt, outer sep=0pt}]
  \node[ball,text width=.20cm,very thick,color=red!50!black] {}; 
 \end{tikzpicture}
$
clearly identify the structure of a lower dimensional scattering facet. Considering the contour integral representation for canonical forms of this type of facets, as for the scattering facet, the delta-functions localise $n_e+n_v-1$ integration leaving exactly $L$ free integration. Of the $L$ free $c_j$, it is convenient to choose $L-1$ of them related to $\mathfrak{g}$ and one external to it. Said more geometrically, we choose those hyperplanes which do not include $L-1$ vertices identified by $\mathfrak{g}$ and one related to one of the edges external to it. %{\it Contrarily} to the scattering facet case, not all the linear propagators in the integrand organise nicely into quadratic Lorentz invariant ones, but only the ones related to the vertices belonging to the subgraph $\mathfrak{g}$. 
Notably, the denominators depending on the $L-1$ variables related to the vertices in the lower-dimensional facet identified by the subgraph $\mathfrak{g}$ do not depend on the other free variable $\hat{c}$ related to one of the edges external to $\mathfrak{g}$, and vice versa, the denominators depending on $\hat{c}$ do not depend on the other free variables. Hence, the canonical form of the full facet decouples into two sets of separate integrals:
\begin{equation}\label{eq:CFg2ne}
 \begin{split}
 \Omega\:&\sim\:\left(\int\frac{d\hat{c}}{\left[\left(\hat{c}-\frac{\mathfrak{y}_a}{4}\right)^2-\left(\frac{\mathfrak{y}_a}{4}-i\hat{\varepsilon}\right)^2\right]
                                          \left[\left(\hat{c}-\frac{\mathfrak{y}_b}{4}\right)^2-\left(\frac{\mathfrak{y}_c}{4}-i\hat{\varepsilon}\right)^2\right]}\right)\times\\
     &\times\left(\int\prod_{j=1}^{L-1}\frac{dc_j}{\left(c_j-\frac{y_{e_j}}{2}\right)^2-\left(\frac{y_{e_j}}{2}-i\varepsilon_j\right)^2}\prod_{k=1}^{2n_e-1+L}\frac{1}{\left(\sum_r\sigma_rc_r-\mathfrak{y}_r/2\right)^2-\left(\frac{y_k}{2}-i\varepsilon_k\right)^2}\right)
 \end{split}
\end{equation}
where the first line is related to the subgraph external to $\mathfrak{g}$, the second line represents the subgraph $\mathfrak{g}$ itself, the $\mathfrak{y}$'s are linear combinations of $x$'s and $y$'s determined by higher codimension subgraphs, and $\sigma_r$ are suitable signs. Importantly, both expressions could be recast into manifestly Lorentz invariant form, with the Feynman $i\varepsilon$ prescription dictated by the canonical prescription for the canonical form, as it happens in the scattering amplitude facet. While the second line is clearly the amplitude $\mathcal{A}[\mathfrak{g}]$ associated to the subgraph $\mathfrak{g}$ as expected -- the facet we are analysing is identified by the condition which restores energy conservation for the subgraph $\mathfrak{g}$ --, the first line is isomorphic to a one-loop scattering facet. The contour integral expression \eqref{eq:CFg2ne} for the canonical form of this class of faces makes both its Lorenz invariance and its factorisation in two lower point amplitude manifest.

A similar discussion holds for higher codimension facets: the way the delta functions localise some of the integrations makes again the factorisation manifest, while for Lorentz invariance one would have to resort to specific triangulations, as for the higher codimensions faces at tree level discussed in Section \ref{subsec:TCP}.

%%%%%%%%%%%%%%%%%%%%%%
%%%%%%%%%%%%%%%%%%%%%%

\subsection{Combinatorial automorphisms}\label{subsec:CA}

The analysis carried out in the previous subsections showed that there exists a subset of facets which are isomorphic to each other, {\it i.e.} they have the same number of vertices and their canonical forms are the same (up to a sign): these facets can thus be mapped into each other. We can ask now how such a mapping can be defined and whether it can be part of a symmetry group of the full cosmological polytope. More generally, we look at the group of combinatorial automorphisms of a given cosmological polytopes, {\it i.e.} the symmetry group preserving its face lattice\footnote{A face lattice of a given polytope is a lattice whose vertices are given by the faces of the polytope, including the full polytope and the empty set, with the edges connecting these vertices determined by containment relations.}.

Let us start with considering the triangle, which is the building block for constructing the cosmological polytopes and it encodes the contribution of the two-site graph to the wavefunction of the universe. Its face lattice is a cube and the group of the combinatorial automorphism is given by the transposition of two vertices

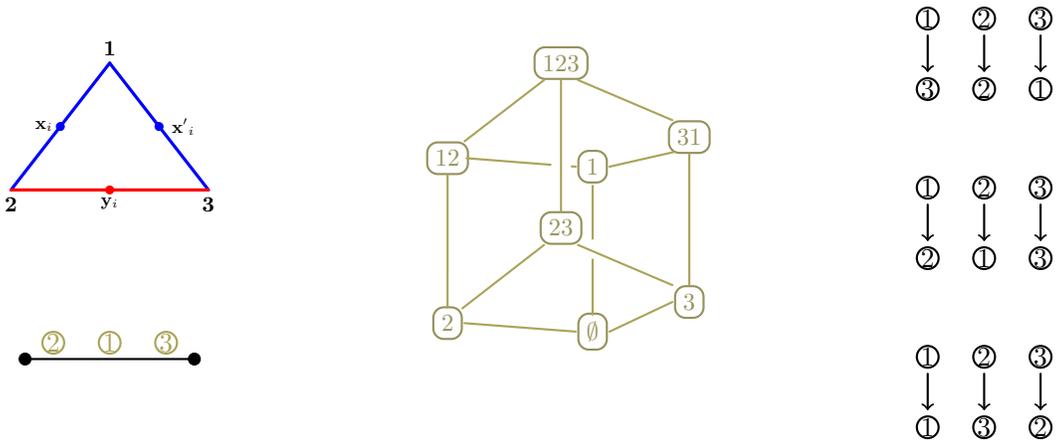
\begin{equation*}
 \begin{tikzpicture}[line join = round, line cap = round, ball/.style = {circle, draw, align=center, anchor=north, inner sep=0}, 
                     axis/.style={very thick, ->, >=stealth'}, pile/.style={thick, ->, >=stealth', shorten <=2pt, shorten>=2pt}, every node/.style={color=black}, scale={1.5}]
  \begin{scope}[scale={.5}, transform shape]
   \coordinate [label=above:{$\displaystyle \mathbf{1}$}] (A) at (0,0);
   \coordinate [label=below:{$\displaystyle \mathbf{2}$}] (B) at (-1.75,-2.25);
   \coordinate [label=below:{$\displaystyle \mathbf{3}$}] (C) at (+1.75,-2.25);
   \coordinate [label=left:{\footnotesize $\displaystyle {\bf x}_i$}] (m1) at ($(A)!0.5!(B)$);
   \coordinate [label=right:{\footnotesize $\displaystyle \;{\bf x'}_i$}] (m2) at ($(A)!0.5!(C)$);
   \coordinate [label=below:{\footnotesize $\displaystyle {\bf y}_i$}] (m3) at ($(B)!0.5!(C)$);
   \tikzset{point/.style={insert path={ node[scale=2.5*sqrt(\pgflinewidth)]{.} }}} 

   \draw[color=blue,fill=blue] (m1) circle (2pt);
   \draw[color=blue,fill=blue] (m2) circle (2pt);
   \draw[color=red,fill=red] (m3) circle (2pt);

   \draw[-, very thick, color=blue] (B) -- (A);
   \draw[-, very thick, color=blue] (A) -- (C);  
   \draw[-, very thick, color=red] (B) -- (C);    
  \end{scope}
  \begin{scope}[shift={(0,-1.5)}, scale={.5}, transform shape]
   \coordinate (v1) at (-1.5,-2.25);
   \coordinate (v2) at (+1.5,-2.25);
   \draw[fill,color=black] (v1) circle (3pt);
   \draw[fill,color=black] (v2) circle (3pt);
   \draw[thick] (v1) -- (v2);

   \coordinate (l1) at ($(v1)!0.50!(v2)+(0,.5)$);
   \coordinate (l2) at ($(v1)+(.5,.5)$);
   \coordinate (l3) at ($(v2)+(-.5,.5)$);
   \node[ball,thick,color=yellow!60!black, scale=1.25] at (l1) {$\displaystyle 1$};
   \node[ball,thick,color=yellow!60!black, scale=1.25] at (l2) {$\displaystyle 2$};
   \node[ball,thick,color=yellow!60!black, scale=1.25] at (l3) {$\displaystyle 3$};   
  \end{scope}
  \begin{scope}[shift={(4,0)}, scale={.75}, transform shape]
   \def\r{1.75}
   \def\rr{1.95}
   \pgfmathsetmacro\bx{\r*cos{220}};
   \pgfmathsetmacro\by{\r*sin{220}};
   \pgfmathsetmacro\cx{\r*cos{-30}};
   \pgfmathsetmacro\cy{\r*sin{-30}};
   \coordinate (a) at (0,0);
   \coordinate (b) at (\bx,\by);
   \coordinate (c) at (\cx,\cy);
   \coordinate (d) at (0,-\rr);
   \coordinate (e) at ($(b)+(0,-\rr)$);
   \coordinate (f) at ($(c)+(0,-\rr)$);
   \coordinate (g) at ($(b)!0.60!(c)+(0,-.25)$);
   \coordinate (h) at ($(e)!0.60!(f)+(0,-.25)$);

   \node[draw, rectangle, rounded corners, thick, color=yellow!50!black, scale=.75] at (a) {$\displaystyle 123$}; 
   \node[draw, rectangle, rounded corners, thick, color=yellow!50!black, scale=.75] at (b) {$\displaystyle 12$}; 
   \node[draw, rectangle, rounded corners, thick, color=yellow!50!black, scale=.75] at (c) {$\displaystyle 31$}; 
   \node[draw, rectangle, rounded corners, thick, color=yellow!50!black, scale=.75] at (d) {$\displaystyle 23$};
   \node[draw, rectangle, rounded corners, thick, color=yellow!50!black, scale=.75] at (e) {$\displaystyle 2$}; 
   \node[draw, rectangle, rounded corners, thick, color=yellow!50!black, scale=.75] at (f) {$\displaystyle 3$};
   \node[draw, rectangle, rounded corners, thick, color=yellow!50!black, scale=.75] at (g) {$\displaystyle 1$}; 
   \node[draw, rectangle, rounded corners, thick, color=yellow!50!black, scale=.75] at (h) {$\displaystyle \emptyset$};    

   \coordinate (Int1) at ($(a)!0.65!(d)$);
   \coordinate (Int2) at ($(g)!0.50!(h)$);
   \tikzstyle{interrupt}=[
    postaction={
        decorate,
        decoration={markings,
                    mark= at position 0.5 
                          with
                          {
                            \node[rectangle, color=white, fill=white] at (Int1) {};
                            \node[rectangle, color=white, fill=white] at (Int2) {};                            
                          }}}
   ]
   \draw[interrupt, thick, color=yellow!60!black] ($(b)+(.225,0)$) -- ($(g)+(-.2,0)$);
   \draw[interrupt, thick, color=yellow!60!black] ($(g)+(0,-.225)$) -- ($(h)+(0,.2)$);

   \draw[thick, color=yellow!60!black] ($(a)+(-.2,-.2)$) -- ($(b)+(.2,.2)$);
   \draw[thick, color=yellow!60!black] ($(a)+(0,-.2)$) -- ($(d)+(0,.2)$);
   \draw[thick, color=yellow!60!black] ($(a)+(.2,-.2)$) -- ($(c)+(-.2,.2)$);
   \draw[thick, color=yellow!60!black] ($(d)+(-.2,-.2)$) -- ($(e)+(.175,.175)$);
   \draw[thick, color=yellow!60!black] ($(d)+(.2,-.2)$) -- ($(f)+(-.2,.2)$);
   \draw[thick, color=yellow!60!black] ($(b)+(0,-.2)$) -- ($(e)+(0,.2)$);
   \draw[thick, color=yellow!60!black] ($(c)+(0,-.2)$) -- ($(f)+(0,.2)$);
   \draw[thick, color=yellow!60!black] ($(c)+(-.175,-.175)$) -- ($(g)+(.2,0)$);
   \draw[thick, color=yellow!60!black] ($(e)+(.2,0)$) -- ($(h)+(-.2,0)$);
   \draw[thick, color=yellow!60!black] ($(f)+(-.2,0)$) -- ($(h)+(.2,0)$);
  \end{scope}
  \begin{scope}[shift={(7,+.25)}, scale={.5}, transform shape]
   \coordinate (v1) at (0,0);
   \coordinate (v2) at (3,0);
   \coordinate (l2) at ($(v1)!0.50!(v2)+(0,.5)$);
   \coordinate (l1) at ($(v1)+(.5,.5)$);
   \coordinate (l3) at ($(v2)+(-.5,.5)$);
   \node[ball,thick, scale=1.25] (a1) at (l1) {$\displaystyle 1$};
   \node[ball,thick, scale=1.25] (a2) at (l2) {$\displaystyle 2$};
   \node[ball,thick, scale=1.25] (a3) at (l3) {$\displaystyle 3$};
   \node[ball,thick, below=1.25cm of l1, scale=1.25] (b1) {$\displaystyle 3$};
   \node[ball,thick, below=1.25cm of l2, scale=1.25] (b2) {$\displaystyle 2$};
   \node[ball,thick, below=1.25cm of l3, scale=1.25] (b3) {$\displaystyle 1$};
   \draw[->, thick] ($(a1)+(0,-.3)$) -- ($(b1)+(0,.3)$);
   \draw[->, thick] ($(a2)+(0,-.3)$) -- ($(b2)+(0,.3)$);
   \draw[->, thick] ($(a3)+(0,-.3)$) -- ($(b3)+(0,.3)$);
  \end{scope}
  \begin{scope}[shift={(7,-1.25)}, scale={.5}, transform shape]
   \coordinate (v1) at (0,0);
   \coordinate (v2) at (3,0);
   \coordinate (l2) at ($(v1)!0.50!(v2)+(0,.5)$);
   \coordinate (l1) at ($(v1)+(.5,.5)$);
   \coordinate (l3) at ($(v2)+(-.5,.5)$);
   \node[ball,thick, scale=1.25] (a1) at (l1) {$\displaystyle 1$};
   \node[ball,thick, scale=1.25] (a2) at (l2) {$\displaystyle 2$};
   \node[ball,thick, scale=1.25] (a3) at (l3) {$\displaystyle 3$};
   \node[ball,thick, below=1.25cm of l1, scale=1.25] (b1) {$\displaystyle 2$};
   \node[ball,thick, below=1.25cm of l2, scale=1.25] (b2) {$\displaystyle 1$};
   \node[ball,thick, below=1.25cm of l3, scale=1.25] (b3) {$\displaystyle 3$};
   \draw[->, thick] ($(a1)+(0,-.3)$) -- ($(b1)+(0,.3)$);
   \draw[->, thick] ($(a2)+(0,-.3)$) -- ($(b2)+(0,.3)$);
   \draw[->, thick] ($(a3)+(0,-.3)$) -- ($(b3)+(0,.3)$);
  \end{scope}
  \begin{scope}[shift={(7,-2.75)}, scale={.5}, transform shape]
   \coordinate (v1) at (0,0);
   \coordinate (v2) at (3,0);
   \coordinate (l2) at ($(v1)!0.50!(v2)+(0,.5)$);
   \coordinate (l1) at ($(v1)+(.5,.5)$);
   \coordinate (l3) at ($(v2)+(-.5,.5)$);
   \node[ball,thick, scale=1.25] (a1) at (l1) {$\displaystyle 1$};
   \node[ball,thick, scale=1.25] (a2) at (l2) {$\displaystyle 2$};
   \node[ball,thick, scale=1.25] (a3) at (l3) {$\displaystyle 3$};
   \node[ball,thick, below=1.25cm of l1, scale=1.25] (b1) {$\displaystyle 1$};
   \node[ball,thick, below=1.25cm of l2, scale=1.25] (b2) {$\displaystyle 3$};
   \node[ball,thick, below=1.25cm of l3, scale=1.25] (b3) {$\displaystyle 2$};
   \draw[->, thick] ($(a1)+(0,-.3)$) -- ($(b1)+(0,.3)$);
   \draw[->, thick] ($(a2)+(0,-.3)$) -- ($(b2)+(0,.3)$);
   \draw[->, thick] ($(a3)+(0,-.3)$) -- ($(b3)+(0,.3)$);
  \end{scope}
 \end{tikzpicture}
\end{equation*}
where the numbers on the graph represent the vertices as related to the marking indicating them.
The group of the combinatorial automorphism of a triangle is actually two-dimensional, and its generators are given by two of the three transpositions above. Indeed, such transpositions can be seen as linear transformations in energy space, with two of the energies $(x_i,\,y_i,\,x_i')$ exchange while the third one is kept fixed.

Given the combinatorial automorphisms of a triangle, it is possible to find the group of combinatorial automorphisms of any cosmological polytope generated by intersecting $n_e$ triangles as {\it induced} by the latter: the intersections among triangles impose relations among vertices and such relations must be preserved. For a generic tree graph, then 
\begin{equation*}
 \begin{tikzpicture}[ball/.style = {circle, draw, align=center, anchor=north, inner sep=0}, cross/.style={cross out, draw, minimum size=2*(#1-\pgflinewidth), inner sep=0pt, outer sep=0pt}]
  \begin{scope}
   \node[ball,text width=.18cm,fill,color=black,label=above:{$x_1$}] at (0,0) (x1) {};    
   \node[ball,text width=.18cm,fill,color=black,right=1.2cm of x1.east, label=above:{$x_2$}] (x2) {};    
   \node[ball,text width=.18cm,fill,color=black,right=1.2cm of x2.east, label=above:{$x_3$}] (x3) {};
   \node[ball,text width=.18cm,fill,color=black, label=left:{$x_4$}] at (-1,.8) (x4) {};    
   \node[ball,text width=.18cm,fill,color=black, label=above:{$x_5$}] at (-1,-.8) (x5) {};    
   \node[ball,text width=.18cm,fill,color=black, label=below:{$x_6$}] at (-1.7,-2) (x6) {};    
   \node[ball,text width=.18cm,fill,color=black, label=below:{$x_7$}] at (-.3,-2) (x7) {};

   \node[above=.35cm of x5.north] (ref2) {};
   \coordinate (Int2) at (intersection of x5--x1 and ref2--x2);  

   \coordinate (t1) at (x3.east);
   \coordinate (t2) at (x4.west);
   \coordinate (t3) at (x1.south west);
   \coordinate (t4) at (x2.south);

   \draw[-,thick,color=black] (x1) -- (x2) -- (x3); 
   \draw[-,thick,color=black] (x1) -- (x4);
   \draw[-,thick,color=black] (x5) -- (x1);
   \draw[-,thick,color=black] (x5) -- (x7);   
   \draw[-,thick,color=black] (x5) -- (x6); 

   \coordinate (x12) at ($(x1)!0.50!(x2)$);
   \coordinate (x23) at ($(x2)!0.50!(x3)$);
   \coordinate (x14) at ($(x1)!0.25!(x4)$);
   \coordinate (x14u) at ($(x1)!0.75!(x4)$);
   \coordinate (x14c) at ($(x1)!0.50!(x4)$) {};
   \coordinate (x15) at ($(x1)!0.50!(x5)$);
   \coordinate (x56) at ($(x5)!0.50!(x6)$);
   \coordinate (x57) at ($(x5)!0.50!(x7)$);
   
   \node[ball, thick, color=yellow!50!black, scale=.9] at ($(x1)!0.50!(x2)+(0,.4)$) {$\displaystyle 1$};
   \node[ball, thick, color=yellow!50!black, scale=.9] at ($(x1)+(.3,.4)$) {$\displaystyle 2$};
   \node[ball, thick, color=yellow!50!black, scale=.9] at ($(x2)+(-.3,.4)$) {$\displaystyle 3$};
   
   \node[ball, thick, color=yellow!50!black, scale=.9] at ($(x2)!0.50!(x3)+(0,.4)$) {$\displaystyle 4$};
   \node[ball, thick, color=yellow!50!black, scale=.9] at ($(x2)+(.3,.4)$) {$\displaystyle 5$};
   \node[ball, thick, color=yellow!50!black, scale=.9] at ($(x3)+(-.3,.4)$) {$\displaystyle 6$};

   \node[ball, thick, color=yellow!50!black, scale=.9] at ($(x1)!0.50!(x4)+(-.15,0)$) {$\displaystyle 7$};
   \node[ball, thick, color=yellow!50!black, scale=.9] at ($(x1)+(-.35,.2)$) {$\displaystyle 8$};
   \node[ball, thick, color=yellow!50!black, scale=.9] at ($(x4)+(0,-.1)$) {$\displaystyle 9$};

   \node[ball, thick, color=yellow!50!black, scale=.9] at ($(x1)!0.50!(x5)+(.2,.1)$) {\footnotesize $\displaystyle 10$};
   \node[ball, thick, color=yellow!50!black, scale=.9] at ($(x1)+(.05,-.05)$) {\footnotesize $\displaystyle 11$};
   \node[ball, thick, color=yellow!50!black, scale=.9] at ($(x5)+(.35,.2)$) {\footnotesize $\displaystyle 12$};
 
   \node[ball, thick, color=yellow!50!black, scale=.9] at ($(x5)!0.35!(x6)+(-.35,.1)$) {\footnotesize $\displaystyle 13$};
   \node[ball, thick, color=yellow!50!black, scale=.9] at ($(x5)!0.1!(x6)+(-.325,.15)$) {\footnotesize $\displaystyle 14$};
   \node[ball, thick, color=yellow!50!black, scale=.9] at ($(x5)!0.65!(x6)+(-.325,.15)$) {\footnotesize $\displaystyle 15$};  

   \node[ball, thick, color=yellow!50!black, scale=.9] at ($(x5)!0.35!(x7)+(-.1,-.15)$) {\footnotesize $\displaystyle 16$};
   \node[ball, thick, color=yellow!50!black, scale=.9] at ($(x5)!0.1!(x7)+(-.1,-.1)$) {\footnotesize $\displaystyle 17$};
   \node[ball, thick, color=yellow!50!black, scale=.9] at ($(x5)!0.65!(x7)+(-.1,-.1)$) {\footnotesize $\displaystyle 18$};  
  \end{scope}
  \begin{scope}
   \coordinate (t1) at (8,.8) {};
   \node[align=center] (eq1) at (t1) {$\displaystyle 1+3\,=\,4+5$};
   \node[align=center, below=1cm of t1] (eq2) {$\displaystyle 1+2\,=\,7+8\,=\,10+11$};
   \node[align=center, below=1cm of eq2] (eq3) {$\displaystyle 10+12\,=\,13+14\,=\,16+17$};
  \end{scope}
 \end{tikzpicture}
\end{equation*}
and most of the transpositions of the vertices related to the internal edges are no longer allowed given that they do not leave the hyperplanes identified by the vertex relations invariant. However, the relations above make clear the vertex exchanges which do not affect the above relations. First, none of the most external vertices appear in such a relation: all the transpositions between any two of these vertices belongs to the group of combinatorial automorphisms (in the graph above, such vertices are $\left\{6,\,9,\,15,\,18\right\}$ ). Secondly, there are pairs of vertices related to the same edge ({\it i.e.} to the same generating triangle) which appear just in one of the above relations: any transposition within each of such pairs is a combinatorial automorphism -- in the case of the graph above, such pairs are given by $\left\{(4,5),\,(7,8),\,(13,14),\,(16,17)\right\}$. Such pairs are again related to the most outer edges of the graph. Thus, for a {\it generic} tree graph, the group of combinatorial automorphisms involves only the vertices of the cosmological polytope related to the most outer edges of the graph.

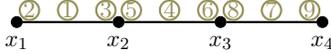
\begin{wrapfigure}{l}{4.5cm}
 \begin{tikzpicture}[ball/.style = {circle, draw, align=center, anchor=north, inner sep=0}, cross/.style={cross out, draw, minimum size=2*(#1-\pgflinewidth), inner sep=0pt, outer sep=0pt}, scale=.8, transform shape]
  \begin{scope}
   \node[ball,text width=.18cm,fill,color=black,label=below:{$x_1$}] at (0,0) (x1) {};    
   \node[ball,text width=.18cm,fill,color=black,right=1.5cm of x1.east, label=below:{$x_2$}] (x2) {};    
   \node[ball,text width=.18cm,fill,color=black,right=1.5cm of x2.east, label=below:{$x_3$}] (x3) {};
   \node[ball,text width=.18cm,fill,color=black,right=1.5cm of x3.east, label=below:{$x_4$}] (x4) {};   

   \draw[thick] (x1) -- (x2) -- (x3) -- (x4);

   \node[ball, thick, color=yellow!50!black, scale=1.25] at ($(x1)!0.50!(x2)+(0,.375)$) {\footnotesize $\displaystyle 1$};
   \node[ball, thick, color=yellow!50!black, scale=1.25] at ($(x1)!0.125!(x2)+(0,.375)$) {\footnotesize $\displaystyle 2$};
   \node[ball, thick, color=yellow!50!black, scale=1.25] at ($(x1)!0.875!(x2)+(0,.375)$) {\footnotesize $\displaystyle 3$};

   \node[ball, thick, color=yellow!50!black, scale=1.25] at ($(x2)!0.50!(x3)+(0,.375)$) {\footnotesize $\displaystyle 4$};
   \node[ball, thick, color=yellow!50!black, scale=1.25] at ($(x2)!0.125!(x3)+(0,.375)$) {\footnotesize $\displaystyle 5$};
   \node[ball, thick, color=yellow!50!black, scale=1.25] at ($(x2)!0.875!(x3)+(0,.375)$) {\footnotesize $\displaystyle 6$};

   \node[ball, thick, color=yellow!50!black, scale=1.25] at ($(x3)!0.50!(x4)+(0,.375)$) {\footnotesize $\displaystyle 7$};
   \node[ball, thick, color=yellow!50!black, scale=1.25] at ($(x3)!0.125!(x4)+(0,.375)$) {\footnotesize $\displaystyle 8$};
   \node[ball, thick, color=yellow!50!black, scale=1.25] at ($(x3)!0.875!(x4)+(0,.375)$) {\footnotesize $\displaystyle 9$};

   \node[align=center] (s1) at ($(x2)!0.5!(x3)+(0,-1.25)$) {$\displaystyle 1+3\,=\,4+5$};
   \node[align=center, below=.5cm of s1] (s2) {$\displaystyle 4+6\,=\,7+8$};
  \end{scope}
 \end{tikzpicture}
\end{wrapfigure}

Were the graph be more symmetric, the group would contain additional symmetry given by the reflection of the vertices with respect the axis of symmetry -- in the simple case of the four-site line graph, together with the transpositions $2\,\longleftrightarrow\,9$ and $\{(1\longleftrightarrow3),\,(7\longleftrightarrow8)\}$, there is a further transformation $(1,3,5)\,\longleftrightarrow\,(7,8,6)$, with $(2,4,9)$ kept frozen.

A loop graph with $n_e$ edges has a lower number of symmetries with respect with any tree graph with the same number of edges. This is just a consequence of the higher number of constraints at its vertices. 

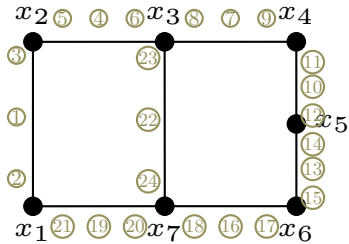
\begin{wrapfigure}{l}{4.75cm}
 \begin{tikzpicture}[ball/.style = {circle, draw, align=center, anchor=north, inner sep=0}, cross/.style={cross out, draw, minimum size=2*(#1-\pgflinewidth), inner sep=0pt, outer sep=0pt}]
  \begin{scope}[scale={1.75}, transform shape]
   \coordinate[label=below:{\tiny $x_1$}] (v1) at (0,0);
   \coordinate[label=above:{\tiny $x_2$}] (v2) at ($(v1)+(0,1.25)$);
   \coordinate[label=above:{\tiny $x_3$}] (v3) at ($(v2)+(1,0)$);
   \coordinate[label=above:{\tiny $x_4$}] (v4) at ($(v3)+(1,0)$);
   \coordinate[label=right:{\tiny $x_5$}] (v5) at ($(v4)-(0,.625)$);   
   \coordinate[label=below:{\tiny $x_6$}] (v6) at ($(v5)-(0,.625)$);
   \coordinate[label=below:{\tiny $x_7$}] (v7) at ($(v6)-(1,0)$);

   \draw[thick] (v1) -- (v2) -- (v3) -- (v4) -- (v5) -- (v6) -- (v7) -- cycle;
   \draw[thick] (v3) -- (v7);

   \draw[fill=black] (v1) circle (2pt);
   \draw[fill=black] (v2) circle (2pt);
   \draw[fill=black] (v3) circle (2pt);
   \draw[fill=black] (v4) circle (2pt);
   \draw[fill=black] (v5) circle (2pt);
   \draw[fill=black] (v6) circle (2pt);
   \draw[fill=black] (v7) circle (2pt);   

   \node[ball, thick, color=yellow!50!black, scale=.5] at ($(v1)!0.60!(v2)+(-.125,0)$) {\footnotesize $\displaystyle 1$};
   \node[ball, thick, color=yellow!50!black, scale=.5] at ($(v1)!0.225!(v2)+(-.125,0)$) {\footnotesize $\displaystyle 2$};
   \node[ball, thick, color=yellow!50!black, scale=.5] at ($(v1)!0.975!(v2)+(-.125,0)$) {\footnotesize $\displaystyle 3$};
 
   \node[ball, thick, color=yellow!50!black, scale=.5] at ($(v2)!0.50!(v3)+(0,.25)$) {\footnotesize $\displaystyle 4$};
   \node[ball, thick, color=yellow!50!black, scale=.5] at ($(v2)!0.225!(v3)+(0,.25)$) {\footnotesize $\displaystyle 5$};
   \node[ball, thick, color=yellow!50!black, scale=.5] at ($(v2)!0.775!(v3)+(0,.25)$) {\footnotesize $\displaystyle 6$};

   \node[ball, thick, color=yellow!50!black, scale=.5] at ($(v3)!0.50!(v4)+(0,.25)$) {\footnotesize $\displaystyle 7$};
   \node[ball, thick, color=yellow!50!black, scale=.5] at ($(v3)!0.225!(v4)+(0,.25)$) {\footnotesize $\displaystyle 8$};
   \node[ball, thick, color=yellow!50!black, scale=.5] at ($(v3)!0.775!(v4)+(0,.25)$) {\footnotesize $\displaystyle 9$};

   \node[ball, thick, color=yellow!50!black, scale=.45] at ($(v5)!0.60!(v4)+(.125,0)$) {\footnotesize $\displaystyle 10$};
   \node[ball, thick, color=yellow!50!black, scale=.45] at ($(v5)!0.25!(v4)+(.125,0)$) {\footnotesize $\displaystyle 12$};
   \node[ball, thick, color=yellow!50!black, scale=.45] at ($(v5)!0.9!(v4)+(.125,0)$) {\footnotesize $\displaystyle 11$};

   \node[ball, thick, color=yellow!50!black, scale=.45] at ($(v6)!0.60!(v5)+(.125,0)$) {\footnotesize $\displaystyle 13$};
   \node[ball, thick, color=yellow!50!black, scale=.45] at ($(v6)!0.25!(v5)+(.125,0)$) {\footnotesize $\displaystyle 15$};
   \node[ball, thick, color=yellow!50!black, scale=.45] at ($(v6)!0.9!(v5)+(.125,0)$) {\footnotesize $\displaystyle 14$};

   \node[ball, thick, color=yellow!50!black, scale=.45] at ($(v6)!0.50!(v7)+(0,-.0625)$) {\footnotesize $\displaystyle 16$};
   \node[ball, thick, color=yellow!50!black, scale=.45] at ($(v6)!0.225!(v7)+(0,-.0625)$) {\footnotesize $\displaystyle 17$};
   \node[ball, thick, color=yellow!50!black, scale=.45] at ($(v6)!0.775!(v7)+(0,-.0625)$) {\footnotesize $\displaystyle 18$};

   \node[ball, thick, color=yellow!50!black, scale=.45] at ($(v7)!0.50!(v1)+(0,-.0625)$) {\footnotesize $\displaystyle 19$};
   \node[ball, thick, color=yellow!50!black, scale=.45] at ($(v7)!0.225!(v1)+(0,-.0625)$) {\footnotesize $\displaystyle 20$};
   \node[ball, thick, color=yellow!50!black, scale=.45] at ($(v7)!0.775!(v1)+(0,-.0625)$) {\footnotesize $\displaystyle 21$};

   \node[ball, thick, color=yellow!50!black, scale=.45] at ($(v7)!0.60!(v3)+(-.125,0)$) {\footnotesize $\displaystyle 22$};
   \node[ball, thick, color=yellow!50!black, scale=.45] at ($(v7)!0.225!(v3)+(-.125,0)$) {\footnotesize $\displaystyle 24$};
   \node[ball, thick, color=yellow!50!black, scale=.45] at ($(v7)!0.975!(v3)+(-.125,0)$) {\footnotesize $\displaystyle 23$};

%   \node[align=center, scale=.5] (s1) at ($(v7)+(0,-.75)$) {$\displaystyle 1+3\,=\,4+5$};
%   \node[align=center, scale=.5, below=.125cm of s1] (s2) {$\displaystyle 4+6\,=\,7+8\,=\,22+23$};
%   \node[align=center, scale=.5, below=.125cm of s2] (s3) {$\displaystyle 7+9\,=\,10+11$};   
%   \node[align=center, scale=.5, below=.125cm of s3] (s4) {$\displaystyle 10+12\,=\,13+14$};
%   \node[align=center, scale=.5, below=.125cm of s4] (s5) {$\displaystyle 13+15\,=\,16+17$};   
%   \node[align=center, scale=.5, below=.125cm of s5] (s6) {$\displaystyle 16+18\,=\,19+20\,=\,22+24$};   
%   \node[align=center, scale=.5, below=.125cm of s6] (s7) {$\displaystyle 1+2\,=\,19+21$};     
  \end{scope}
  \end{tikzpicture} 
\end{wrapfigure}

In general, unless the graph is endowed with any external tree-like structure, the combinatorial automorphisms of a loop cosmological polytope reflect into reflection-like symmetries of the related graph. For the concrete example on the left, the combinatorial automorphism group is one-dimensional and its generator is given by the reflection of the polytope vertices with-respect to the axis passing by the site $x_5$ of the graph $(3,5,4,6,23,8,7,9,11,10)\,\longrightarrow\,(2,21,19,20,24,18,16,17,15,13)$, with the vertices $(1,22)$ are kept frozen.

As they have been discussed so far, the combinatorial automorphisms directly act on the space of vertices. However, our picture of the cosmological polytopes keeps the vertices fixed while any transformation acts on the hyperplane at infinity. It is thus convenient to translate the combinatorial automorphism group into (discrete) transformations directly on the $\mathcal{Y}$ space. 

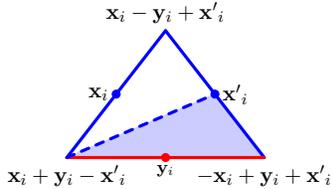
\begin{wrapfigure}{l}{4.75cm}
 \begin{tikzpicture}[line join = round, line cap = round, ball/.style = {circle, draw, align=center, anchor=north, inner sep=0}]
  \begin{scope}[scale={.75}, transform shape]
   \coordinate [label=above:${\bf x}_i-{\bf y}_i+{\bf x'}_i$] (A) at (0,0);
   \coordinate [label=below:${\bf x}_i+{\bf y}_i-{\bf x'}_i$] (B) at (-1.75,-2.25);
   \coordinate [label=below:$-{\bf x}_i+{\bf y}_i+{\bf x'}_i$] (C) at (+1.75,-2.25);
   \coordinate [label=left:${\bf x}_i$] (mx) at ($(A)!0.5!(B)$);
   \coordinate [label=right:${\bf x'}_i$] (mxp) at ($(A)!0.5!(C)$);
   \coordinate [label=below:{\footnotesize ${\bf y}_i$}] (my) at ($(B)!0.5!(C)$);
   \tikzset{point/.style={insert path={ node[scale=2.5*sqrt(\pgflinewidth)]{.} }}} 

   \draw[color=blue!20, fill=blue!20] (B) -- (mxp) -- (C) -- cycle;

   \draw[-, very thick, color=blue] (B) -- (A);
   \draw[-, very thick, color=blue] (A) -- (C);  
   \draw[-, very thick, color=red] (B) -- (C);    
   \draw[-, dashed, very thick, color=blue] (B) -- (mxp);

    \draw[color=blue, fill] (mx) circle (2pt);
   \draw[color=blue, fill] (mxp) circle (2pt);
   \draw[color=red, fill] (my) circle (2pt);
  \end{scope}
 \end{tikzpicture}
\end{wrapfigure}

It is straightforward to see that, for a cosmological polytope $\mathcal{P}_{\mathcal{G}}$ related to a graph $\mathcal{G}$ with an external tree structure, those automorphisms exchanging two vertices of $\mathcal{P}$ related to any of the most outer edges is just the reflection $x_i\,\longleftrightarrow\,y_{i}$, with $x_i$ and $y_{i}$ being the energy of the unconstrained graph node and of the relevant edge respectively: the cosmological polytope $\mathcal{P}$ is the convex hull of the vertices $\{{\bf x}_i-{\bf y}_i+{\bf x'}_i,\,{\bf x}_i+{\bf y}_i-{\bf x'}_i,\,-{\bf x}_i+{\bf y}_i+{\bf x'}_i\}$, with some of the midpoints $x'_i$ identified. Then the vertices related any of the most outer edges $e_i$, which are just the vertices of one of the generating triangles constrained to one of its midpoints ${\bf x}'_i$ only, can be mapped into each other by taking $x'_i$ fixed and exchanging the other two midpoints, {\it i.e.} $x_i\,\longleftrightarrow\,y_i$. 

As far as the other combinatorial automorphisms are concerned, above we mentioned that they corresponds to reflections along the symmetry axis of the graph: the vertices of $\mathcal{P}_{\mathcal{G}}$ along such an axis are kept frozen, while the others are exchanged with the symmetric one with respect to the axis itself. In the $\mathcal{Y}$ space, it corresponds to a map between the ($x_i,\,y_i,\,x'_i$)'s leaving the vertices along a given symmetry axis invariant.

%%%%%%%%%%%%%%%%%%%%%%
%%%%%%%%%%%%%%%%%%%%%%
%%%%%%%%%%%%%%%%%%%%%%

\section{From the scattering amplitudes to the wavefunction of the universe}\label{sec:SAWU}

The cosmological polytope encodes the flat-space S-matrix in its {\it scattering} facet identified by the hyperplane $\mathcal{W}_I\:=\:\sum_{v}x_{vI}$. The analysis of the other facets showed how actually the canonical forms of {\it all} of them encode again the flat space scattering amplitudes -- in the case of the facets corresponding to highest codimension graphs, they admit one and only one triangulation with each terms encoding products of lower point scattering amplitude. This holds both for the tree and loop cosmological polytopes. 

The fact that all the facets of the cosmological polytope encode flat-space processes means that the physics encoded in the residues of the wavefunction of the universe, is nothing but the physics of flat-space scattering! By itself, this is a quite remarkable result, because there is no obvious reason for which it ought to be the case. Indeed a universal feature of our observables is the absence of energy conservation, which manifest itself in their analytic structure through the appearance of a singularity in the total energy. Performing an analytic continuation of the energies we can access the region where the total energy can vanish: in this region of energy space, the full Lorentz symmetry is restored and our observables reduce to the ordinary flat-space S-matrix. What is far less intuitive is that reaching the location of any other pole, the related residue had anything to do with the S-matrix physics {\it at all}: the wavefunctions show further structures which are absent in the S-matrix, such as poles that disappear at the point in energy space where energy conservation is restored. These poles are a specific feature of the wavefunctions, while they do not contribute to the S-matrix.

A question we can now ask is whether our wavefunction of the universe can be generated from the S-matrix. This would imply that the cosmological evolution for our class of models can be reconstructed from the flat-space information! Jumping a little bit ahead, even if true for our class of models, this will unlikely be a universal feature of all cosmological models: a prototypical example is provided by Yang-Mills theory in which the three-point all-plus/minus helicity wavefunctions are non-zero, while the related flat-space S-matrix vanishes \cite{Maldacena:2011nz}. Interestingly, the all-plus/minus helicity wavefunctions in Yang-Mills do not show a singularity in the total energy, which is consistent with the vanishing of the related flat-space S-matrix. However, this observation suggests a connection between the flat-space S-matrices which are zero and the structure of the wavefunction: the existence of vanishing flat-space S-matrices corresponds to polynomial-like terms in the sum of the energies which cannot be eliminated via field redefinition.

Here we will keep focusing on our toy model and answer the question of how flat-space scattering processes determine the wavefunction.

%%%%%%%%%%%%%%%%%%%%%%
%%%%%%%%%%%%%%%%%%%%%%

\subsection{Cosmological polytopes and the wavefunction representative}\label{sec:CPWFr}

As reviewed in Section \ref{sec:CP}, given a space of $n_e$ triangles $\triangle_i$, a cosmological polytope can be defined by intersecting them on their midpoints, imposing relations among their vertices. These triangles can be intersected along two out of their three edges, reflecting the space-time causal structure. 

However, there is a further special point on which triangles can be intersected: it is the vertex which is common to the two intersectable edges. Thus, when we construct polytopes by intersecting triangles, we can also intersect them in this vertex or one can identify a midpoint of a triangle with a vertex of an other one. Notice that two triangles intersected in this way cannot be intersected in a second point only: once a triangle intersects another one in its intersectable vertex, requiring a further intersection between the two identifies an intersectable edge. When a triangle is going to be intersected in its intersectable vertex, we will identify it via the vectors $\left\{{\bf x}+{\bf y}-{\bf x}',\,{\bf x}',\, -{\bf x}+{\bf y}+{\bf x}'\right\}$, with ${\bf x'}$ being the vector of the intersectable vertex, while the other two vertices identify the non-intersectable edge with midpoint ${\bf y}$:

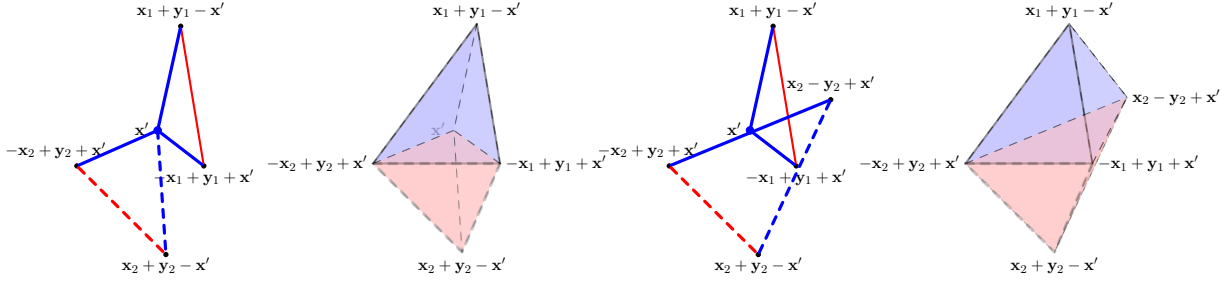
\begin{figure}[H]
 \centering
 \begin{tikzpicture}[line join = round, line cap = round, ball/.style = {circle, draw, align=center, anchor=north, inner sep=0}, scale={.75}, transform shape]
  \begin{scope}[scale={.75}, shift={(7.,.5)}, transform shape]
   \pgfmathsetmacro{\factor}{1/sqrt(2)};  
   \coordinate (B2) at (1.5,-3,-1.5*\factor);
   \coordinate (A1) at (-1.5,-3,-1.5*\factor);
   \coordinate [label=below:{$-{\bf x}_1+{\bf y}_1+{\bf x'}$}] (B1) at (1.5,-3.75,1.5*\factor);
   \coordinate [label=above:{$\hspace{-1cm}-{\bf x}_2+{\bf y}_2+{\bf x'}$}] (A2) at (-1.5,-3.75,1.5*\factor);  
   \coordinate [label=above:{${\bf x}_1+{\bf y}_1-{\bf x'}$}] (C1) at (0.75,-.65,.75*\factor);
   \coordinate [label=below:{${\bf x}_2+{\bf y}_2-{\bf x'}$}] (C2) at (0.4,-6.05,.75*\factor);
   \coordinate (Int) at (intersection of A2--B2 and B1--C1);
   \coordinate [label=left:{${\bf x'_{\phantom{2}}}$}] (Int2) at (intersection of A1--B1 and A2--B2);

   \node[ball,text width=.1cm,fill,color=black] at (B1) (B1b) {};
   \node[ball,text width=.1cm,fill,color=black] at (A2) (A2b) {};
   \node[ball,text width=.1cm,fill,color=black] at (C1) (C1b) {};
   \node[ball,text width=.1cm,fill,color=black] at (C2) (C2b) {};

   \draw[thick,color=red] (B1b) -- (C1b);
   \draw[-,very thick,color=blue] (Int2) -- (B1b);
   \draw[-,very thick,color=blue] (A2b) -- (Int2);
   \draw[-,very thick,color=blue] (Int2) -- (C1b);
   \draw[-, dashed, very thick, color=red] (A2b) -- (C2b);
   \draw[-, dashed, very thick, color=blue] (Int2) -- (C2b);

   \draw [fill,color=blue] (Int2) circle (2.5pt); 
  \end{scope}
  \begin{scope}[scale={.75}, shift={(14,.5)}, transform shape]
   \pgfmathsetmacro{\factor}{1/sqrt(2)}; 
   \coordinate (B2) at (1.5,-3,-1.5*\factor);
   \coordinate (A1) at (-1.5,-3,-1.5*\factor);
   \coordinate [label=right:{$-{\bf x}_1+{\bf y}_1+{\bf x'}$}] (B1) at (1.5,-3.75,1.5*\factor);
   \coordinate [label=left:{$\hspace{-1cm}-{\bf x}_2+{\bf y}_2+{\bf x'}$}] (A2) at (-1.5,-3.75,1.5*\factor);  
   \coordinate [label=above:{${\bf x}_1+{\bf y}_1-{\bf x'}$}] (C1) at (0.75,-.65,.75*\factor);
   \coordinate [label=below:{${\bf x}_2+{\bf y}_2-{\bf x'}$}] (C2) at (0.4,-6.05,.75*\factor);
   \coordinate (Int) at (intersection of A2--B2 and B1--C1);
   \coordinate [label=left:{${\bf x'_{\phantom{2}}}$}] (Int2) at (intersection of A1--B1 and A2--B2);

   \draw[-,dashed,fill=blue!25, opacity=.6] (A2) -- (Int2) -- (C1) -- cycle;
   \draw[-,dashed,fill=blue!20, opacity=.6] (B1) -- (Int2) -- (C1) -- cycle;
   \draw[-,thick,fill=blue!30, opacity=.4] (B1) -- (A2) -- (C1) -- cycle;

   \draw[-,dashed,fill=red!50, opacity=.3] (A2) -- (Int2) -- (C2) -- cycle;
   \draw[-,dashed, fill=red!45, opacity=.3] (B1) -- (Int2) -- (C2) -- cycle;
   \draw[-,dashed, very thick,fill=red!35, opacity=.2] (A2) -- (B1) -- (C2) -- cycle;

  \end{scope}
  \begin{scope}[scale={.75}, shift={(21,.5)}, transform shape]
   \pgfmathsetmacro{\factor}{1/sqrt(2)};  
   \coordinate [label=above:{${\bf x}_2-{\bf y}_2+{\bf x'}$}] (B2) at (1.5,-3,-1.5*\factor);
   \coordinate (A1) at (-1.5,-3,-1.5*\factor);
   \coordinate [label=below:{$-{\bf x}_1+{\bf y}_1+{\bf x'}$}] (B1) at (1.5,-3.75,1.5*\factor);
   \coordinate [label=above:{$\hspace{-1cm}-{\bf x}_2+{\bf y}_2+{\bf x'}$}] (A2) at (-1.5,-3.75,1.5*\factor);  
   \coordinate [label=above:{${\bf x}_1+{\bf y}_1-{\bf x'}$}] (C1) at (0.75,-.65,.75*\factor);
   \coordinate [label=below:{${\bf x}_2+{\bf y}_2-{\bf x'}$}] (C2) at (0.4,-6.05,.75*\factor);
   \coordinate (Int) at (intersection of A2--B2 and B1--C1);
   \coordinate [label=left:{${\bf x'_{\phantom{2}}}$}] (Int2) at (intersection of A1--B1 and A2--B2);

   \node[ball,text width=.1cm,fill,color=black] at (B1) (B1b) {};
   \node[ball,text width=.1cm,fill,color=black] at (A2) (A2b) {};
   \node[ball,text width=.1cm,fill,color=black] at (C1) (C1b) {};
   \node[ball,text width=.1cm,fill,color=black] at (C2) (C2b) {};
   \node[ball,text width=.1cm,fill,color=black] at (B2) (B2b) {};

   \draw[thick,color=red] (B1b) -- (C1b);
   \draw[-,very thick,color=blue] (Int2) -- (B1b);
   \draw[-,very thick,color=blue] (A2b) -- (B2b);
   \draw[-,very thick,color=blue] (Int2) -- (C1b);
   \draw[-, dashed, very thick, color=red] (A2b) -- (C2b);
   \draw[-, dashed, very thick, color=blue] (B2b) -- (C2b);

   \draw [fill,color=blue] (Int2) circle (2.5pt); 
  \end{scope}
  \begin{scope}[scale={.75}, shift={(28,.5)}, transform shape]
   \pgfmathsetmacro{\factor}{1/sqrt(2)}; 
   \coordinate [label=right:{${\bf x}_2-{\bf y}_2+{\bf x'}$}] (B2) at (1.5,-3,-1.5*\factor);
   \coordinate (A1) at (-1.5,-3,-1.5*\factor);
   \coordinate [label=right:{$-{\bf x}_1+{\bf y}_1+{\bf x'}$}] (B1) at (1.5,-3.75,1.5*\factor);
   \coordinate [label=left:{$\hspace{-1cm}-{\bf x}_2+{\bf y}_2+{\bf x'}$}] (A2) at (-1.5,-3.75,1.5*\factor);  
   \coordinate [label=above:{${\bf x}_1+{\bf y}_1-{\bf x'}$}] (C1) at (0.75,-.65,.75*\factor);
   \coordinate [label=below:{${\bf x}_2+{\bf y}_2-{\bf x'}$}] (C2) at (0.4,-6.05,.75*\factor);
   \coordinate (Int) at (intersection of A2--B2 and B1--C1);
   \coordinate (Int2) at (intersection of A1--B1 and A2--B2);

   \draw[-,dashed,fill=blue!25, opacity=.6] (A2) -- (B2) -- (C1) -- cycle;
   \draw[-,dashed,fill=blue!20, opacity=.6] (B1) -- (B2) -- (C1) -- cycle;
   \draw[-,thick,fill=blue!30, opacity=.4] (B1) -- (A2) -- (C1) -- cycle;

   \draw[-,dashed,fill=red!50, opacity=.3] (A2) -- (B2) -- (C2) -- cycle;
   \draw[-,dashed, fill=red!45, opacity=.3] (B1) -- (B2) -- (C2) -- cycle;
   \draw[-,dashed, very thick,fill=red!35, opacity=.2] (A2) -- (B1) -- (C2) -- cycle;

   \end{scope}
 \end{tikzpicture}
\end{figure}

Some comments are now in order. Given a collection of $n_e$ triangles, there is just one polytope $\mathcal{P}_{\mbox{\tiny $\mathcal{A}$}}$ which can be generated by intersecting {\it all of them} in their intersectable vertex: it lives in $\mathbb{P}^{2n_e}$ (the total number of constraints identifying the vertices is $n_e-1$) and it is the convex hull of the vertices
\begin{equation*}
 \left\{{\bf x}_i+{\bf y}_i-{\bf x'},\,-{\bf x}_i+{\bf y}_i+{\bf x'},\,{\bf x'}\right\}\qquad i\,=\,1,\ldots,n_e
\end{equation*}
Furthermore, it has $2n_e+1$ vertices: being in $\mathbb{P}^{2n_e}$, such a polytope is thus a simplex. Finally, knowing the vertices, it is straightforward to compute its canonical form using, for example, the contour integral representation \eqref{eq:CFgF}:
\begin{equation}\label{eq:CFwfr}
 \Omega(\mathcal{P}_{\mbox{\tiny $\mathcal{A}$}})\:=\:dx'\left(\prod_{j=1}^{2n_e}\frac{dx_j dy_j}{\mbox{Vol\{GL(1)\}}}\right)\,\frac{1}{\displaystyle x'+\sum_{j=1}^{2n_e}x_j}\prod_{j=1}^{2n_e}\frac{1}{y_j^2-x_j^2}.
\end{equation}
If we identify the variables $x_j$'s and $x'$ to the energies at the vertices $v_j$'s and $v'$ of a certain graph and $y_j$ to the energy of the edge between $v_j$ and $v'$, then $x'+\sum_{j=1}^{2n_e}x_j$ is the total energy of the graph, while $(y_j^2-x_j^2)^{-1}$ is a Lorentz invariant propagator associated to the edges connecting the vertices $v_j$ and $v'$  -- the putative graph has a star topology, with $n_e$ edges glued together at the vertex $v'$\footnote{ 
There is a special (a larger) class of polytopes that we can construct exploiting the intersectable vertex. Concretely, given $n_e$ triangles, we can intersect $n_e^{\mbox{\tiny (I)}}$ of them in their intersectable midpoints, and $n_e^{\mbox{\tiny (E)}}$ in their intersectable vertex, with overall $r\,=\,n_e-1\,\equiv\,n_e^{\mbox{\tiny (I)}}+n_e^{\mbox{\tiny (E)}}-1$ identifications, in such a way that the $n_e^{\mbox{\tiny (I)}}$ triangles are {\it internal} and the resulting object is connected. This procedure defines a polytope $\mathcal{P}_{\mbox{\tiny $\mathcal{A}$}}$ in $\mathbb{P}^{2n_e}$ and $3n_e^{\mbox{\tiny (I)}}+2n_e^{\mbox{\tiny (E)}}$ vertices:
\begin{equation*}
 \begin{split}
  &\left\{{\bf x}_i-{\bf y}_i+{\bf x'}_i,\,{\bf x}_i+{\bf y}_i-{\bf x'}_i,\,-{\bf x}_i+{\bf y}_i+{\bf x'}_i\right\},\qquad i=1,\ldots,n_e^{\mbox{\tiny (I)}},\\
  &\qquad\left\{\,{\bf x}_j+{\bf y}_j-{\bf x'}_j,\,-{\bf x}_j+{\bf y}_j+{\bf x'}_j\right\},\hspace{1.75cm} j=1,\ldots,n_e^{\mbox{\tiny (E)}}.
 \end{split}
\end{equation*}
}.

Notice that a polytope constructed this way is just a simplex in $\mathbb{P}^{2n_e}$ built by lifting the scattering amplitude facet $\{{\bf x}_i+{\bf y}_i-{\bf x'},\,-{\bf x}_i+{\bf y}_i+{\bf x'}\}$ ($i\,=\,1,\ldots,\,2n_e$) up of one dimension via the extra vertex ${\bf x'}$. Such an extra vertex induces a further pole in the canonical form, which is the total energy pole: the canonical form is nothing but the flat-space scattering amplitude times the total energy pole, or, in other words, it is a representative\footnote{It is {a representative} of the wavefunction of the universe because the latter coincide with the coefficient of our canonical form \eqref{eq:CFwfr}, where the residue is the flat space scattering amplitude. Furthermore, in \eqref{eq:CFwfr} a specific realisation of the flat-space scattering amplitude appears, {\it i.e.} its form depends on a specific way of solving the energy conservation condition (the variable $x'$ has been integrated out).} of the wavefunction of the universe in a neighbourhood of the total energy pole. This is a special case of a larger set of polytopes constructing by uplifting the scattering amplitude facet by adding an extra vertex corresponding of {\it any} of the special point ${\bf x}_i$.

A general scattering facet is the convex hull of $2n_e$ vertices $\{{\bf x}_i+{\bf y}_{ij}-{\bf x'}_j,\,-{\bf x}_i+{\bf y}_{ij}+{\bf x'}_j\}$ in $\mathbb{P}^{2n_e-L-1}$, with $r\,=\,n_e+L-1$ identifications on the vectors ${\bf x}$'s, and it is a simplex only for $L\,=\,0$, as remarked previously. It is possible to consider a polytope $\mathcal{P}_{\mathcal{A}}$ as the convex hull in $\mathbb{P}^{2n_e-L}$ defined by the very same vertices $\{{\bf x}_i+{\bf y}_{ij}-{\bf x'}_j,\,-{\bf x}_i+{\bf y}_{ij}+{\bf x'}_j\}$ and the extra vertex ${\bf x}_{k}$, ${\bf x}_k$ being {\it any} of the vectors parametrising the other vertices. The related canonical form can be again computed via the contour integral representation \eqref{eq:CFgF}. Notably, the integration variable associated to the vertex ${\bf x}_k$ is independent of the free integration variables (irrespectively of how they are chosen), and it turns out to be proportional to the total energy. Hence, the canonical form of this polytope is just the canonical form of the scattering facet times the total energy pole:
\begin{equation}\label{eq:CF1lp}
 \Omega\left(\mathcal{P}_{\mathcal{A}}\right)\:\sim\:\frac{1}{\displaystyle x_k+\sum_{v\in\mathcal{V}\setminus\{v_k\}}x_v}\times\Omega[\mathcal{A}].
\end{equation}
A straightforward computation shows that the canonical form $\Omega[\mathcal{A}]$ is independent of $x_k$: as anticipated, it returns a representation for the scattering amplitude with $x_k$ integrated out by the energy conservation condition. 

For the sake of clarity, let us provide a specific example, starting with the polytope in $\mathbb{P}^3$ identified by the vertices $\{{\bf x}_i+{\bf y}_{i,i+1}-{\bf x}_{i+1},\,-{\bf x}_{i}+{\bf y}_{i,i+1}+{\bf x}_{i+1}\}$ ($i\,=\,1,\,2$) -- it is simply a tetrahedron. We can construct three (isomorphic) polytopes in $\mathbb{P}^4$ by adding, in turn, one of the ${\bf x}_i$ as a further vertex:
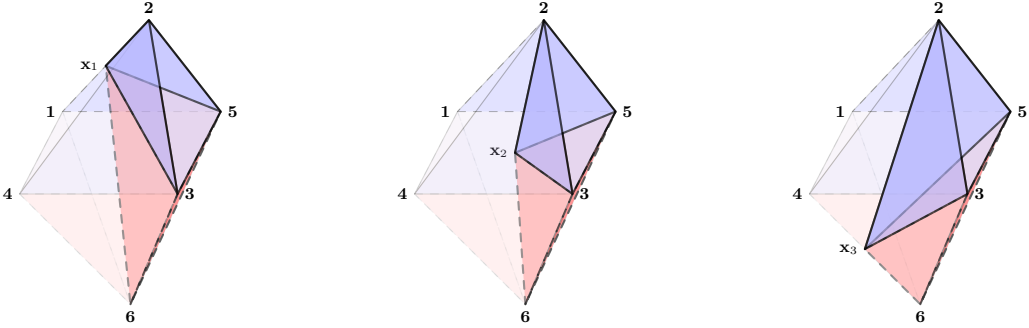
\begin{equation*}
 \begin{tikzpicture}[line join = round, line cap = round, ball/.style = {circle, draw, align=center, anchor=north, inner sep=0}]
  \begin{scope}[scale={.7}, transform shape]
   \pgfmathsetmacro{\factor}{1/sqrt(2)}; 
   \coordinate [label=left:{\footnotesize ${\bf 4}$}] (A2c) at (-1.5,-3.75,1.5*\factor);
   \coordinate [label=below:{\footnotesize ${\bf 6}$}] (B2c) at (0.4,-6.05,.75*\factor);  
   \coordinate [label=left:{\footnotesize ${\bf 1}$}] (A1c) at (-1.5,-3,-1.5*\factor);
   \coordinate [label=above:{\footnotesize ${\bf 2}$}] (B1c) at (0.75,-.65,.75*\factor);
   \coordinate [label=right:{\footnotesize ${\bf 3}$}] (C1c) at (1.5,-3.75,1.5*\factor);
   \coordinate [label=right:{\footnotesize ${\bf 5}$}] (C2c) at (1.5,-3,-1.5*\factor);

   \coordinate [label=left:{\footnotesize ${\bf x}_1$}] (x1) at ($(A1c)!.5!(B1c)$);   

   \node at (A1c) (A1d) {};
   \node at (B2c) (B2d) {};
   \node at (B1c) (B1d) {};
   \node at (A2c) (A2d) {};
   \node at (C1c) (C1d) {};
   \node at (C2c) (C2d) {};

   \draw[-,dashed,fill=blue!25, opacity=.3] (A1c) -- (C2c) -- (B1c) -- cycle;
   \draw[-,fill=blue!15, opacity=.15] (A1c) -- (A2c) -- (B1c) -- cycle;
   \draw[-,fill=blue!15, opacity=.15] (C1c) -- (C2c) -- (B1c) -- cycle;
   \draw[-,fill=blue!30, opacity=.15] (A2c) -- (C1c) -- (B1c) -- cycle;

   \draw[-,dashed, fill=red!25, opacity=.05] (A1c) -- (C2c) -- (B2c) -- cycle;
   \draw[-,dashed, fill=red!45, opacity=.1] (C2c) -- (C1c) -- (B2c) -- cycle;
   \draw[-,dashed, fill=red!35, opacity=.05] (A1c) -- (A2c) -- (B2c) -- cycle;
   \draw[-,dashed, fill=red!40, opacity=.1] (A2c) -- (C1c) -- (B2c) -- cycle;

   \draw[-,dashed, thick, fill=red!50, opacity=.3] (x1) -- (C2c) -- (C1c) -- cycle;
   \draw[-,dashed, thick, fill=red!50, opacity=.4] (x1) -- (C1c) -- (B2c) -- cycle;
   \draw[-,dashed, thick, fill=red!65, opacity=.6] (B2c) -- (C1c) -- (C2c) -- cycle;
   \draw[-, thick, fill=blue!15, opacity=.6] (C1c) -- (C2c) -- (B1c) -- cycle;
   \draw[-,thick,fill=blue!25, opacity=.6] (x1) -- (B1c) -- (C2c) -- cycle;
   \draw[-,thick,fill=blue!30, opacity=.6] (x1) -- (B1c) -- (C1c) -- cycle; 

   \coordinate (ref) at ($(C2c)!0.5!(C1c)$);
   \node[right=.75cm of ref] (pl) {$\displaystyle\phantom{+}$};
  \end{scope}
  \begin{scope}[scale={.7}, shift={(7.5,0)},transform shape]
   \pgfmathsetmacro{\factor}{1/sqrt(2)}; 
   \coordinate [label=left:{\footnotesize ${\bf 4}$}] (A2c) at (-1.5,-3.75,1.5*\factor);
   \coordinate [label=below:{\footnotesize ${\bf 6}$}] (B2c) at (0.4,-6.05,.75*\factor);  
   \coordinate [label=left:{\footnotesize ${\bf 1}$}] (A1c) at (-1.5,-3,-1.5*\factor);
   \coordinate [label=above:{\footnotesize ${\bf 2}$}] (B1c) at (0.75,-.65,.75*\factor);
   \coordinate [label=right:{\footnotesize ${\bf 3}$}] (C1c) at (1.5,-3.75,1.5*\factor);
   \coordinate [label=right:{\footnotesize ${\bf 5}$}] (C2c) at (1.5,-3,-1.5*\factor);

   \coordinate [label=left:{\footnotesize ${\bf x}_2$}](x2) at (intersection of A2c--C2c and A1c--C1c);   

   \node at (A1c) (A1d) {};
   \node at (B2c) (B2d) {};
   \node at (B1c) (B1d) {};
   \node at (A2c) (A2d) {};
   \node at (C1c) (C1d) {};
   \node at (C2c) (C2d) {};

   \draw[-,dashed,fill=blue!25, opacity=.3] (A1c) -- (C2c) -- (B1c) -- cycle;
   \draw[-,fill=blue!15, opacity=.15] (A1c) -- (A2c) -- (B1c) -- cycle;
   \draw[-,fill=blue!15, opacity=.15] (C1c) -- (C2c) -- (B1c) -- cycle;
   \draw[-,fill=blue!30, opacity=.15] (A2c) -- (C1c) -- (B1c) -- cycle;

   \draw[-,dashed, fill=red!25, opacity=.05] (A1c) -- (C2c) -- (B2c) -- cycle;
   \draw[-,dashed, fill=red!45, opacity=.1] (C2c) -- (C1c) -- (B2c) -- cycle;
   \draw[-,dashed, fill=red!35, opacity=.05] (A1c) -- (A2c) -- (B2c) -- cycle;
   \draw[-,dashed, fill=red!40, opacity=.1] (A2c) -- (C1c) -- (B2c) -- cycle;

   \draw[-,dashed, thick, fill=red!50, opacity=.3] (x2) -- (C2c) -- (C1c) -- cycle;
   \draw[-,dashed, thick, fill=red!50, opacity=.4] (x2) -- (C1c) -- (B2c) -- cycle;
   \draw[-,dashed, thick, fill=red!65, opacity=.6] (B2c) -- (C1c) -- (C2c) -- cycle;
   \draw[-, thick, fill=blue!15, opacity=.6] (C1c) -- (C2c) -- (B1c) -- cycle;
   \draw[-,thick,fill=blue!25, opacity=.6] (x2) -- (B1c) -- (C2c) -- cycle;
   \draw[-,thick,fill=blue!30, opacity=.6] (x2) -- (B1c) -- (C1c) -- cycle; 

   \coordinate (ref) at ($(C2c)!0.5!(C1c)$);
   \node[right=.75cm of ref] (pl) {$\displaystyle\phantom{+}$};
  \end{scope}
  \begin{scope}[scale={.7}, shift={(15,0)},transform shape]
   \pgfmathsetmacro{\factor}{1/sqrt(2)}; 
   \coordinate [label=left:{\footnotesize ${\bf 4}$}] (A2c) at (-1.5,-3.75,1.5*\factor);
   \coordinate [label=below:{\footnotesize ${\bf 6}$}] (B2c) at (0.4,-6.05,.75*\factor);  
   \coordinate [label=left:{\footnotesize ${\bf 1}$}] (A1c) at (-1.5,-3,-1.5*\factor);
   \coordinate [label=above:{\footnotesize ${\bf 2}$}] (B1c) at (0.75,-.65,.75*\factor);
   \coordinate [label=right:{\footnotesize ${\bf 3}$}] (C1c) at (1.5,-3.75,1.5*\factor);
   \coordinate [label=right:{\footnotesize ${\bf 5}$}] (C2c) at (1.5,-3,-1.5*\factor);

   \coordinate [label=left:{\footnotesize ${\bf x}_3$}] (x3) at ($(A2c)!0.5!(B2c)$);   

   \node at (A1c) (A1d) {};
   \node at (B2c) (B2d) {};
   \node at (B1c) (B1d) {};
   \node at (A2c) (A2d) {};
   \node at (C1c) (C1d) {};
   \node at (C2c) (C2d) {};

   \draw[-,dashed,fill=blue!25, opacity=.3] (A1c) -- (C2c) -- (B1c) -- cycle;
   \draw[-,fill=blue!15, opacity=.15] (A1c) -- (A2c) -- (B1c) -- cycle;
   \draw[-,fill=blue!15, opacity=.15] (C1c) -- (C2c) -- (B1c) -- cycle;
   \draw[-,fill=blue!30, opacity=.15] (A2c) -- (C1c) -- (B1c) -- cycle;

   \draw[-,dashed, fill=red!25, opacity=.05] (A1c) -- (C2c) -- (B2c) -- cycle;
   \draw[-,dashed, fill=red!45, opacity=.1] (C2c) -- (C1c) -- (B2c) -- cycle;
   \draw[-,dashed, fill=red!35, opacity=.05] (A1c) -- (A2c) -- (B2c) -- cycle;
   \draw[-,dashed, fill=red!40, opacity=.1] (A2c) -- (C1c) -- (B2c) -- cycle;

   \draw[-,dashed, thick, fill=red!50, opacity=.3] (x3) -- (C2c) -- (C1c) -- cycle;
   \draw[-,dashed, thick, fill=red!50, opacity=.4] (x3) -- (C1c) -- (B2c) -- cycle;
   \draw[-,dashed, thick, fill=red!65, opacity=.6] (B2c) -- (C1c) -- (C2c) -- cycle;
   \draw[-, thick, fill=blue!15, opacity=.6] (C1c) -- (C2c) -- (B1c) -- cycle;
   \draw[-,thick,fill=blue!25, opacity=.6] (x3) -- (B1c) -- (C2c) -- cycle;
   \draw[-,thick,fill=blue!30, opacity=.6] (x3) -- (B1c) -- (C1c) -- cycle; 

   \coordinate (ref) at ($(C2c)!0.5!(C1c)$);
   \node[right=.75cm of ref] (pl) {$\displaystyle\phantom{+}$};
  \end{scope}
 \end{tikzpicture}
\end{equation*}
It is again a simplex with canonical form given respectively by
\begin{equation}
 \begin{split}
  &\Omega_1\:\sim\:\prod_{i=1}^{3}dx_i\prod_{i=1}^{2}dy_{ij}\frac{1}{\mbox{Vol}\{GL(1)\}}\,\frac{1}{x_1+x_2+x_3}\frac{1}{(y_{12}^2-(x_2+x_3)^2)(y_{23}^2-x_3^2)},\\
  &\Omega_2\:\sim\:\prod_{i=1}^{3}dx_i\prod_{i=1}^{2}dy_{ij}\frac{1}{\mbox{Vol}\{GL(1)\}}\,\frac{1}{x_1+x_2+x_3}\frac{1}{(y_{12}^2-x_1^2)(y_{23}^2-x_3^2)},\\
  &\Omega_3\:\sim\:\prod_{i=1}^{3}dx_i\prod_{i=1}^{2}dy_{ij}\frac{1}{\mbox{Vol}\{GL(1)\}}\,\frac{1}{x_1+x_2+x_3}\frac{1}{(y_{12}^2-x_1^2)(y_{23}^2-(x_1+x_2)^2)}.
 \end{split}
\end{equation}
with the polytope generated by adding the $x_2$ vertex can be seen, as discussed earlier, as generated by intersecting two triangles at the intersectable vertex and with canonical form $\Omega_2$ which his the $n_e\,=\,2$ case of \eqref{eq:CFwfr}. Actually, all the polytopes constructed by adding a vertex ${\bf x}_k$ at the scattering facet, can be generated via triangles with a common vertex ${\bf x}_k$ and the other vertices $\{{\bf x}_i+{\bf y}_{ij}-{\bf x}_j,\,-{\bf x}_i+{\bf y}_{ij}+{\bf x}_j\}$.

Finally, given a full-fledge cosmological polytope generated by intersecting $n_e$ triangles in some of their midpoints, one can explore its triangulations using internal points. There are special point that one can consider, which are precisely the midpoints of the generating triangles. Let us consider the triangulation of the cosmological polytope including one of such a point. Among all the possible triangulations, there is at least one such that all the simplices has this extra vertex and one of the simplices is a polytope of the type we have been discussing so far, whose canonical form provides a representative of the wavefunction of the universe in a neighbourhood of the total energy conservation. In other words, any polytope obtained by the adding a vertex ${\bf x}_k$ to the scattering facet is a simplex belonging to a triangulation of a certain cosmological polytope.

%%%%%%%%%%%%%%%%%%%%%%
%%%%%%%%%%%%%%%%%%%%%%

\subsection{Reconstructing the wavefunction}\label{subsec:WFrec}

The analysis of the cosmological polytope facets allowed us to understand the residue of {\it any} pole of the wavefunction of the universe in terms of scattering processes: the canonical form of each of the facet can be nicely written in a Lorentz invariant way. In particular, there is a subset of facets that is {\it isomorphic} to the scattering facets. We also learnt that the wavefunction enjoys a group of discrete symmetries in energy space, which, from the cosmological polytope perspective corresponds to the group of combinatorial automorphisms. 

We would now like to turn the table around and ask: if we were given the flat-space scattering amplitude, can we reconstruct the  wavefunction of the universe? Or how much information can we extract about this cosmological observable? Which is the minimal set of information that we need?

Let us assume that, together with a tree-level (contribution to the) scattering amplitude $\mathcal{A}_{\mathcal{G}}$, corresponding to a certain graph $\mathcal{G}$, we are provided with the information that the object we want to construct has to be invariant under the exchange $x_i\,\longleftrightarrow\,y_i$ of the energies related to the most outer edges $e_i$ of $\mathcal{G}$:

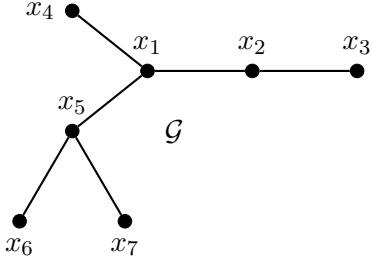
\begin{equation}\label{eq:AG}
 \begin{tikzpicture}[ball/.style = {circle, draw, align=center, anchor=north, inner sep=0}, cross/.style={cross out, draw, minimum size=2*(#1-\pgflinewidth), inner sep=0pt, outer sep=0pt}]
  \begin{scope}
   \node[ball,text width=.18cm,fill,color=black,label=above:{$x_1$}] at (0,0) (x1) {};    
   \node[ball,text width=.18cm,fill,color=black,right=1.2cm of x1.east, label=above:{$x_2$}] (x2) {};    
   \node[ball,text width=.18cm,fill,color=black,right=1.2cm of x2.east, label=above:{$x_3$}] (x3) {};
   \node[ball,text width=.18cm,fill,color=black, label=left:{$x_4$}] at (-1,.8) (x4) {};    
   \node[ball,text width=.18cm,fill,color=black, label=above:{$x_5$}] at (-1,-.8) (x5) {};    
   \node[ball,text width=.18cm,fill,color=black, label=below:{$x_6$}] at (-1.7,-2) (x6) {};    
   \node[ball,text width=.18cm,fill,color=black, label=below:{$x_7$}] at (-.3,-2) (x7) {};

   \node[above=.35cm of x5.north] (ref2) {};
   \coordinate (Int2) at (intersection of x5--x1 and ref2--x2);  

   \coordinate (t1) at (x3.east);
   \coordinate (t2) at (x4.west);
   \coordinate (t3) at (x1.south west);
   \coordinate (t4) at (x2.south);

   \draw[-,thick,color=black] (x1) -- (x2) -- (x3); 
   \draw[-,thick,color=black] (x1) -- (x4);
   \draw[-,thick,color=black] (x5) -- (x1);
   \draw[-,thick,color=black] (x5) -- (x7);   
   \draw[-,thick,color=black] (x5) -- (x6); 

   \coordinate (x12) at ($(x1)!0.50!(x2)$);
   \coordinate (x23) at ($(x2)!0.50!(x3)$);
   \coordinate (x14) at ($(x1)!0.25!(x4)$);
   \coordinate (x14u) at ($(x1)!0.75!(x4)$);
   \coordinate (x14c) at ($(x1)!0.50!(x4)$) {};
   \coordinate (x15) at ($(x1)!0.50!(x5)$);
   \coordinate (x56) at ($(x5)!0.50!(x6)$);
   \coordinate (x57) at ($(x5)!0.50!(x7)$);
   
   \node[right=1cm of x5, align=center] {$\displaystyle\mathcal{G}$};
  \end{scope}
  \begin{scope}
   \coordinate (t1) at (8,-.8) {};
   \node[align=center] (eq1) at (t1) {$\displaystyle\mathcal{A}_{\mathcal{G}}\,=\,\prod_{e\in\mathcal{E}_{\mbox{\tiny ext}}}\frac{1}{y_{e}^2-x_{v_e}^2}\prod_{I\in\mathcal{E}_{\mbox{\tiny int}}}\frac{1}{y_{I}^2-x_{v_I}^2}$};
  \end{scope}
 \end{tikzpicture}
\end{equation}
where $x_{v_e}$ are the energies at the most outer vertices $v_e$ and $x_{v_I}\,=\,\sum_{\bar{v}\in\mathcal{R}_I}x_{\bar{v}}\:=\:-\sum_{\bar{v}\in\mathcal{L}_I}x_{\bar{v}}$, $\mathcal{R}_I$ ($\mathcal{L}_I$) being the set of vertices on the right (left) of the edge $e_{I}$. Notice that $\mathcal{A}$ depends on all but one $x_v$'s because of the energy conservation condition $\sum_v x_v\,=\,0$: when the Lorentz invariant propagators are expressed in terms of the energies $x_v$ and $y_e$, each of them can acquire different forms depending on which $x_v$ is integrated out. 

Given a representation, such as \eqref{eq:AG}, for an amplitude $\mathcal{A}_{\mathcal{G}}$, we can construct a representative of the wavefunction of the universe in a neighbourhood of the energy conservation sheet:
\begin{equation}\label{eq:PAG}
 \Psi_{\mathcal{A}_{\mathcal{G}}}\:=\:\frac{\mathcal{A}_{\mathcal{G}}}{\sum_{v\in\mathcal{V}}x_v}.
\end{equation}
This is a quantity defined out of the energy conservation sheet, on which it has a pole whose residue is the scattering amplitude. It is a representative of the wavefunction of the universe $\Psi_{\mathcal{G}}$ related to the graph $\mathcal{G}$ because the latter has to acquire the form $\mathcal{A}_{\mathcal{G}}/\sum_{v\in\mathcal{V}}x_v$ as $\sum_{v\in\mathcal{V}}x_v\,\longrightarrow\,0$. We can construct the representative \eqref{eq:PAG} out of any of the possible representations of $\mathcal{A}_{\mathcal{G}}$: while the latter are all equivalent to each other, the various $\Psi_{\mathcal{A}_{\mathcal{G}}}$ we can define are not, except on the energy conservation sheet.

Already now, we can associate a polytope $\tilde{\mathcal{P}}_{\mathcal{A}_{\mathcal{G}}}$ to $\Psi_{\mathcal{A}_{\mathcal{G}}}$: in the projective space $\mathbb{P}^{n_v+n_e-1}$ defined by the hyperplane at infinity $\mathcal{Y}\,=\,(\{x_v\},\,\{y_e\})$ of the energies, it is the convex hull of vectors $Z_{\mbox{\tiny $(i)$}}$ such that $\langle\mathcal{Y},Z_{\mbox{\tiny $(i)$}}\rangle\,\equiv\,\mathcal{Y}_IZ_{\mbox{\tiny $(i)$}}^I$ corresponds to a linear propagator in \eqref{eq:PAG} for any $i$ -- {\it i.e.} each vertex correspond to a pole of the wavefunction of the universe. The wavefunction of the universe is then the volume of $\tilde{\mathcal{P}}_{\mathcal{A}_{\mathcal{G}}}$. Notice that, except for the total energy pole, there are two poles for each edge of the graph $\mathcal{G}$: the polytope $\tilde{\mathcal{P}}_{\mathcal{A}_{\mathcal{G}}}$ has $2n_e+1$ vertices. Given that for any three graphs the number of vertices is $n_v\,=\,n_e+1$, then $\tilde{\mathcal{P}}_{\mathcal{A}_{\mathcal{G}}}$ has $n_e+n_v$ vertices in $\mathbb{P}^{n_v+n_e-1}$ and, hence, it is a simplex. As usual, given a polytope, it is possible to define its dual by mapping its codimension-$k$ boundaries into codimension-$(N-k+1)$ ones ($N\,\equiv\,n_v+n_e-1$). In the case of the polytope $\tilde{\mathcal{P}}_{\mathcal{A}_{\mathcal{G}}}$ we are concerned with, its dual $\mathcal{P}_{\mathcal{A}_{\mathcal{G}}}$ is the convex hull defined by the vertices ${\bf\left\{\right.}{\bf x}_{\bar{v}},\,\{{\bf x}_{v_{e_i}}+{\bf y}_{e_i}-{\bf x}_{v'_{e_i}},\,-{\bf x}_{v_{e_i}}+{\bf y}_{e_i}+{\bf x}_{v'_{e_i}}\}_{i=1}^{n_e}{\left.\bf\right\}}$, and its canonical form $\Omega(\mathcal{Y},\mathcal{P}_{\mathcal{A}_{\mathcal{G}}})$ is related to the volume of $\tilde{\mathcal{P}}_{\mathcal{A}_{\mathcal{G}}}$ -- this is nothing but the polytope \eqref{eq:CF1lp} constructed from the scattering facet of a cosmological polytope by adding an extra vertex.

Secondly, let us consider again the scattering amplitude $\mathcal{A}_{\mathcal{G}}$: it is mapped onto itself, up to an overall sign, by a class of discrete transformations in energy space. A subset of these transformations are symmetries of $\mathcal{A}_{\mathcal{G}}$  and it is inherited by the wavefunction representative $\Psi_{\mathcal{A}_{\mathcal{G}}}$: such a subset constitutes the group of combinatorial automorphisms of $\mathcal{P}_{\mathcal{A}_{\mathcal{G}}}$. Let us focus on such transformations modulo the group of combinatorial automorphisms of $\mathcal{P}_{\mathcal{A}_{\mathcal{G}}}$. In energy space, these are all those transformations exchanging $y_{e}$ and $x_{v_e}$ related to the same edge $e$. More precisely, given the specific representation of the scattering amplitude $\mathcal{A}_{\mathcal{G}}$, they are given by all the exchanges $y_{I}\,\longleftrightarrow\,x_{v_I}$ related to each internal edge $e_I$ as well as those exchanges $y_e\,\longleftrightarrow\,x_{v_e}$ related to the most outer edges, such that no energy $x_{v_e}$ is included into the sum $x_{v_I}$. Importantly these two sets of transformations commute among each other.

A comment is now in order. The energy exchanges on the outer edges are symmetries of the wavefunction of the universe. This is a piece of information that we have assumed to be given, together with the flat-space scattering amplitude. Furthermore, if on one side the amplitude can be written in different representations because of energy conservation, the wavefunction representative takes a given representation of the amplitude out of the energy conservation sheet so that, at a generic point in energy space, the wavefunction representatives $\Psi_{\mathcal{A}_{\mathcal{G}}}$ and $\Psi_{\mathcal{A'}_{\mathcal{G}}}$ defined as in \eqref{eq:PAG} but using two different representations $\mathcal{A}_{\mathcal{G}}$ and $\mathcal{A'}_{\mathcal{G}}$ for the flat-space amplitude are not equivalent. However, the final object we would like to construct should not know about the specific representation chosen for the amplitude -- {\it i.e.} how the energy conservation is implemented before taking the flat-space amplitude out of the energy conservation sheet and defining $\Psi_{\mathcal{A}_{\mathcal{G}}}$. This can be achieved by constructing first an object which is invariant under the exchanges $y_{I}\,\longleftrightarrow\,x_{v_I}$ related to each internal edge $e_I$.

An object $\tilde{\Psi}_{\mathcal{A}_{\mathcal{G}}}$ which enjoys such transformations as symmetries is then given by the sum of the wavefunction representative $\Psi_{\mathcal{A}_{\mathcal{G}}}$ with its images under each of these transformations and their inequivalent compositions:
\begin{equation}\label{eq:PAG2}
 \tilde{\Psi}_{\mathcal{A}_{\mathcal{G}}}\:=\:\sum_{t\in\tau}\Img_t\left\{\Psi_{\mathcal{A}_{\mathcal{G}}}\right\},
\end{equation}
where $\tau$ is the set of transformations, including their inequivalent compositions as well as the identity, and $\Img_t\{\Psi_{\mathcal{A}_{\mathcal{G}}}\}$ is the image of $\Psi_{\mathcal{A}_{\mathcal{G}}}$ under the transformation $t\,\in\,\tau$. Interestingly, given that each of these transformations map the scattering amplitude into itself (up to a sign), each term in \eqref{eq:PAG2} which is not $\Psi_{\mathcal{A}_{\mathcal{G}}}$, differs with respect to $\Psi_{\mathcal{A}_{\mathcal{G}}}$ itself by the extra denominator, which now is no longer the total energy but a sum of $x$'s and $y$'s, {\it e.g.}
\begin{equation}\label{eq:Exe}
 \sum_{v\in\mathcal{V}}x_v\:\equiv\:x_{v_I}+\sum_{v\in\mathcal{V}\setminus\{v_{I}\}}x_{v}\:\xrightarrow{\mbox{\tiny $y_{e_I}\,\longleftrightarrow\,x_{v_I}$}}\:y_{e_I}+\sum_{v\in\mathcal{V}\setminus\{v_{I}\}}x_{v},
\end{equation}
with multiple $y$'s appearing in \eqref{eq:Exe} as a consequence of the composition of such transformations. Each of these transformations is related to a certain internal edge of the graph $\mathcal{G}$. Let $\tilde{n}_e$ be the number of such transformations. Then, $\tilde{\Psi}_{\mathcal{A}_{\mathcal{G}}}$ has $2^{\tilde{n}_e}$ terms. Each of the terms in \eqref{eq:PAG2} has a polytope interpretation: given the polytope $\mathcal{P}_{\mathcal{A}_{\mathcal{G}}}$, each of these transformations maps a vertex of $\mathcal{P}_{\mathcal{A}_{\mathcal{G}}}$ into its mirror with respect to a (fixed) codimension-$1$ hyperplane identified by all the other vertices and, thus, each image of $\mathcal{P}_{\mathcal{A}_{\mathcal{G}}}$ is a new simplex in $\mathbb{P}^{n_e+n_v-1}$ with $\Img_t\left\{\Psi_{\mathcal{A}_{\mathcal{G}}}\right\}$ being the coefficient of its canonical form. Importantly, these maps leave the vertex ${\bf x}_{\bar{v}}$ invariant.

Finally, we would like to have an object which is invariant under the exchange $y_e\,\longleftrightarrow\,x_{v_e}$, for all the most outer edges $e$, which are the symmetries we ascribe to the wavefunction of the universe. It can be constructed as a sum of images of $ \tilde{\Psi}_{\mathcal{A}_{\mathcal{G}}}$ under these transformations, including the identity:
\begin{equation}\label{eq:PAG3}
 \hat{\Psi}_{\mathcal{A}_{\mathcal{G}}}\:=\:\sum_{s\in\sigma}\Img_s\left\{\tilde{\Psi}_{\mathcal{A}_{\mathcal{G}}}\right\},
\end{equation}
where $\sigma$ is the set of discrete transformations $y_{e}\,\longleftrightarrow\,x_{v_e}$ related to the most outer edges of the graph $\mathcal{G}$.
The new object $\hat{\Psi}_{\mathcal{A}_{\mathcal{G}}}$ has $2^{n_e}$ terms. Again, {\it any} transformation $s\in\sigma$ maps a simplex into another simplex. Those transformations which commute with the $y_{e_I}\,\longleftrightarrow\,x_{v_I}$ implemented earlier, also map the amplitude $\mathcal{A}_{\mathcal{G}}$ onto itself (up to a sign) changing the extra pole:
\begin{equation}\label{eq:Exe2}
 \begin{split}
  &\sum_{v\in\mathcal{V}}x_v\:\equiv\:x_{v_e}+\sum_{v\in\mathcal{V}\setminus\{v_{e}\}}x_{v}\:\xrightarrow{\mbox{\tiny $y_{e}\,\longleftrightarrow\,x_{v_e}$}}\:y_{e}+\sum_{v\in\mathcal{V}\setminus\{v_{e}\}}x_{v},\\
  &y_{e_I}+x_{v_e}+\sum_{v\in\mathcal{V}\setminus\{v_{I},v_e\}}x_{v}\:\xrightarrow{\mbox{\tiny $y_{e}\,\longleftrightarrow\,x_{v_e}$}}\:y_{e_I}+y_{e}+\sum_{v\in\mathcal{V}\setminus\{v_{I},v_e\}}x_{v}
 \end{split}
\end{equation}
There are also transformations such that $v_e\,\subset\,v_I$. In this case, the linear pole $y_{e_I}+\sum_{v\in\mathcal{V}\setminus\{v_{I}\}}x_{v}$ is left invariant, while the amplitude $\mathcal{A}_{\mathcal{G}}$ is mapped into another object $\mathcal{A}'_{\mathcal{G}}$
\begin{equation}\label{eq:Exe3}
 \begin{tikzpicture}[ball/.style = {circle, draw, align=center, anchor=north, inner sep=0}, cross/.style={cross out, draw, minimum size=2*(#1-\pgflinewidth), inner sep=0pt, outer sep=0pt}]
  \node (A) at (0,0) {$\displaystyle\mathcal{A}_{\mathcal{G}}\,=\,\prod_{e\in\mathcal{E}_{\mbox{\tiny ext}}}\frac{1}{y_{e}^2-x_{v_e}^2}\prod_{I\in\mathcal{E}_{\mbox{\tiny int}}}\frac{1}{y_{I}^2-x_{v_I}^2}$};
  \node (B) at ($(A)+(5,-1.5)$) {$\displaystyle\mathcal{A}'_{\mathcal{G}}\,=\,\underbrace{\frac{-1}{y_{e}^2-x_{v_e}^2}}_{-\mathcal{A}_{e}}
          			  \underbrace{\prod_{\bar{e}\in\mathcal{E}_{\mbox{\tiny ext}}\setminus\{e\}}\frac{1}{y_{\bar{e}}^2-x_{v_{\bar{e}}}^2}
				  \prod_{I\in\mathcal{E}'_{\mbox{\tiny int}}}\frac{1}{y_{I}^2-x_{v_I}^2}
				  \prod_{I\in\mathcal{E}''_{\mbox{\tiny int}}}\frac{1}{y_{I}^2-(x_{v'_I}+y_{e})^2}}_{\mathcal{A}_{\mathcal{G}\setminus\{e\}}}$};
  \draw[->] (A.south west) edge[bend right] node[pos=0.5, below, sloped, scale=.75] {$y_{e}\,\longleftrightarrow\,x_{v_e}$} (B.west);
 \end{tikzpicture}
\end{equation}
where now the sum $x_{v'_I}$ does not involve $x_{v_e}$, and $\mathcal{E}'_{\mbox{\tiny int}}\,\cup\,\mathcal{E}''_{\mbox{\tiny int}}\:=\:\mathcal{E}_{\mbox{\tiny int}}$: $\mathcal{A}'_{\mathcal{G}}$ factorises into (minus) the product of a two vertex amplitude $\mathcal{A}_{e}$ and an $n_v-1$ vertex one $\mathcal{A}_{\mathcal{G}\setminus\{e\}}$. The images under compositions of $n_{\sigma}$ such transformations have still the same linear propagator $y_{e_I}+\sum_{v\in\mathcal{V}\setminus\{v_{I}\}}x_{v}$ times a product of $n_{\sigma}+1$ lower vertex amplitudes.

From a polytope perspective, these transformations on $\mathcal{P}_{\mathcal{A}_{\mathcal{G}}}$ takes a codimension-$1$ hyperplane identified by all but one of its vertices, and maps the remaining one into its mirror with respect such an hyperplane. Thus, given that the original polytope  $\mathcal{P}_{\mathcal{A}_{\mathcal{G}}}$ is a simplex in $\mathbb{P}^{n_e+n_v-1}$, all its images under these reflections (and their compositions) are still simplices. The sum of all these simplices is another polytope $\hat{\mathcal{P}}_{\mathcal{G}}$ with canonical form  $\Omega(\hat{\mathcal{P}}_{\mathcal{G}})\:=\:\prod_{v,e}dx_v dy_e\,\hat{\Psi}_{\mathcal{A}_{\mathcal{G}}}(x,y)$. We now need to prove that $\hat{\Psi}_{\mathcal{A}_{\mathcal{G}}}$ is actually the wavefunction of the universe $\Psi_{\mathcal{G}}$ related to the graph $\mathcal{G}$, or equivalently, that $\hat{\mathcal{P}}_{\mathcal{G}}$ is the actual cosmological polytope $\mathcal{P}_{\mathcal{G}}$ related to the graph $\mathcal{G}$.

Let us start with the polytope $\mathcal{P}_{\mathcal{A}_{\mathcal{G}}}$ in $\mathbb{P}^{n_e+n_v-1}$, whose canonical form is given by $\Omega(\mathcal{P}_{\mathcal{A}_{\mathcal{G}}};\,x,y)\:=\:\Psi_{\mathcal{A}_{\mathcal{G}}}(x,y)\prod_{v,e}dx_vdy_e$. It is defined as the convex hull of the vertices $\left\{{\bf x}_{\bar{v}},\,\{{\bf x}_i+{\bf y}_{e_i}-{\bf x}_i,\,-{\bf x}_i+{\bf y}_{e_i}+{\bf x'}_i\}_{i=1}^{n_e}\right\}$. In $\mathcal{P}_{\mathcal{A}_{\mathcal{G}}}$ in $\mathbb{P}^{n_e+n_v-1}$, the discrete transformations we have been studied, can be viewed as morphisms on $\mathcal{P}_{\mathcal{A}_{\mathcal{G}}}$ and can be realised as a matrix action of the form
\begin{equation}\label{eq:ExeA}
 T_e\:=\:
 \resizebox{.35\hsize}{!}{
 \bordermatrix{
  ~         & \bar{v} & 1      & \ldots & v_e    	     & \ldots & n_v    & e_1    & \ldots & e      	     & \ldots & n_e \cr
  {\bar{v}} & 1       & 0      & \ldots & 0      	     & \ldots & 0      & 0      & \ldots & 0      	     & \ldots & 0\cr
  1         & 0       & 1      & \ldots & 0      	     & \ldots & 0      & 0      & \ldots & 0      	     & \ldots & 0\cr
  \ldots    & \ldots  & \ldots & \ldots & \ldots 	     & \ldots & \ldots & \ldots & \ldots & \ldots	     & \ldots & \ldots\cr 
  v_e       & 0       & 0      & \ldots & 0      	     & \ldots & 0      & 0      & \ldots & {\bf\color{red} 1} & \ldots & 0\cr
  \ldots    & \ldots  & \ldots & \ldots & \ldots 	     & \ldots & \ldots & \ldots & \ldots & \ldots 	     & \ldots & \ldots\cr 
  n_v       & 0       & 0      & \ldots & 0      	     & \ldots & 1      & 0      & \ldots & 0      	     & \ldots & 0\cr
  \ldots    & \ldots  & \ldots & \ldots & \ldots 	     & \ldots & \ldots & \ldots & \ldots & \ldots 	     & \ldots & \ldots\cr 
  e_1       & 0       & 0      & \ldots & 0      	     & \ldots & 0      & 1      & \ldots & 0      	     & \ldots & 0\cr
  \ldots    & \ldots  & \ldots & \ldots & \ldots 	     & \ldots & \ldots & \ldots & \ldots & \ldots 	     & \ldots & \ldots\cr 
  e         & 0       & 0      & \ldots & {\bf\color{red} 1} & \ldots & 0      & 0      & \ldots & 0                  & \ldots & 0\cr
  \ldots    & \ldots  & \ldots & \ldots & \ldots 	     & \ldots & \ldots & \ldots & \ldots & \ldots 	     & \ldots & \ldots\cr 
  n_e       & 0       & 0      & \ldots & 0      	     & \ldots & 0      & 0      & \ldots & 0      	     & \ldots & 1\cr  
 }
 },\quad
 T_I\:=\:
 \resizebox{.35\hsize}{!}{
 \bordermatrix{
  ~         & \bar{v} & 1    	           & \ldots & v_e    & \ldots & n_v    		    & e_1    	         & \ldots & e      & \ldots & n_e \cr
  {\bar{v}} & 1       & 0      	           & \ldots & 0      & \ldots & 0      		    & 0      	         & \ldots & 0      & \ldots & 0\cr
  1         & 0       & {\bf\color{red} 0} & \ldots & 0      & \ldots & {\bf\color{red} -1} & {\bf\color{red} 1} & \ldots & 0      & \ldots & 0\cr
  \ldots    & \ldots  & \ldots 	           & \ldots & \ldots & \ldots & \ldots 		    & \ldots 	         & \ldots & \ldots & \ldots & \ldots\cr 
  v_e       & 0       & 0      	           & \ldots & 0      & \ldots & 0      		    & 0       	         & \ldots & 1      & \ldots & 0\cr
  \ldots    & \ldots  & \ldots 	           & \ldots & \ldots & \ldots & \ldots 		    & \ldots 	         & \ldots & \ldots & \ldots & \ldots\cr 
  n_v       & 0       & 0      	           & \ldots & 0      & \ldots & 1      		    & 0      	         & \ldots & 0      & \ldots & 0\cr
  \ldots    & \ldots  & \ldots 	           & \ldots & \ldots & \ldots & \ldots 		    & \ldots 	         & \ldots & \ldots & \ldots & \ldots\cr 
  e_1       & 0       & {\bf\color{red} 1} & \ldots & 0      & \ldots & {\bf\color{red} 1}  & 0      	         & \ldots & 0      & \ldots & 0\cr
  \ldots    & \ldots  & \ldots             & \ldots & \ldots & \ldots & \ldots 		    & \ldots 	         & \ldots & \ldots & \ldots & \ldots\cr 
  e         & 0       & 0      	           & \ldots & 1      & \ldots & 0      		    & 0      	         & \ldots & 0      & \ldots & 0\cr
  \ldots    & \ldots  & \ldots 	           & \ldots & \ldots & \ldots & \ldots 		    & \ldots 	         & \ldots & \ldots & \ldots & \ldots\cr 
  n_e       & 0       & 0       	   & \ldots & 0      & \ldots & 0      		    & 0      	         & \ldots & 0      & \ldots & 1\cr  
 }
 } 
\end{equation}
where $T_e$ and $T_I$ are the transformation related to an outer edge $e$ and an internal one $I$\footnote{Here, as a matter of example, $T_I$ is taken to represent the transformation $y_{e_1}\,\longleftrightarrow\,x_1+x_{n_v}+\ldots$.} respectively. When applied to $\mathcal{P}_{\mathcal{A}_{\mathcal{G}}}$, each of these transformations keep all the vertices fixed but one:
\begin{equation}\label{eq:Exe4}
 -{\bf x}_e+{\bf y}_e+{\bf x'}_e\:\longrightarrow\: {\bf x}_e-{\bf y}_e+{\bf x'}_e,
\end{equation}
so that $T_e\left(\mathcal{P}_{\mathcal{A}_{\mathcal{G}}}\right)\:=\:\mathcal{P'}_{\mathcal{A'}_{\mathcal{G}}}$. Notice that the new vertex \eqref{eq:Exe4} belongs exactly to the class of vertices which characterise the cosmological polytope $\mathcal{P}_{\mathcal{G}}$ but which are absent in $\hat{\mathcal{P}}_{\mathcal{A}_{\mathcal{G}}}$. These transformations get applied to all edges so that the vertices of type \eqref{eq:Exe4} are generated for {\it all} edges: being $\hat{\mathcal{P}_{\mathcal{G}}}$ a sum of all the $T_{e/I}(\mathcal{P}_{\mathcal{A}_{\mathcal{G}}})$ and their composition, it has in principle all the same vertices as $\mathcal{P}_{\mathcal{G}}$, plus ${\bf x}_v$, which is common to all the $T_{e/I}(\mathcal{P}_{\mathcal{A}_{\mathcal{G}}})$'s. However, ${\bf x}_{\bar{v}}$ is the midpoint of $\{{\bf x}_{\bar{v}}-{\bf y}_{e'}+{\bf x}_{v'},{\bf x}_{\bar{v}}+{\bf y}_{e'}-{\bf x}_{v'},\}$, both of which are vertices of $\hat{\mathcal{P}}_{\mathcal{G}}$ (and $\mathcal{P}_{\mathcal{G}}$). We now need to show that the sum constituting $\hat{\mathcal{P}}_{\mathcal{A}_{\mathcal{G}}}$ is a triangulation of $\mathcal{P}_{\mathcal{A}_{\mathcal{G}}}$ through the {\it internal} point ${\bf x}_{\bar{v}}$. 

\begin{figure}
 \centering
 \begin{tikzpicture}[ball/.style = {circle, draw, align=center, anchor=north, inner sep=0}, cross/.style={cross out, draw, minimum size=2*(#1-\pgflinewidth), inner sep=0pt, outer sep=0pt}, scale=.625, transform shape]
  \begin{scope}
   \node[ball,text width=.18cm,fill,color=black,label=above:{$x_1$}] at (0,0) (x1) {};    
   \node[ball,text width=.18cm,fill,color=black,right=1.2cm of x1.east, label=above:{$x_2$}] (x2) {};    
   \node[ball,text width=.18cm,fill,color=black,right=1.2cm of x2.east, label=above:{$x_3$}] (x3) {};
   \node[ball,text width=.18cm,fill,color=black, label=left:{$x_4$}] at (-1,.8) (x4) {};    
   \node[ball,text width=.18cm,fill,color=black, label=left:{$x_5$}] at (-1,-.8) (x5) {};    
   \node[ball,text width=.18cm,fill,color=black, label=below:{$x_6$}] at (-1.7,-2) (x6) {};    
   \node[ball,text width=.18cm,fill,color=black, label=below:{$x_7$}] at (-.3,-2) (x7) {};

   \node[above=.35cm of x5.north] (ref2) {};
   \coordinate (Int2) at (intersection of x5--x1 and ref2--x2);  

   \coordinate (t1) at (x3.east);
   \coordinate (t2) at (x4.west);
   \coordinate (t3) at (x1.south west);
   \coordinate (t4) at (x2.south);

   \draw[-,thick,color=black] (x1) -- (x2) -- (x3); 
   \draw[-,thick,color=black] (x1) -- (x4);
   \draw[-,thick,color=black] (x5) -- (x1);
   \draw[-,thick,color=black] (x5) -- (x7);   
   \draw[-,thick,color=black] (x5) -- (x6); 

   \coordinate (x12) at ($(x1)!0.50!(x2)$);
   \coordinate (x23) at ($(x2)!0.50!(x3)$);
   \coordinate (x14) at ($(x1)!0.25!(x4)$);
   \coordinate (x14u) at ($(x1)!0.75!(x4)$);
   \coordinate (x14c) at ($(x1)!0.50!(x4)$) {};
   \coordinate (x15) at ($(x1)!0.50!(x5)$);
   \coordinate (x56) at ($(x5)!0.50!(x6)$);
   \coordinate (x57) at ($(x5)!0.50!(x7)$);
   \node[very thick, cross=4pt, rotate=0, color=blue] at (x14c) {};   
   \node[very thick, cross=4pt, rotate=0, color=blue] at (x12) {};   
   \node[very thick, cross=4pt, rotate=0, color=blue] at (x23) {};   
   \node[very thick, cross=4pt, rotate=0, color=blue] at (x15) {};   
   \node[very thick, cross=4pt, rotate=0, color=blue] at (x56) {};   
   \node[very thick, cross=4pt, rotate=0, color=blue] at (x57) {};  
   \node[align=center, below=2.5cm of x12] (T) {$\displaystyle T_{\mathbb{I}}(\mathcal{P}_{\mathcal{A}_{\mathcal{G}}})\,=\,\mathcal{P}_{\mathcal{A}_{\mathcal{G}}}$}; 
   \node[align=center, below=.1cm of T] {$\displaystyle x_5+\sum_{v\in\mathcal{V}\setminus\{x_5\}}x_v\,=\,0$};
  \end{scope}
  \begin{scope}[shift={(6,0)}, transform shape]
   \node[ball,text width=.18cm,fill,color=black,label=above:{$x_1$}] at (0,0) (x1) {};    
   \node[ball,text width=.18cm,fill,color=black,right=1.2cm of x1.east, label=above:{$x_2$}] (x2) {};    
   \node[ball,text width=.18cm,fill,color=black,right=1.2cm of x2.east, label=above:{$x_3$}] (x3) {};
   \node[ball,text width=.18cm,fill,color=black, label=left:{$x_4$}] at (-1,.8) (x4) {};    
   \node[ball,text width=.18cm,fill,color=black, label=left:{$x_5$}] at (-1,-.8) (x5) {};    
   \node[ball,text width=.18cm,fill,color=black, label=below:{$x_6$}] at (-1.7,-2) (x6) {};    
   \node[ball,text width=.18cm,fill,color=black, label=below:{$x_7$}] at (-.3,-2) (x7) {};

   \node[above=.35cm of x5.north] (ref2) {};
   \coordinate (Int2) at (intersection of x5--x1 and ref2--x2);  

   \coordinate (t1) at (x3.east);
   \coordinate (t2) at (x4.west);
   \coordinate (t3) at (x1.south west);
   \coordinate (t4) at (x2.south);

   \draw[-,thick,color=black] (x1) -- (x2) -- (x3); 
   \draw[-,thick,color=black] (x1) -- (x4);
   \draw[-,thick,color=black] (x5) -- (x1);
   \draw[-,thick,color=black] (x5) -- (x7);   
   \draw[-,thick,color=black] (x5) -- (x6); 

   \coordinate (x12) at ($(x1)!0.50!(x2)$);
   \coordinate (x15l) at ($(x1)!0.75!(x5)$);
   \coordinate (x23) at ($(x2)!0.50!(x3)$);
   \coordinate (x14d) at ($(x1)!0.25!(x4)$);
   \coordinate (x14u) at ($(x1)!0.75!(x4)$);
   \coordinate (x14) at ($(x1)!0.50!(x4)$);   
   \coordinate (x15) at ($(x1)!0.50!(x5)$);
   \coordinate (x56) at ($(x5)!0.50!(x6)$);
   \coordinate (x57) at ($(x5)!0.50!(x7)$);
   \node[very thick, cross=4pt, rotate=0, color=blue] at (x12) {};   
   \node[very thick, cross=4pt, rotate=0, color=blue] at (x14) {};      
   \node[very thick, cross=4pt, rotate=0, color=red] at (x15l) {};   
   \node[very thick, cross=4pt, rotate=0, color=blue] at (x23) {};   
   \node[very thick, cross=4pt, rotate=0, color=blue] at (x56) {};   
   \node[very thick, cross=4pt, rotate=0, color=blue] at (x57) {};   
   \coordinate (x12r) at ($(x1)!0.75!(x2)$) {};
   \coordinate (x14c) at ($(x1)!0.50!(x4)$) {};
   \coordinate (x23r) at ($(x2)!0.75!(x3)$) {};   
   \coordinate (x15b) at ($(x1)!0.75!(x5)$) {};   
   \coordinate (x56b) at ($(x5)!0.75!(x6)$) {};
   \coordinate (x57b) at ($(x5)!0.75!(x7)$) {}; 
   \node[below=2.5cm of x12] (T) {$\displaystyle T_{e_{15}}(\mathcal{P}_{\mathcal{A}_{\mathcal{G}}})\,=\,\mathcal{P}'_{\mathcal{A}_{\mathcal{G}}}$};
   \node[align=center, below=.1cm of T] {$\displaystyle x_5+x_6+x_7+y_{15}\,=\,0$};  
  \end{scope}
  \begin{scope}[shift={(12,0)}, transform shape]
   \node[ball,text width=.18cm,fill,color=black,label=above:{$x_1$}] at (0,0) (x1) {};    
   \node[ball,text width=.18cm,fill,color=black,right=1.2cm of x1.east, label=above:{$x_2$}] (x2) {};    
   \node[ball,text width=.18cm,fill,color=black,right=1.2cm of x2.east, label=above:{$x_3$}] (x3) {};
   \node[ball,text width=.18cm,fill,color=black, label=left:{$x_4$}] at (-1,.8) (x4) {};    
   \node[ball,text width=.18cm,fill,color=black, label=left:{$x_5$}] at (-1,-.8) (x5) {};    
   \node[ball,text width=.18cm,fill,color=black, label=below:{$x_6$}] at (-1.7,-2) (x6) {};    
   \node[ball,text width=.18cm,fill,color=black, label=below:{$x_7$}] at (-.3,-2) (x7) {};

   \node[above=.35cm of x5.north] (ref2) {};
   \coordinate (Int2) at (intersection of x5--x1 and ref2--x2);  

   \coordinate (t1) at (x3.east);
   \coordinate (t2) at (x4.west);
   \coordinate (t3) at (x1.south west);
   \coordinate (t4) at (x2.south);

   \draw[-,thick,color=black] (x1) -- (x2) -- (x3); 
   \draw[-,thick,color=black] (x1) -- (x4);
   \draw[-,thick,color=black] (x5) -- (x1);
   \draw[-,thick,color=black] (x5) -- (x7);   
   \draw[-,thick,color=black] (x5) -- (x6); 

   \coordinate (x12) at ($(x1)!0.50!(x2)$);
   \coordinate (x15l) at ($(x1)!0.75!(x5)$);
   \coordinate (x23) at ($(x2)!0.50!(x3)$);
   \coordinate (x12l) at ($(x1)!0.25!(x2)$);
   \coordinate (x14u) at ($(x1)!0.75!(x4)$);
   \coordinate (x14) at ($(x1)!0.50!(x4)$);   
   \coordinate (x15) at ($(x1)!0.50!(x5)$);
   \coordinate (x56) at ($(x5)!0.50!(x6)$);
   \coordinate (x57) at ($(x5)!0.50!(x7)$);
   \node[very thick, cross=4pt, rotate=0, color=red] at (x12l) {};   
   \node[very thick, cross=4pt, rotate=0, color=blue] at (x14) {};      
   \node[very thick, cross=4pt, rotate=0, color=blue] at (x15) {};   
   \node[very thick, cross=4pt, rotate=0, color=blue] at (x23) {};   
   \node[very thick, cross=4pt, rotate=0, color=blue] at (x56) {};   
   \node[very thick, cross=4pt, rotate=0, color=blue] at (x57) {};   
   \coordinate (x12r) at ($(x1)!0.75!(x2)$) {};
   \coordinate (x14c) at ($(x1)!0.50!(x4)$) {};
   \coordinate (x23r) at ($(x2)!0.75!(x3)$) {};   
   \coordinate (x15b) at ($(x1)!0.75!(x5)$) {};   
   \coordinate (x56b) at ($(x5)!0.75!(x6)$) {};
   \coordinate (x57b) at ($(x5)!0.75!(x7)$) {}; 
   \node[below=2.5cm of x12] (T) {$\displaystyle T_{e_{12}}(\mathcal{P}_{\mathcal{A}_{\mathcal{G}}})\,=\,\mathcal{P}''_{\mathcal{A}_{\mathcal{G}}}$};
   \node[align=center, below=.1cm of T] {$\displaystyle x_5+\sum_{j=6}^1x_j+x_4+y_{12}\,=\,0$};  
  \end{scope}
  \begin{scope}[shift={(18,0)}, transform shape]
   \node[ball,text width=.18cm,fill,color=black,label=above:{$x_1$}] at (0,0) (x1) {};    
   \node[ball,text width=.18cm,fill,color=black,right=1.2cm of x1.east, label=above:{$x_2$}] (x2) {};    
   \node[ball,text width=.18cm,fill,color=black,right=1.2cm of x2.east, label=above:{$x_3$}] (x3) {};
   \node[ball,text width=.18cm,fill,color=black, label=left:{$x_4$}] at (-1,.8) (x4) {};    
   \node[ball,text width=.18cm,fill,color=black, label=left:{$x_5$}] at (-1,-.8) (x5) {};    
   \node[ball,text width=.18cm,fill,color=black, label=below:{$x_6$}] at (-1.7,-2) (x6) {};    
   \node[ball,text width=.18cm,fill,color=black, label=below:{$x_7$}] at (-.3,-2) (x7) {};

   \node[above=.35cm of x5.north] (ref2) {};
   \coordinate (Int2) at (intersection of x5--x1 and ref2--x2);  

   \coordinate (t1) at (x3.east);
   \coordinate (t2) at (x4.west);
   \coordinate (t3) at (x1.south west);
   \coordinate (t4) at (x2.south);

   \draw[-,thick,color=black] (x1) -- (x2) -- (x3); 
   \draw[-,thick,color=black] (x1) -- (x4);
   \draw[-,thick,color=black] (x5) -- (x1);
   \draw[-,thick,color=black] (x5) -- (x7);   
   \draw[-,thick,color=black] (x5) -- (x6); 

   \coordinate (x12) at ($(x1)!0.50!(x2)$);
   \coordinate (x15l) at ($(x1)!0.75!(x5)$);
   \coordinate (x23) at ($(x2)!0.50!(x3)$);
   \coordinate (x12l) at ($(x1)!0.25!(x2)$);
   \coordinate (x14u) at ($(x1)!0.75!(x4)$);
   \coordinate (x14) at ($(x1)!0.50!(x4)$);   
   \coordinate (x15) at ($(x1)!0.50!(x5)$);
   \coordinate (x56) at ($(x5)!0.50!(x6)$);
   \coordinate (x57) at ($(x5)!0.50!(x7)$);
   \node[very thick, cross=4pt, rotate=0, color=red] at (x12l) {};   
   \node[very thick, cross=4pt, rotate=0, color=blue] at (x14) {};      
   \node[very thick, cross=4pt, rotate=0, color=red] at (x15l) {};   
   \node[very thick, cross=4pt, rotate=0, color=blue] at (x23) {};   
   \node[very thick, cross=4pt, rotate=0, color=blue] at (x56) {};   
   \node[very thick, cross=4pt, rotate=0, color=blue] at (x57) {};   
   \coordinate (x12r) at ($(x1)!0.75!(x2)$) {};
   \coordinate (x14c) at ($(x1)!0.50!(x4)$) {};
   \coordinate (x23r) at ($(x2)!0.75!(x3)$) {};   
   \coordinate (x15b) at ($(x1)!0.75!(x5)$) {};   
   \coordinate (x56b) at ($(x5)!0.75!(x6)$) {};
   \coordinate (x57b) at ($(x5)!0.75!(x7)$) {}; 
   \node[below=2.5cm of x12] (T) {$\displaystyle T_{e_{12}}\circ T_{e_{15}}(\mathcal{P}_{\mathcal{A}_{\mathcal{G}}})\,=\,\mathcal{P}^{\mbox{\tiny $(3)$}}_{\mathcal{A}^{\mbox{\tiny $(3)$}}_{\mathcal{G}}}$};
   \node[align=center, below=.1cm of T] {$\displaystyle x_5+x_6+x_7+y_{15}\,=\,0$};  
  \end{scope}
  \begin{scope}[shift={(0,-6)}, transform shape]
   \node[ball,text width=.18cm,fill,color=black,label=above:{$x_1$}] at (0,0) (x1) {};    
   \node[ball,text width=.18cm,fill,color=black,right=1.2cm of x1.east, label=above:{$x_2$}] (x2) {};    
   \node[ball,text width=.18cm,fill,color=black,right=1.2cm of x2.east, label=above:{$x_3$}] (x3) {};
   \node[ball,text width=.18cm,fill,color=black, label=left:{$x_4$}] at (-1,.8) (x4) {};    
   \node[ball,text width=.18cm,fill,color=black, label=left:{$x_5$}] at (-1,-.8) (x5) {};    
   \node[ball,text width=.18cm,fill,color=black, label=below:{$x_6$}] at (-1.7,-2) (x6) {};    
   \node[ball,text width=.18cm,fill,color=black, label=below:{$x_7$}] at (-.3,-2) (x7) {};

   \node[above=.35cm of x5.north] (ref2) {};
   \coordinate (Int2) at (intersection of x5--x1 and ref2--x2);  

   \coordinate (t1) at (x3.east);
   \coordinate (t2) at (x4.west);
   \coordinate (t3) at (x1.south west);
   \coordinate (t4) at (x2.south);

   \draw[-,thick,color=black] (x1) -- (x2) -- (x3); 
   \draw[-,thick,color=black] (x1) -- (x4);
   \draw[-,thick,color=black] (x5) -- (x1);
   \draw[-,thick,color=black] (x5) -- (x7);   
   \draw[-,thick,color=black] (x5) -- (x6); 

   \coordinate (x12) at ($(x1)!0.50!(x2)$);
   \coordinate (x15l) at ($(x1)!0.75!(x5)$);
   \coordinate (x23) at ($(x2)!0.50!(x3)$);
   \coordinate (x12l) at ($(x1)!0.25!(x2)$);
   \coordinate (x23l) at ($(x2)!0.25!(x3)$);
   \coordinate (x14) at ($(x1)!0.50!(x4)$);   
   \coordinate (x15) at ($(x1)!0.50!(x5)$);
   \coordinate (x56) at ($(x5)!0.50!(x6)$);
   \coordinate (x57) at ($(x5)!0.50!(x7)$);
   \node[very thick, cross=4pt, rotate=0, color=blue] at (x12) {};   
   \node[very thick, cross=4pt, rotate=0, color=red] at (x23l) {};      
   \node[very thick, cross=4pt, rotate=0, color=blue] at (x14) {};      
   \node[very thick, cross=4pt, rotate=0, color=blue] at (x15) {};   
   \node[very thick, cross=4pt, rotate=0, color=blue] at (x56) {};   
   \node[very thick, cross=4pt, rotate=0, color=blue] at (x57) {};   
   \coordinate (x12r) at ($(x1)!0.75!(x2)$) {};
   \coordinate (x14c) at ($(x1)!0.50!(x4)$) {};
   \coordinate (x23r) at ($(x2)!0.75!(x3)$) {};   
   \coordinate (x15b) at ($(x1)!0.75!(x5)$) {};   
   \coordinate (x56b) at ($(x5)!0.75!(x6)$) {};
   \coordinate (x57b) at ($(x5)!0.75!(x7)$) {}; 
   \node[below=2.5cm of x12] (T) {$\displaystyle T_{e_{23}}(\mathcal{P}_{\mathcal{A}_{\mathcal{G}}})\,=\,\mathcal{P}^{\mbox{\tiny $(4)$}}_{\mathcal{A}_{\mathcal{G}}}$};
   \node[align=center, below=.1cm of T] {$\displaystyle x_5+\sum_{v\in\mathcal{V}\setminus\{x_5,x_3\}}x_j+y_{23}\,=\,0$};  
  \end{scope}
  \begin{scope}[shift={(6,-6)}, transform shape]
   \node[ball,text width=.18cm,fill,color=black,label=above:{$x_1$}] at (0,0) (x1) {};    
   \node[ball,text width=.18cm,fill,color=black,right=1.2cm of x1.east, label=above:{$x_2$}] (x2) {};    
   \node[ball,text width=.18cm,fill,color=black,right=1.2cm of x2.east, label=above:{$x_3$}] (x3) {};
   \node[ball,text width=.18cm,fill,color=black, label=left:{$x_4$}] at (-1,.8) (x4) {};    
   \node[ball,text width=.18cm,fill,color=black, label=left:{$x_5$}] at (-1,-.8) (x5) {};    
   \node[ball,text width=.18cm,fill,color=black, label=below:{$x_6$}] at (-1.7,-2) (x6) {};    
   \node[ball,text width=.18cm,fill,color=black, label=below:{$x_7$}] at (-.3,-2) (x7) {};

   \node[above=.35cm of x5.north] (ref2) {};
   \coordinate (Int2) at (intersection of x5--x1 and ref2--x2);  

   \coordinate (t1) at (x3.east);
   \coordinate (t2) at (x4.west);
   \coordinate (t3) at (x1.south west);
   \coordinate (t4) at (x2.south);

   \draw[-,thick,color=black] (x1) -- (x2) -- (x3); 
   \draw[-,thick,color=black] (x1) -- (x4);
   \draw[-,thick,color=black] (x5) -- (x1);
   \draw[-,thick,color=black] (x5) -- (x7);   
   \draw[-,thick,color=black] (x5) -- (x6); 

   \coordinate (x12) at ($(x1)!0.50!(x2)$);
   \coordinate (x15l) at ($(x1)!0.75!(x5)$);
   \coordinate (x23) at ($(x2)!0.50!(x3)$);
   \coordinate (x12l) at ($(x1)!0.25!(x2)$);
   \coordinate (x23l) at ($(x2)!0.25!(x3)$);
   \coordinate (x14) at ($(x1)!0.50!(x4)$);   
   \coordinate (x15) at ($(x1)!0.50!(x5)$);
   \coordinate (x56) at ($(x5)!0.50!(x6)$);
   \coordinate (x57) at ($(x5)!0.50!(x7)$);
   \node[very thick, cross=4pt, rotate=0, color=blue] at (x12) {};   
   \node[very thick, cross=4pt, rotate=0, color=red] at (x23l) {};      
   \node[very thick, cross=4pt, rotate=0, color=blue] at (x14) {};      
   \node[very thick, cross=4pt, rotate=0, color=red] at (x15l) {};   
   \node[very thick, cross=4pt, rotate=0, color=blue] at (x56) {};   
   \node[very thick, cross=4pt, rotate=0, color=blue] at (x57) {};   
   \coordinate (x12r) at ($(x1)!0.75!(x2)$) {};
   \coordinate (x14c) at ($(x1)!0.50!(x4)$) {};
   \coordinate (x23r) at ($(x2)!0.75!(x3)$) {};   
   \coordinate (x15b) at ($(x1)!0.75!(x5)$) {};   
   \coordinate (x56b) at ($(x5)!0.75!(x6)$) {};
   \coordinate (x57b) at ($(x5)!0.75!(x7)$) {}; 
   \node[below=2.5cm of x12] (T) {$\displaystyle T_{e_{23}}\circ T_{e_{15}}(\mathcal{P}_{\mathcal{A}_{\mathcal{G}}})\,=\,\mathcal{P}^{\mbox{\tiny $(5)$}}_{\mathcal{A}^{\mbox{\tiny $(5)$}}_{\mathcal{G}}}$};
   \node[align=center, below=.1cm of T] {$\displaystyle x_5+x_6+x_7+y_{15}\,=\,0$};  
  \end{scope}
  \begin{scope}[shift={(12,-6)}, transform shape]
   \node[ball,text width=.18cm,fill,color=black,label=above:{$x_1$}] at (0,0) (x1) {};    
   \node[ball,text width=.18cm,fill,color=black,right=1.2cm of x1.east, label=above:{$x_2$}] (x2) {};    
   \node[ball,text width=.18cm,fill,color=black,right=1.2cm of x2.east, label=above:{$x_3$}] (x3) {};
   \node[ball,text width=.18cm,fill,color=black, label=left:{$x_4$}] at (-1,.8) (x4) {};    
   \node[ball,text width=.18cm,fill,color=black, label=left:{$x_5$}] at (-1,-.8) (x5) {};    
   \node[ball,text width=.18cm,fill,color=black, label=below:{$x_6$}] at (-1.7,-2) (x6) {};    
   \node[ball,text width=.18cm,fill,color=black, label=below:{$x_7$}] at (-.3,-2) (x7) {};

   \node[above=.35cm of x5.north] (ref2) {};
   \coordinate (Int2) at (intersection of x5--x1 and ref2--x2);  

   \coordinate (t1) at (x3.east);
   \coordinate (t2) at (x4.west);
   \coordinate (t3) at (x1.south west);
   \coordinate (t4) at (x2.south);

   \draw[-,thick,color=black] (x1) -- (x2) -- (x3); 
   \draw[-,thick,color=black] (x1) -- (x4);
   \draw[-,thick,color=black] (x5) -- (x1);
   \draw[-,thick,color=black] (x5) -- (x7);   
   \draw[-,thick,color=black] (x5) -- (x6); 

   \coordinate (x12) at ($(x1)!0.50!(x2)$);
   \coordinate (x15l) at ($(x1)!0.75!(x5)$);
   \coordinate (x23) at ($(x2)!0.50!(x3)$);
   \coordinate (x12l) at ($(x1)!0.25!(x2)$);
   \coordinate (x23l) at ($(x2)!0.25!(x3)$);
   \coordinate (x14) at ($(x1)!0.50!(x4)$);   
   \coordinate (x15) at ($(x1)!0.50!(x5)$);
   \coordinate (x56) at ($(x5)!0.50!(x6)$);
   \coordinate (x57) at ($(x5)!0.50!(x7)$);
   \node[very thick, cross=4pt, rotate=0, color=red] at (x12l) {};   
   \node[very thick, cross=4pt, rotate=0, color=red] at (x23l) {};      
   \node[very thick, cross=4pt, rotate=0, color=blue] at (x14) {};      
   \node[very thick, cross=4pt, rotate=0, color=blue] at (x15) {};   
   \node[very thick, cross=4pt, rotate=0, color=blue] at (x56) {};   
   \node[very thick, cross=4pt, rotate=0, color=blue] at (x57) {};   
   \coordinate (x12r) at ($(x1)!0.75!(x2)$) {};
   \coordinate (x14c) at ($(x1)!0.50!(x4)$) {};
   \coordinate (x23r) at ($(x2)!0.75!(x3)$) {};   
   \coordinate (x15b) at ($(x1)!0.75!(x5)$) {};   
   \coordinate (x56b) at ($(x5)!0.75!(x6)$) {};
   \coordinate (x57b) at ($(x5)!0.75!(x7)$) {}; 
   \node[below=2.5cm of x12] (T) {$\displaystyle T_{e_{23}}\circ T_{e_{12}}(\mathcal{P}_{\mathcal{A}_{\mathcal{G}}})\,=\,\mathcal{P}^{\mbox{\tiny $(6)$}}_{\mathcal{A}^{\mbox{\tiny $(6)$}}_{\mathcal{G}}}$};
   \node[align=center, below=.1cm of T] {$\displaystyle x_5+\sum_{j=7}^1x_j+x_4+y_{12}\,=\,0$};  
  \end{scope}
  \begin{scope}[shift={(18,-6)}, transform shape]
   \node[ball,text width=.18cm,fill,color=black,label=above:{$x_1$}] at (0,0) (x1) {};    
   \node[ball,text width=.18cm,fill,color=black,right=1.2cm of x1.east, label=above:{$x_2$}] (x2) {};    
   \node[ball,text width=.18cm,fill,color=black,right=1.2cm of x2.east, label=above:{$x_3$}] (x3) {};
   \node[ball,text width=.18cm,fill,color=black, label=left:{$x_4$}] at (-1,.8) (x4) {};    
   \node[ball,text width=.18cm,fill,color=black, label=left:{$x_5$}] at (-1,-.8) (x5) {};    
   \node[ball,text width=.18cm,fill,color=black, label=below:{$x_6$}] at (-1.7,-2) (x6) {};    
   \node[ball,text width=.18cm,fill,color=black, label=below:{$x_7$}] at (-.3,-2) (x7) {};

   \node[above=.35cm of x5.north] (ref2) {};
   \coordinate (Int2) at (intersection of x5--x1 and ref2--x2);  

   \coordinate (t1) at (x3.east);
   \coordinate (t2) at (x4.west);
   \coordinate (t3) at (x1.south west);
   \coordinate (t4) at (x2.south);

   \draw[-,thick,color=black] (x1) -- (x2) -- (x3); 
   \draw[-,thick,color=black] (x1) -- (x4);
   \draw[-,thick,color=black] (x5) -- (x1);
   \draw[-,thick,color=black] (x5) -- (x7);   
   \draw[-,thick,color=black] (x5) -- (x6); 

   \coordinate (x12) at ($(x1)!0.50!(x2)$);
   \coordinate (x15l) at ($(x1)!0.75!(x5)$);
   \coordinate (x23l) at ($(x2)!0.25!(x3)$);
   \coordinate (x12l) at ($(x1)!0.25!(x2)$);
   \coordinate (x14u) at ($(x1)!0.75!(x4)$);
   \coordinate (x14) at ($(x1)!0.50!(x4)$);   
   \coordinate (x15) at ($(x1)!0.50!(x5)$);
   \coordinate (x56) at ($(x5)!0.50!(x6)$);
   \coordinate (x57) at ($(x5)!0.50!(x7)$);
   \node[very thick, cross=4pt, rotate=0, color=red] at (x12l) {};   
   \node[very thick, cross=4pt, rotate=0, color=blue] at (x14) {};      
   \node[very thick, cross=4pt, rotate=0, color=red] at (x15l) {};   
   \node[very thick, cross=4pt, rotate=0, color=red] at (x23l) {};   
   \node[very thick, cross=4pt, rotate=0, color=blue] at (x56) {};   
   \node[very thick, cross=4pt, rotate=0, color=blue] at (x57) {};   
   \coordinate (x12r) at ($(x1)!0.75!(x2)$) {};
   \coordinate (x14c) at ($(x1)!0.50!(x4)$) {};
   \coordinate (x23r) at ($(x2)!0.75!(x3)$) {};   
   \coordinate (x15b) at ($(x1)!0.75!(x5)$) {};   
   \coordinate (x56b) at ($(x5)!0.75!(x6)$) {};
   \coordinate (x57b) at ($(x5)!0.75!(x7)$) {}; 
   \node[below=2.5cm of x12] (T) {$\displaystyle T_{e_{23}}\circ T_{e_{12}}\circ T_{e_{15}}(\mathcal{P}_{\mathcal{A}_{\mathcal{G}}})\,=\,\mathcal{P}^{\mbox{\tiny $(7)$}}_{\mathcal{A}^{\mbox{\tiny $(7)$}}_{\mathcal{G}}}$};
   \node[align=center, below=.1cm of T] {$\displaystyle x_5+x_6+x_7+y_{15}\,=\,0$};  
  \end{scope}
 \end{tikzpicture}
 \caption{Facets of images of $\mathcal{P}_{\mathcal{A}_{\mathcal{G}}}$. The facets identified by an hyperplane $\mathcal{W}_{\bar{v}I}\,=\,\tilde{x}_{\bar{v}}{\bf\tilde{X}}_{\bar{v}I}+\ldots$ are shown for some of the polytopes generated as morphisms on $\mathcal{P}_{\mathcal{A}_{\mathcal{G}}}$. A given hyperplane of this type identified the facet of different polytopes: their sum provides a triangulation of the facet of the sum $\hat{\mathcal{P}}_{\mathcal{G}}$. In this Figure, a complete example is given by third pictures in both lines: they provide a triangulation of the facets at $\sum_{j=4}^{7}x_j+x_1+y_{12}\,=\,0$. The sum of the second and the fourth facets in each line are part of a triangulation of the facet $x_5+x_6+x_7+y_{15}\,=\,0$.}
 \label{fig:T}
\end{figure}
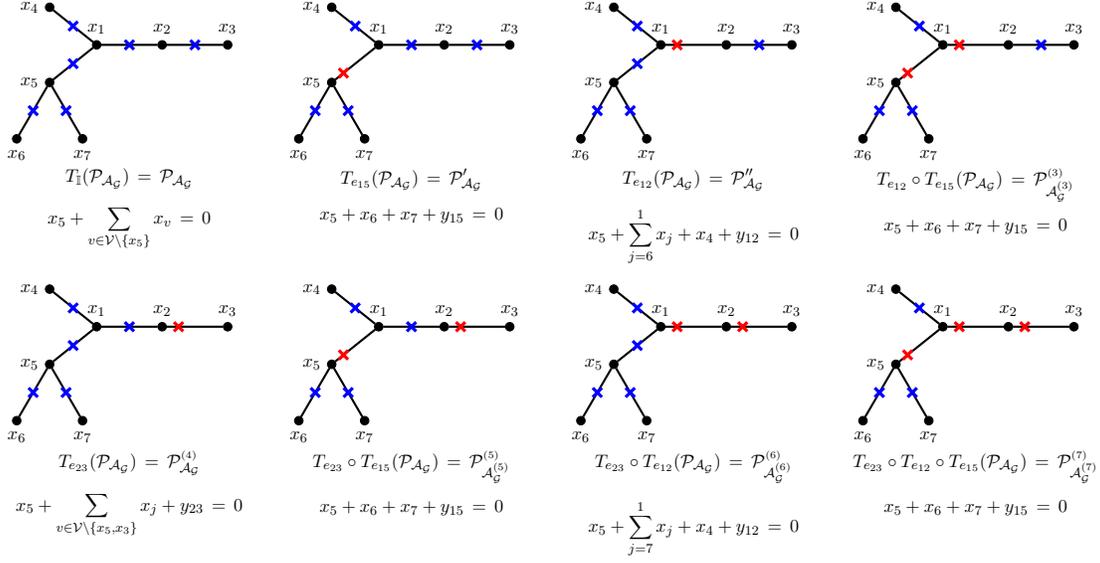

Let us consider all the facets of each summand of $\hat{\mathcal{P}}_{\mathcal{G}}$ corresponding to the hyperplanes $\mathcal{W}_{\bar{v}I}\,=\,\tilde{x}_{\bar{v}}{\bf\tilde{X}}_{\bar{v}I}+\ldots$. First, notice that these hyperplanes identifies some of the facets of the cosmological polytope $\mathcal{P}_{\mathcal{G}}$. Given one of such hyperplane it can identify a facet for a subset (with dimension greater than one) of such polytopes (see Figure \ref{fig:T}). 

%First, notice that these hyperplanes identifies some of the facets of the cosmological polytope $\mathcal{P}_{\mathcal{G}}$. Given one of such hyperplane it can a facet for a subset of such polytopes with dimension greater than one (see Figure \ref{fig:T}). 
As argued earlier, the action of each of the morphisms $T$ generates a new vertex at a time which are also vertices of $\mathcal{P}_{\mathcal{G}}$ and thus, the vertices of any polytope generated in this way are a subset of the vertices of $\mathcal{P}_{\mathcal{G}}$, plus an extra vertex. For any these polytopes, the boundary identified by the hyperplane $\mathcal{W}_{\bar{v}I}\,=\,\tilde{x}_{\bar{v}}{\bf\tilde{X}}_{\bar{v}I}+\ldots$ does not contain such an extra vertex. Let us now take the facet of the cosmological polytope $\mathcal{P}_{\mathcal{G}}$ identified by one of such hyperplane, and consider its canonical form in its contour integral representation. Let us perform the integration in such a way that the resulting triangulation is made out by simplices with two vertices for each edge, as we discussed in Section \ref{subsec:TCP}: such a triangulation is exactly the sum of the relevant facets of these polytopes\footnote{An explicit example is given by the sum of the third graphs in both lines of Figure \ref{fig:T}.}.

We can consider the cosmological polytope $\mathcal{P}_{\mathcal{A}_{\mathcal{G}}}$ and its triangulations through an additional point ${\bf x}_{\bar{v}}$, taken to be one of the midpoints of the generating triangles: they can be analysed via the contour integral representation of its canonical form now defined including ${\bf x}_{\bar{v}}$ as a vertex as well
\begin{equation}\label{eq:CFev}
 \Omega\:\sim\:\int\frac{dc_{\bar{v}}}{c_{\bar{v}}-i\varepsilon_{\bar{v}}}\int\prod_{j=1}^{3n_e}\frac{dc_j}{c_j-i\varepsilon_j}\delta^{\mbox{\tiny $(n_v+n_e)$}}\left(\mathcal{Y}-c_{\bar{v}}{\bf V}^{\mbox{\tiny $(\bar{v})$}}-\sum_{j=1}^{\nu}c_j{\bf V}^{\mbox{\tiny $(j)$}}\right).
\end{equation}
Let us take $c_{\bar{v}}$ as one of the free variables. The fact that ${\bf V}^{\mbox{\tiny $(\bar{v})$}}$ is an actual internal point of the polytope manifests itself in the location of the poles in the complex $c_{\bar{v}}$-plane, {\it after}\footnote{In line of principle, one is allowed to change order of integrations and perform the integration in $c_{\bar{v}}$ before (some of) the other integrations. In this case, some of the triangulations obtained are the same that one would obtain integration ${\bf c}_{\bar{v}}$ at the end, while others are such that just some of the simplices have ${\bf x}_{\bar{v}}$ as a vertex.} the delta functions localised $n_e+n_v$ integrations and the contour integration over the other free variable have been performed: the pole at zero appearing in \eqref{eq:CFev} is the only one appearing in the UHP, while all the others are in the LHP. Thus, closing the integration contour for $c_{\bar{v}}$ in the UHP simply returns the triangulation given by the choice of the integration contours for the other variables, while closing it in the LHP returns triangulations through the point ${\bf x}_{\bar{v}}$. Choosing the integration contour for the $c_j$'s such that the hyperplanes picked have two vertices for each edge of the graph (as explained in Section \ref{subsec:TCP}), then the integration of $c_{\bar{v}}$ in the LHP returns a triangulations whose simplices are exactly the ones generated from $\mathcal{P}_{\mathcal{A}_{\mathcal{G}}}$ via morphisms $T$ (and their compositions): concretely, given a graph $\mathcal{G}$, each of these simplices are given by a facet of the cosmological polytope $\mathcal{P}_{\mathcal{G}}$ related to a subgraph $\mathfrak{g}\,\subseteq\,\mathcal{G}$ containing the vertex ${\bar{v}}$ in $\mathcal{G}$, and the vertex ${\bf x}_{\bar{v}}$ -- which is exactly what the summands in $\hat{\mathcal{P}}_{\mathcal{G}}$ are. Hence, $\hat{\mathcal{P}}_{\mathcal{G}}\,\equiv\,\mathcal{P}_{\mathcal{G}}$.

As a final comment, this way of generating the wavefunction of the universe is the inverse procedure of the recursion relation discussed proved in Section 2.4 of \cite{Arkani-Hamed:2017fdk}: the wavefunction representative is nothing but the only term with the total energy pole that one obtains if the recursion relation is iterated until the graph reduces to sums of vertices, and our discrete transformations on it generates all the other terms. Our discussion is indeed valid a tree level only. What about the loop wavefunctions? We can generate them from the tree ones using the tree-loop map discussed in Section \ref{subsec:TL}: more generally, it generates an $(L+1)$-loop wavefunction from an $L$-loop one, which, from the geometrical perspective, it is as simple as a projection.

%%%%%%%%%%%%%%%%%%%%%%
%%%%%%%%%%%%%%%%%%%%%%
%%%%%%%%%%%%%%%%%%%%%%

\section{Conclusion}\label{sec:Concl}

Our understanding of quantum mechanical observables in cosmology is quite primitive. Leaving on a side the unsolved long standing issue about how to define such quantities in an universe with accelerated expansion, and contenting ourselves with working in the approximation in which the universe becomes infinitely large at sufficiently late times, even in such an approximation and in perturbation theory we do not have a deep understanding of them. Drawing a parallel with the better known perturbative scattering amplitude in flat-space, in the case of the wavefunction of the universe we do not have a real understanding of its singularities, their physical meaning, and what are the properties that it ought to satisfy to come from a consistent cosmological evolution, {\it e.g.} whether there is a cosmological equivalent of the Cutkosky rules: at the end of the day, the wavefunction of the universe is a boundary observable --  it just depends on the data at the time when we are performing the measurements --, while the time evolution is integrated out. One piece of information that we do have at our disposal is that the lack of invariance under time translation and, consequently, of energy conservation, reflects itself into the analytic structure of the wavefunction via the presence of a singularity in the total energy. However, when such a point is approached, the wavefunction returns the flat-space scattering amplitude.

The singularity structure of the wavefunction is beautifully encoded into the boundary structure of the cosmological polytope: its scattering facet describes the flat space scattering amplitude and allows us to observe the emergence of both Lorentz invariance and unitarity. In this paper we have analysed the physical content of the other facets of the cosmological polytope: surprisingly enough, a subset of these boundaries are isomorphic to the scattering facet, with their canonical forms encoding again the flat-space scattering amplitude itself or products of lower point scattering amplitudes. Furthermore, as far as the remaining facets are concerned, they admit a triangulation such that the canonical form of each of these simplices is a product of flat space processes. This is really striking given that there is no {\it a priori} reason to expect that all the residues of the wavefunction of the universe can be expressed in terms of flat-space scattering amplitudes, beyond the simple fact that, when the location of any of these poles is reached, a local energy conservation is restored and the wavefunction naively factorises into a flat-space amplitude and lower point wavefunction computed at that point.

Given that all the residues could be related to flat-space processes, we have asked ourselves whether (and to which extent) the knowledge of the flat-space scattering amplitudes could fix the cosmological wavefunction. Surprisingly enough, if one is given a flat-space scattering amplitude as well as a set of symmetries that the wavefunction is supposed to have, then the wavefunction can be determined. This seems to suggest that the flat-space physics can determine cosmological observables to a higher extent that one would have ever expected, and it is indeed true for the class of toy models the cosmological polytope describe. On the other hand, the cosmological polytope captures features which are universal and thus it becomes compelling to ask whether and how this can be a more general feature rather than a peculiarity of our toy models. There are indeed a few counter-examples that one can readily mention, {\it e.g.} Yang-Mills and gravity: in de Sitter space, the three-point wavefunction having all modes with the same helicities are non-trivial, while the flat-space counterpart is zero. This is a consequence of the fact that the de Sitter wavefunction has zero residue in the total energy pole. It is certainly unplausible to reconstruct a non-zero object starting from a vanishing one. However, we are still left with the general question that begs to be answered: how much of the structure of the wavefunction of the universe is fixed by the flat-space physics? Addressing such a question is left for future work, for which Yang-Mills theory seems the most suitable playground. More broadly, it would be interesting to extend this combinatorial and geometrical description beyond the class of toy models discussed so far.

Interestingly, in reconstructing the cosmological polytope from the simplex encoding a representative of the wavefunction itself in a neighbourhood of the total energy pole, we obtained a new class of triangulations, which decompose the canonical form of the cosmological polytope into a sum of Lorentz-like expressions times a linear pole. An interesting feature of these triangulations is that they go through a single internal point, which is one of the midpoints ${\bf x}_i$ of the generating triangles. Such points, thus, acquire a new physical meaning -- they are the ones allowing to make explicit the relation between the residues of the wavefunction of the universe and flat-space scattering processes. Actually, they will play a crucial role in extracting the cosmological correlators from the wavefunction \cite{Arkani-Hamed:2019abc}.

The cosmological polytope is showing to have an extremely rich structure, returning us also universal information -- a good example is given by the scattering facet whose vertex structure provides a new (and simpler) picture for the cutting rules, valid in any quantum field theory \cite{Arkani-Hamed:2018ahb}. However, the cosmological polytope allowed us so far to formulate graph by graph statements, while a complete understanding of the perturbative wavefunction of the universe demands for a picture which is capable to capture all the relevant graphs, in a similar fashion as the amplituhedron \cite{Arkani-Hamed:2013jha} for maximally supersymmetric Yang-Mills theory and the associahedra for bi-adjoint scalars \cite{Arkani-Hamed:2017mur, Frost:2018djd, Salvatori:2018aha}. This is indeed an open problem that we hope to solve in the near future.

%%%%%%%%%%%%%%%%%%%%%%
%%%%%%%%%%%%%%%%%%%%%%
%%%%%%%%%%%%%%%%%%%%%%

\section*{Acknowledgements}

I am in debt with Nima Arkani-Hamed and Freddy Cachazo for insightful discussions, and with Nima Arkani-Hamed for reading the manuscript. I am grateful to the Institute for Advanced Study in Princeton for hospitality during several stages of this work, as well as Perimeter Institute where this work was initiated. I would also like to thank Claudio Corian{\`o} for hospitality at the Universit{\`a} del Salento and INFN Lecce, where this work was terminated. P.B. would also like to thank the developers of SageMath \cite{sage}, Maxima \cite{maxima}, and Tikz \cite{tantau:2013a}. Some of the polytope analysis has been performed with the aid of {\tt polymake} \cite{polymake:2000}, and the triangulation studies with {\tt TOPCOM} \cite{Rambau:TOPCOM-ICMS:2002}. P.B. is supported in part by a grant from the Villum Fonden, an ERC-StG grant (N. 757978) and the Danish National Research Foundation (DNRF91).

%%%%%%%%%%%%%%%%%%%%%%
%%%%%%%%%%%%%%%%%%%%%%
%%%%%%%%%%%%%%%%%%%%%%

\appendix

%%%%%%%%%%%%%%%%%%%%%%
%%%%%%%%%%%%%%%%%%%%%%

\section{From trees to loops via contour integrals}\label{app:TL}

In Section \ref{subsec:TL} we have seen how tree cosmological polytopes contain the information about the loop ones, or, more generally, from a cosmological polytope related to an $L$-loop graph, it is possible to read-off a cosmological polytope related to lower loop graphs: it is a simple projection of a given polytope through a cone with origin in ${\bf O}\,\equiv\,{\bf x}_i-{\bf x}_j$, ${\bf x}_i$ and ${\bf x}_j$ being two midpoints of edges of two different generating triangles. From a graph perspective, the projection through such a cone is equivalent to merge the two sites of the graph corresponding to the midpoints ${\bf x}_i$ and ${\bf x}_j$. This can be nicely understood via a contour integration. 

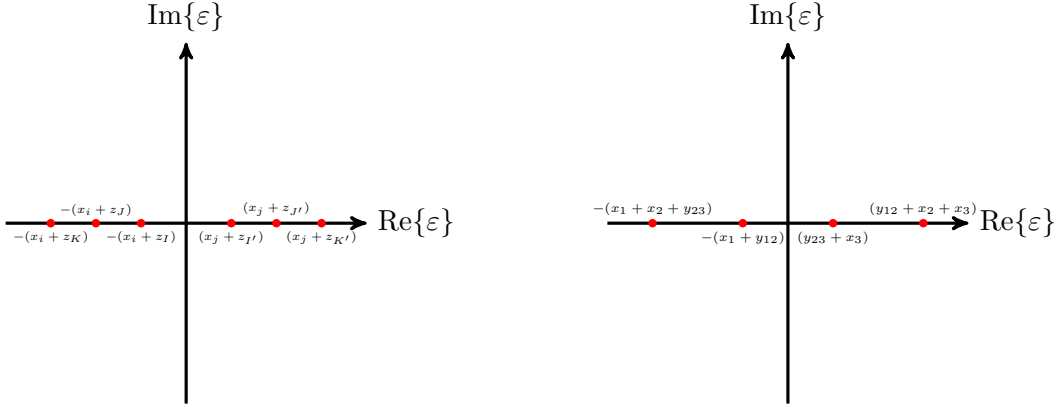
\begin{figure}
 \centering
 \begin{tikzpicture}[
    scale=4,
    axis/.style={very thick, ->, >=stealth'},
    pile/.style={thick, ->, >=stealth', shorten <=2pt, shorten
    >=2pt},
    every node/.style={color=black}
    ]
  \begin{scope}
   \draw[axis] (-0.6,0) -- (+0.6,0) node(xline)[right]{Re$\{\varepsilon\}$};
   \draw[axis] (0,-0.6) -- (0,+0.6) node(yline)[above]{Im$\{\varepsilon\}$};
   \fill[red] (-0.15,0) circle (.4pt) node(pole5)[below, scale=.75] {\tiny $-(x_i+z_I)$};
   \fill[red] (-0.30,0) circle (.4pt) node(pole1)[above, scale=.75] {\tiny $-(x_i+z_J)$};
   \fill[red] (-0.45,0) circle (.4pt) node(pole2)[below, scale=.75] {\tiny $-(x_i+z_K)$};  
   \fill[red] (+0.15,0) circle (.4pt) node(pole3)[below, scale=.75] {\tiny $(x_j+z_{I'})$};  
   \fill[red] (+0.30,0) circle (.4pt) node(pole4)[above, scale=.75] {\tiny $(x_j+z_{J'})$};    
   \fill[red] (+0.45,0) circle (.4pt) node(pole4)[below, scale=.75] {\tiny $(x_j+z_{K'})$};      
  \end{scope}
  \begin{scope}[shift={(2,0)}, transform shape]
   \draw[axis] (-0.6,0) -- (+0.6,0) node(xline)[right, scale=.25]{Re$\{\varepsilon\}$};
   \draw[axis] (0,-0.6) -- (0,+0.6) node(yline)[above, scale=.25]{Im$\{\varepsilon\}$};
   \fill[red] (-0.15,0) circle (.4pt) node(pole5)[below, scale=.2] {\tiny $-(x_1+y_{12})$};
   \fill[red] (-0.45,0) circle (.4pt) node(pole2)[above, scale=.2] {\tiny $-(x_1+x_2+y_{23})$};  
   \fill[red] (+0.15,0) circle (.4pt) node(pole3)[below, scale=.2] {\tiny $(y_{23}+x_3)$};  
   \fill[red] (+0.45,0) circle (.4pt) node(pole4)[above, scale=.2] {\tiny $(y_{12}+x_2+x_3)$};
  \end{scope}
 \end{tikzpicture}
 \caption{Poles location in the $\varepsilon$-plane for a generic one-parameter family of wavefunctions $\Psi_{\mbox{\tiny $\mathcal{G}$}}(\varepsilon)$ (left) and for the three-site graph $\Psi_3(\varepsilon)$ at tree level. The contour can be closed either in the left half-plane, enclosing with counter-clockwise orientation the poles which are function of $x_i$; or in the right half-plane, enclosing the poles which are function of $x_j$ with clockwise orientation. The quantities $z$'s are different sums of $x$'s and $y$'s, which do not involve neither $x_i$ nor $x_j$.}
 \label{fig:poles}
\end{figure}

Let $\Psi_{\mbox{\tiny $\mathcal{G}$}}(x_v,y_e)$ the wavefunction associated to a graph $\mathcal{G}$. In order to merge the sites $v_i$ and $v_j$, let us introduce a one-parameter family of deformations on the energies $x_i$ and $x_j$ related to the two sites, preserving the total energy:
\begin{equation}\label{eq:LTdef}
 x_i\,\longrightarrow\,x_i+\varepsilon,\qquad
 x_j\,\longrightarrow\,x_j-\varepsilon,\qquad
 x_k\,\longrightarrow\,x_k,\:\forall\,k\neq i,j,\qquad
 \varepsilon\,\in\,\mathbb{C}.
\end{equation}
Such a deformation maps the original wavefunction $\Psi_{\mathcal{G}}(x_v,y_e)$ into a one-parameter family of wavefunctions $\Psi_{\mathcal{G}}(x_v,y_e;\,\varepsilon)$, which can now be examined as a function of the deformation parameter $\varepsilon$. There are two classes of poles, whose location depends either on $x_i$ or $x_j$ -- when they appear together, they always appear as a sum so that this type of terms never gets a dependence on $\varepsilon$. Furthermore, because of the Cauchy theorem, the integrations along a contour which encloses the poles of one class only are equal (up to a sign). In the complex $\varepsilon$-plane, let us consider a contour along the imaginary axis, with all the energies $x_v$ and $y_e$, so that the poles which depends on $x_i$ and the ones which depends on $x_j$ are separated -- because the energies are kept all real and positives and because of the way that the deformation is defined, all the poles which depend on $x_i$ lie on the negative real axis $\mathbb{R}_{-}$, while the ones which depend on $x_j$ lie on the positive real axis $\mathbb{R}_{+}$ (see Figure \ref{fig:poles}). We refer to these poles as left and right poles, respectively.
The residue of a given pole is a function of two classes of combination of the $x_v$'s and $y_e$'s: It is either given by sums where $x_i$ and $x_j$ appear in the combination $x_i+x_j$ only, or as sum and subtractions of $x_v$'s and $y_e$'s and they are independent on $x_i$ and $x_j$. It is easy to show that, upon summation of all the (left or right pole) residues, the latter disappear: they are actually spurious poles. The reason is that the very same combination of $x_v$'s and $y_e$'s of the second type appears in two residues and when they are summed a numerator with a factor with this combination is produced.

The result of this procedure is a function $\Psi_{\mbox{\tiny $\mathcal{G}'$}}(x,y)$ depending just on the sums of $x_v$'s and $y_e$'s, and with $x_i$ and $x_j$ appearing as $x_i+x_j$ only: it is associated to a graph $\mathcal{G}'$ having the same number of edges as $\mathcal{G}$ but one site less and obtained from the latter by merging two sites. Again, this map is completely generic and can be used to obtain tadpole graphs.

As an example, let us consider the three-site graph and the energy space deformation $(x_1,\,x_3)\,\longrightarrow\,(x_1+\varepsilon,\,x_3-\varepsilon)$ for the associated wavefunction $\Psi(x,y)$. Then the one-parameter family of wavefunctions $\Psi_3(\varepsilon)$ shows two left and two right poles (see Fig. \ref{fig:poles}):
\begin{equation}\label{eq:p3poles}
 \begin{split}
  &\mathcal{P}_{\mbox{\tiny L}}\:=\:\left\{\varepsilon\,\in\,\mathbb{C}\,|\,\varepsilon_1\,=\,-(x_1+y_{12}),\:\varepsilon_2\,=\,-(x_1+x_2+y_{23})\right\},\\
  &\mathcal{P}_{\mbox{\tiny R}}\:=\:\left\{\varepsilon\,\in\,\mathbb{C}\,|\,\varepsilon_3\,=\,+(y_{23}+x_3),\:\varepsilon_4\,=\,+(y_{12}+x_2+x_{3})\right\}.
 \end{split}
\end{equation}

Performing the integration in the complex $\varepsilon$-space by choosing the contour along the imaginary axis we obtain:
\begin{equation}\label{eq:p3int}
  \Psi'_3(x,y)\:\equiv\:\int_{-i\infty}^{+\infty}d\varepsilon\,\Psi_3(\varepsilon)\:=\:
     	                \sum_{i=1}^{2}\mbox{Res}\left\{\Psi_3(\varepsilon),\,\varepsilon_i\right\}\:=\:
		        -\sum_{i=3}^{4}\mbox{Res}\left\{\Psi_3(\varepsilon),\,\varepsilon_i\right\},
\end{equation}
where the first equality is obtained by closing the integration contour on the left (so that it reduces to two counter-clockwise circles around $\varepsilon_1$ and $\varepsilon_2$), and the second one by closing the contour on the right (and the pole encircled, now clock-wise, are $\varepsilon_3$ and $\varepsilon_4$). For definiteness, let us discuss take the first choice for the integration contour. The explicit expressions for the residues of the left poles are
\begin{equation}\label{eq:psi3pl}
 \begin{split}
  &%\resizebox{.35\hsize}{!}{
    \left.\mbox{Res}\left\{\Psi_3(\varepsilon)\right\}\right|_{\mbox{\tiny $\varepsilon_1$}}\:=\:\frac{1}{(x_2+x_3+x_1)(y_{12}+x_2+y_{23})(y_{23}+x_3+x_1+y_{12})}\times\\
  &%\resizebox{.35\hsize}{!}{
    \hspace{4cm}\times\left[\frac{1}{x_2+y_{23}-y_{12}}+\frac{1}{x_3+x_1+2y_{12}+x_2}\right],\\
  &\left.\mbox{Res}\left\{\Psi_3(\varepsilon)\right\}\right|_{\mbox{\tiny $\varepsilon_2$}}\:=\:\frac{1}{(x_2+x_3+x_1)(y_{12}-x_2-y_{23})(y_{12}+x_2+y_{23})(x_3+x_1+x_2+2y_{23})}.%\times\\
 \end{split}
\end{equation}
The two residues have two common poles: one is the total energy pole, while the other one is $x_2+y_{23}-y_{12}$. The latter is actually spurious and it disappears upon summation between the two terms: the result is the two term expressions returned by the OFPT for a one-loop two site graph, with the two sites have energy $\tilde{x}_1\,\equiv\,x_3+x_1$ and $\tilde{x_2}\,\equiv\,x_2$. Thus, this contour integration gives a representation for the one-loop two-site graph which makes use of spurious singularities.

As a last comment, there is a sense in which this contour integration mapping $L$-loop graphs into $(L+1)$-loop ones can be thought of as a generalisation of the Feynman tree theorem in flat space \cite{Feynman:1963ax, CaronHuot:2010zt}. In the usual Feynman tree theorem, one performs the integration in the $l_0$-plane, with the $i\varepsilon$ prescription splitting the poles into the lower- and upper-half plane and the choice of contour picks the positive or negative energy solutions. The $l_0$-integrals are then localised on the relevant poles dictated by the $i\varepsilon$ prescription in the propagators, obtaining an integration over the Lorentz-invariant phase-space. Finally, these terms reorganise themselves into the forward limit of tree amplitudes. In the present case, we proceed (backwards) graph by graph: we introduce a one-parameter ($\varepsilon$) deformation of the energy space of an $L$-loop graph, with the poles in the $\varepsilon$-plane which are naturally split in the left- and right-half plane, {\it i.e.} along the real line. Integrating over $\varepsilon$ along the imaginary axis, one can close the contour on either the left- and right-half $\varepsilon$-plane, picking again the negative/positive energy solutions. The result is the integrand associated to an $(L+1)$-loop graph, which is nothing but an $(L+1)$-loop integrand with the $l_0$-integration performed \cite{Arkani-Hamed:2018ahb}.

%%%%%%%%%%%%%%%%%%%%%%
%%%%%%%%%%%%%%%%%%%%%%

\section{The analytic structure of the wavefunction and the flat-space S-matrix}\label{app:WFSm}

In Section \ref{sec:FCP} we discussed how the flat-space scattering amplitudes are encoded into the facets of the cosmological polytopes: those facets which are simplices encode the contribution of a given graph to the scattering amplitudes, while the facets with more vertices than a simplex can be triangulated in terms of objects whose canonical forms are products of lower point amplitudes. Given that the boundaries of the cosmological polytope are reached whenever we go to the location of a pole of the wavefunction of the universe, this is just the statement that the residues of {\it all} the poles of the wavefunction can be expressed in terms of the flat-space scattering amplitudes. More deeply, one can see Lorentz invariance emerging as one goes to {\it any} of the facets of the cosmological polytopes. This is a quite striking result can be also understood analytically. 

The most direct direct interpretation of the residue of a generic pole $x_{\mbox{\tiny $I$}}+y_{\mbox{\tiny $I$}}\,\longrightarrow\,0$ of the wavefunction is a factorisation into a lower point scattering amplitude and a product of lower point wavefunctions computed at a specific point in energy space: as any of the pole is reached, a local energy conservation is restored, generating a lower point scattering amplitude
\begin{equation*}%\label{eq:Fact1}
 \begin{tikzpicture}[ball/.style = {circle, draw, align=center, anchor=north, inner sep=0}]
  \begin{scope}
   \def\ra{3}
   \pgfmathsetmacro\xa{\ra*cos(-135)}
   \pgfmathsetmacro\ya{\ra*sin(-135)}
   \def\rb{3}
   \pgfmathsetmacro\xb{\rb*cos(-45)}
   \pgfmathsetmacro\yb{\rb*sin(-45)}
   \coordinate (S1c) at (0,0);
   \coordinate (S2c) at ($(S1c)+(\xa,\ya)$);
   \coordinate (S3c) at ($(S1c)+(\xb,\yb)$);
   \def\rip{.45}
   \def\ri{.6}
   \pgfmathsetmacro\xip{\rip*cos(-135)}
   \pgfmathsetmacro\yip{\rip*sin(-135)}
   \pgfmathsetmacro\xj{\rip*cos(-45)}
   \pgfmathsetmacro\yj{\rip*sin(-45)}
   \pgfmathsetmacro\xi{\ri*cos(45)}
   \pgfmathsetmacro\yi{\ri*sin(45)}
   \pgfmathsetmacro\xjp{\ri*cos(135)}
   \pgfmathsetmacro\yjp{\ri*sin(135)}
   \node[ball,text width=1cm,shade] (S1) at (S1c) {$\displaystyle\mathcal{G}_I$};  
   \node[ball,text width=1cm,shade] (S2) at (S2c) {$\displaystyle\mathcal{G}_J$};
   \node[ball,text width=1cm,shade] (S3) at (S3c) {$\displaystyle\mathcal{G}_K$};
   \node[ball,text width=.18cm,fill,color=black,label=left:{\tiny $\displaystyle x_{i+1}$}] (xi1) at ($(S1)+(\xip,\yip)$) {};
   \node[ball,text width=.18cm,fill,color=black,label=right:{\tiny $\displaystyle x_{j}$}] (xj) at ($(S1)+(\xj,\yj)$) {};
   \node[ball,text width=.18cm,fill,color=black,label=right:{\tiny $\displaystyle x_{i}$}] (xi) at ($(S2)+(\xi,\yi)$) {};
   \node[ball,text width=.18cm,fill,color=black,label=left:{\tiny $\displaystyle x_{j+1}$}] (xj1) at ($(S3)+(\xjp,\yjp)$) {};   
   \draw[-,thick,color=black] (xi1) -- (xi) node[midway, left] {\tiny $\displaystyle y_{i,i+1}$};
   \draw[-,thick,color=black] (xj1) -- (xj) node[midway, right] {\tiny $\displaystyle y_{j,j+1}$};
   \draw[dashed, color=red, ultra thick] (S1) circle (.75cm);
   \coordinate (ref1) at ($(S2)!0.5!(S3)$);
   \coordinate (ref2) at ($(S1)!0.5!(ref1)$);
   \node[right=3.5cm of ref2] (eq) {$\displaystyle\equiv\:$};
  \end{scope}
  \begin{scope}[shift={(7.5,0)}, transform shape]
   \def\ra{3}
   \pgfmathsetmacro\xa{\ra*cos(-135)}
   \pgfmathsetmacro\ya{\ra*sin(-135)}
   \def\rb{3}
   \pgfmathsetmacro\xb{\rb*cos(-45)}
   \pgfmathsetmacro\yb{\rb*sin(-45)}
   \coordinate (S1c) at (0,0);
   \coordinate (S2c) at ($(S1c)+(\xa,\ya)$);
   \coordinate (S3c) at ($(S1c)+(\xb,\yb)$);
   \def\rip{.45}
   \def\ri{.6}
   \pgfmathsetmacro\xip{\rip*cos(-135)}
   \pgfmathsetmacro\yip{\rip*sin(-135)}
   \pgfmathsetmacro\xj{\rip*cos(-45)}
   \pgfmathsetmacro\yj{\rip*sin(-45)}
   \pgfmathsetmacro\xi{\ri*cos(45)}
   \pgfmathsetmacro\yi{\ri*sin(45)}
   \pgfmathsetmacro\xjp{\ri*cos(135)}
   \pgfmathsetmacro\yjp{\ri*sin(135)}
   \node[ball,text width=1cm,shade] (S1) at (S1c) {$\displaystyle\mathcal{G}_I$};  
   \node[ball,text width=1cm,shade] (S2) at (S2c) {$\displaystyle\mathcal{G}_J$};
   \node[ball,text width=1cm,shade] (S3) at (S3c) {$\displaystyle\mathcal{G}_K$};
   \node[ball,text width=.18cm,fill,color=black,label=left:{\tiny $\displaystyle y_{i,i+1}+x_{i+1}\:$}] (xi1) at ($(S1)+(\xip,\yip)$) {};
   \node[ball,text width=.18cm,fill,color=black,label=right:{\tiny $\displaystyle\: x_{j}+y_{j,j+1}$}] (xj) at ($(S1)+(\xj,\yj)$) {};
   \node[ball,text width=.18cm,fill,color=black,label=right:{\tiny $\displaystyle\: x_{i}$}] (xi) at ($(S2)+(\xi,\yi)$) {};
   \node[ball,text width=.18cm,fill,color=black,label=left:{\tiny $\displaystyle x_{j+1}\:$}] (xj1) at ($(S3)+(\xjp,\yjp)$) {}; 
   \coordinate[label=left:{\tiny $\displaystyle x_I+y_{j,j+1}\:$}] (xi2) at ($(xi1)!0.5!(xi)$);
   \draw[fill,color=black] (xi2) circle (2.5pt);
   \draw[dashed, color=red, ultra thick] (xi2) circle (5pt);
   \draw[-,thick,color=black] (xi2) -- (xi) node[midway, left] {\tiny $\displaystyle y_{i,i+1}$};
   \coordinate[label=right:{\tiny $\displaystyle\: y_{i,i+1}+x_I$}] (xj2) at ($(xj1)!0.5!(xj)$);
   \draw[fill,color=black] (xj2) circle (2.5pt);
   \draw[dashed, color=red, ultra thick] (xj2) circle (5pt);
   \draw[-,thick,color=black] (xj2) -- (xj1) node[midway, right] {\tiny $\displaystyle\: y_{j,j+1}$};
   \draw[dashed, color=red, ultra thick] (S1) circle (.75cm);
  \end{scope}
 \end{tikzpicture}
\end{equation*}
\begin{equation}\label{eq:Fact2}
 \psi_{\mbox{\tiny $\mathcal{G}$}}\:\overset{\mbox{\tiny $x_I+y_I\,\longrightarrow\,0$}}{\sim}\:\frac{\mathcal{A}_{\mathcal{G}_I}\otimes\hat{\psi}_{\mathcal{G}'_J}\otimes\hat{\psi}_{\mathcal{G}'_K}}{x_I+y_I},
\end{equation}
where the dashed red circle indicates the subgraph whose energy is conserved, $\mathcal{G}'_J\:\equiv\:\mathcal{G}_J\,\cup\,\{y_{i,i+1},\,x_I+y_{j,j+1}\}$, $\mathcal{G}'_K\:\equiv\:\mathcal{G}_K\,\cup\,\{y_{i,i+1}+x_I,\,y_{j,j+1}\}$, and $\hat{\psi}_{\mathcal{G}'}$ is the residue of the wavefunction $\psi_{\mathcal{G}'}$ at the pole $x_I+y_I$. However, not all the residues of the wavefunction are independent, rather they are related via Cauchy theorem. Now, given  a wavefunction $\Psi_{\mbox{\tiny $\mathcal{G}'$}}$ associated to a graph $\mathcal{G}'$
\begin{equation*}
 \begin{tikzpicture}[ball/.style = {circle, draw, align=center, anchor=north, inner sep=0}]
  \node[ball,text width=.18cm,fill,color=black, label=below:$\mbox{\tiny $x_1$}$] at (0,0) (x1) {};
  \node[ball,text width=1cm,shade,right=.8cm of x1.east] (S1)  {$\mathcal{G}$};
  \node[ball,text width=.18cm,fill,color=black,right=.7cm of x1.east, label=below:$\mbox{\tiny $\hspace{-.3cm}x_2$}$] (x2) {};
  \draw[-,thick,color=black] (x1.east) -- (x2.west) node [midway, above] {\tiny $\displaystyle y_{12}$};
%  \draw[dashed, color=red, ultra thick] (x1) circle (.2cm);
  \node[left=.2cm of x1.west] (pn) {$\Psi_{\mathcal{G}'}\:=\:$};
 \end{tikzpicture}
\end{equation*}
we can integrate with respect to any of the energies $x_i$ on the full Riemann sphere. Let us take $x_i$ to be the energy associated to one of the outer vertices of the graph. Then:
\begin{equation}\label{eq:xyres}
 0\:=\:\frac{1}{2\pi i}\oint_{\hat{\mathbb{C}}}dx_1\,\Psi_{\mathcal{G'}}(x_1)\:=\:\sum_{k\in\mathcal{P}_{\mathcal{G}}}\mbox{Res}\left\{\Psi_{\mathcal{G}'},\,x_1\,=\,x_1^{\mbox{\tiny $(k)$}}\right\}+\mbox{Res}\left\{\Psi_{\mathcal{G'}},\,x_1\,=\,-y_{12}\right\},
\end{equation}
where $\mathcal{P}_{\mbox{\tiny $\mathcal{G}$}}$ is the set of poles involving nodes and edges of the subgraph $\mathcal{G}$, $x_1^{\mbox{\tiny $(k)$}}$ indicates their location, and the last term is the residue of the pole identified by the single-vertex subgraph indicated -- in the figure above, it is indicated by the dashed red circle on the other vertex of the graphs $\mathcal{G}'_J$ and $\mathcal{G}'_K$. Hence, the residue of the highest codimension graph can be expressed as sum of the residues of the other poles (which depend on $x_1$). Now, among these poles, there is the total energy one, whose residue is the scattering amplitude associated to the graph. The other poles impose energy conservation on subgraphs and the related residue is again a factorisation between the amplitude related to such a subgraph, and the residue of the lower point wavefunction at a pole identified by a subgraph containing an outer vertex only, {\it i.e.} a residue of the same type of the one we are studying, but related to a lower point wavefunction. We can integrate this procedure until we get to reduce the residue of $\Psi_{\mathcal{G}'}$ at $x_1+y_{12}\,=\,0$ into a sum of products of lower point amplitudes times the residue of the two-site wavefunction at the pole related to one of its vertices. Here we can use the Cauchy theorem once more
\begin{equation}\label{eq:xyres2st}
 \begin{tikzpicture}[ball/.style = {circle, draw, align=center, anchor=north, inner sep=0}]
  \coordinate (x0a) at (0,0);
  \node (eq) at (x0a) {$\displaystyle 0\:=\:$};
  \node[ball,text width=.18cm,fill,color=black,right=.25cm of eq.east, label=below:$\mbox{\tiny $x_1$}$] (x1b) {};  
  \node[ball,text width=.18cm,fill,color=black,right=1cm of x1b.east, label=below:$\mbox{\tiny $x_2$}$] (x2b) {};
  \draw[-,thick,color=black] (x1b.east) -- (x2b.west) node[midway, above] {\tiny $\displaystyle y_{12}$};
  \draw[dashed, color=red, ultra thick] ($(x1b)!0.5!(x2b)$) ellipse (.75cm and 0.4cm);
  \node[right=.5cm of x2b] (pl) {$\displaystyle\:+\:$};
  \node[ball,text width=.18cm,fill,color=black, label=below:$\mbox{\tiny $x_1$}$, right=.5cm of pl] (x1) {};
  \node[ball,text width=.18cm,fill,color=black,right=1cm of x1.east, label=below:$\mbox{\tiny $x_2$}$] (x2) {};
  \draw[-,thick,color=black] (x1.east) -- (x2.west) node[midway, above] {\tiny $\displaystyle y_{12}$};
  \draw[dashed, color=red, ultra thick] (x1) circle (.3cm);
 \end{tikzpicture}
\end{equation}
The residue at the pole $x_1+y_{12}\,=\,0$ is equal to (minus) the Lorentz-invariant propagator $(y_{12}^2-x_2^2)^{-1}$, and thus the two-site wavefunction can be expressed just in terms of the flat-space scattering amplitude. Interestingly, using iteratively \eqref{eq:xyres} and \eqref{eq:xyres2st} on each residue, it is possible to express them just in terms of flat-space scattering amplitudes!

Let us explicitly discuss the residues of the wavefunction poles in the concrete examples of the three-site tree-level graph and for one-loop graphs

%%%%%%%%%%%%%%%%%%%%%%
%%%%%%%%%%%%%%%%%%%%%%

\subsection{The three-site graph}\label{subsec:3sg}

The three-site graph is the simplest non-trivial example that we can examine:

\begin{equation}\label{eq:3sg}
 \begin{tikzpicture}[ball/.style = {circle, draw, align=center, anchor=north, inner sep=0}]
  \node[ball,text width=.18cm,fill,color=black, label=below:$\mbox{\tiny $x_1$}$] at (0,0) (x1) {};
  \node[ball,text width=.18cm,fill,color=black,right=1cm of x1.east, label=below:$\mbox{\tiny $x_2$}$] (x2) {};
  \node[ball,text width=.18cm,fill,color=black,right=1cm of x2.east, label=below:$\mbox{\tiny $x_2$}$] (x3) {};
  \draw[-,thick,color=black] (x1.east) -- (x2.west) node[midway, above] {\tiny $\displaystyle y_{12}$};
  \draw[-,thick,color=black] (x2.east) -- (x3.west) node[midway, above] {\tiny $\displaystyle y_{23}$};
  \node[left=.2cm of x1.west] (pn3) {$\displaystyle\Psi_3(x;y)\:\equiv\:$};
  \node[right=.2cm of x3.east] (eq) {$\displaystyle\:=$};
  \node at (5.15,-1) (wf3) {\footnotesize $\displaystyle=\:\frac{x_1+y_{12}+2x_2+y_{23}+x_3}{(x_1+x_2+x_3)(x_1+x_2+y_{23})(x_1+y_{12})(y_{12}+x_2+y_{23})(y_{23}+x_3)(y_{12}+x_2+x_3)}$};
 \end{tikzpicture}  
\end{equation}

It shows a pole, $y_{12}+x_2+y_{23}$, which is a distinctive feature of the universe wavefunction: In the total energy conservation limit it disappears and thus it is absent in the flat-space scattering amplitude! This graph is the simplest example where physical information not reducible to the flat-space scattering amplitude could be found. Let us examine in detail the residues of the poles, starting with looking at a codimension-one subgraph

\begin{equation}\label{eq:3sgResc1}
 \begin{tikzpicture}[ball/.style = {circle, draw, align=center, anchor=north, inner sep=0}]
  \node[ball,text width=.18cm,fill,color=black, label=below:$\mbox{\tiny $x_1$}$] at (0,0) (x1) {};
  \node[ball,text width=.18cm,fill,color=black,right=1cm of x1.east, label=below:$\mbox{\tiny $x_2$}$] (x2) {};
  \node[ball,text width=.18cm,fill,color=black,right=1cm of x2.east, label=below:$\mbox{\tiny $x_2$}$] (x3) {};
  \draw[-,thick,color=black] (x1.east) -- (x2.west) node[midway, above] {\tiny $\displaystyle y_{12}$};
  \draw[-,thick,color=black] (x2.east) -- (x3.west) node[midway, above] {\tiny $\displaystyle y_{23}$};
  \draw[dashed, color=red, ultra thick] ($(x1)!0.5!(x2)$) ellipse (.75cm and .4cm);
  \node[right=.25cm of x3.east] (eq) {$\displaystyle =\:$};
  \node[ball,text width=.18cm,fill,color=black, right=.25cm of eq.east, label=below:$\mbox{\tiny $x_1$}$] (x1b) {};
  \node[ball,text width=.18cm,fill,color=black,right=1cm of x1b.east, label=below:$\mbox{\tiny $x_2+y_{23}$}$] (x2b) {};
  \draw[-,thick,color=black] (x1b.east) -- (x2b.west) node[midway, above] {\tiny $\displaystyle y_{12}$};
  \draw[dashed, color=red, ultra thick] ($(x1b)!0.5!(x2b)$) ellipse (.75cm and .4cm);
  \node[right=.25cm of x2b.east] (ot) {$\displaystyle\otimes$};  
  \node[ball,text width=.18cm,fill,color=black, right=.25cm of ot.east, label=below:$\mbox{\tiny $x_1+x_2$}$] (x1c) {};
  \node[ball,text width=.18cm,fill,color=black,right=1cm of x1c.east, label=below:$\mbox{\tiny $x_3$}$] (x2c) {};
  \draw[-,thick,color=black] (x1c.east) -- (x2c.west) node[midway, above] {\tiny $\displaystyle y_{23}$};
  \draw[dashed, color=red, ultra thick] (x1c) circle (.2cm);
  \node[right=.25cm of x2c.east] (eq2) {$\displaystyle =\:-$};  
  \node[ball,text width=.18cm,fill,color=black, right=.25cm of eq2.east, label=below:$\mbox{\tiny $x_1$}$] (x1d) {};
  \node[ball,text width=.18cm,fill,color=black,right=1cm of x1d.east, label=below:$\mbox{\tiny $x_2+y_{23}$}$] (x2d) {};
  \draw[-,thick,color=black] (x1d.east) -- (x2d.west) node[midway, above] {\tiny $\displaystyle y_{12}$};
  \draw[dashed, color=red, ultra thick] ($(x1d)!0.5!(x2d)$) ellipse (.75cm and .4cm);
  \node[right=.25cm of x2d.east] (ot2) {$\displaystyle\otimes$};  
  \node[ball,text width=.18cm,fill,color=black, right=.25cm of ot2.east, label=below:$\mbox{\tiny $x_1+x_2$}$] (x1e) {};
  \node[ball,text width=.18cm,fill,color=black,right=1cm of x1e.east, label=below:$\mbox{\tiny $x_3$}$] (x2e) {};
  \draw[-,thick,color=black] (x1e.east) -- (x2e.west) node[midway, above] {\tiny $\displaystyle y_{23}$};
  \draw[dashed, color=red, ultra thick] ($(x1e)!0.5!(x2e)$) ellipse (.75cm and .4cm);
 \end{tikzpicture}   
\end{equation}

As in the previous section, the dashed red ellipses identifies the pole whose residue we are taking, in this case $x_1+x_2+y_{23}\,=\,0$. The very right-hand-side in \eqref{eq:3sgResc1} is obtained via the identity \eqref{eq:xyres2st}. Thus, the residue at a codimension-one pole is given by (minus) the product of two two-site scattering amplitudes, one computed at the location of the codimension-one pole itself, while the other one is computed on the total energy pole! Notice that, in taking the residue of this pole, we can look at the wavefunction $\psi_3$ as a function of either $x_1$ or $x_2$. In the latter case, the product of the two propagators in the very right-hand-side of \eqref{eq:3sgResc1} is actually the contribution to the scattering amplitude of the full three-site graph itself!

Let us now consider the last class of poles of $\psi_3$, which are related the subgraphs containing just one site. It is worth to have a separate discussion for the poles related to the external and internal sites. In the first case, we can use Cauchy theorem \eqref{eq:xyres}:

\begin{equation}\label{eq:3sgResc2}
 \begin{tikzpicture}[ball/.style = {circle, draw, align=center, anchor=north, inner sep=0}]
  \node[ball,text width=.18cm,fill,color=black, label=below:$\mbox{\tiny $x_1$}$] at (0,0) (x1) {};
  \node[ball,text width=.18cm,fill,color=black,right=1cm of x1.east, label=below:$\mbox{\tiny $x_2$}$] (x2) {};
  \node[ball,text width=.18cm,fill,color=black,right=1cm of x2.east, label=below:$\mbox{\tiny $x_3$}$] (x3) {};
  \draw[-,thick,color=black] (x1.east) -- (x2.west) node[midway, above] {\tiny $\displaystyle y_{12}$};
  \draw[-,thick,color=black] (x2.east) -- (x3.west) node[midway, above] {\tiny $\displaystyle y_{23}$};
  \draw[dashed, color=red, ultra thick] (x1) circle (.2cm);
  \node[right=.25cm of x3.east] (eq) {$\displaystyle =\:-$};
  \node[ball,text width=.18cm,fill,color=black, label=below:$\mbox{\tiny $x_1$}$, right=.25cm of eq.east] (x1b) {};
  \node[ball,text width=.18cm,fill,color=black,right=1cm of x1b.east, label=below:$\mbox{\tiny $x_2$}$] (x2b) {};
  \node[ball,text width=.18cm,fill,color=black,right=1cm of x2b.east, label=below:$\mbox{\tiny $x_3$}$] (x3b) {};
  \draw[-,thick,color=black] (x1b.east) -- (x2b.west) node[midway, above] {\tiny $\displaystyle y_{12}$};
  \draw[-,thick,color=black] (x2b.east) -- (x3b.west) node[midway, above] {\tiny $\displaystyle y_{23}$};
  \draw[dashed, color=red, ultra thick] (x2b) ellipse (1.35cm and .4cm);
  \node[right=.25cm of x3b.east] (pl) {$\displaystyle +\:$};
   \node[ball,text width=.18cm,fill,color=black, right=.25cm of pl.east, label=below:$\mbox{\tiny $x_1$}$] (x1d) {};
  \node[ball,text width=.18cm,fill,color=black,right=1cm of x1d.east, label=below:$\mbox{\tiny $x_2+y_{23}$}$] (x2d) {};
  \draw[-,thick,color=black] (x1d.east) -- (x2d.west) node[midway, above] {\tiny $\displaystyle y_{12}$};
  \draw[dashed, color=red, ultra thick] ($(x1d)!0.5!(x2d)$) ellipse (.75cm and .4cm);
  \node[right=.25cm of x2d.east] (ot2) {$\displaystyle\otimes$};  
  \node[ball,text width=.18cm,fill,color=black, right=.25cm of ot2.east, label=below:$\mbox{\tiny $x_1+x_2$}$] (x1e) {};
  \node[ball,text width=.18cm,fill,color=black,right=1cm of x1e.east, label=below:$\mbox{\tiny $x_3$}$] (x2e) {};
  \draw[-,thick,color=black] (x1e.east) -- (x2e.west) node[midway, above] {\tiny $\displaystyle y_{23}$};
  \draw[dashed, color=red, ultra thick] ($(x1e)!0.5!(x2e)$) ellipse (.75cm and .4cm);
 \end{tikzpicture} 
\end{equation}
with the two terms on the right-hand-side being the flat-space scattering amplitude for the three-site graph and a product of two-lower sites scattering amplitudes. Notice that the identity \eqref{eq:3sgResc2} is obtained by considering $\Psi_3$ as a function of $x_1$. With \eqref{eq:3sgResc1} and \eqref{eq:3sgResc2} at hand, we are already in the position of writing $\Psi_3$ in terms of flat-space quantities:
\begin{equation}\label{eq:3sgFSax1} 
 \Psi_3\:=\:\frac{A_3(x_2,x_3)}{x_1+x_2+x_3}-\frac{A_2(x_2+y_{23})\otimes A_2(x_3)}{x_1+x_2+y_{23}}-\frac{A_3(x_2,x_3)-A_2(x_2+y_{23})\otimes A_2(x_3)}{x_1+y_{12}},
\end{equation}
where $A_n(x_{a_1},\ldots\,x_{a_{n-1}})$ is simply the flat-space scattering amplitude related to the $n$-site graph, with $x_{a_1},\,\ldots,\,x_{a_{n-1}}$ being the $n-1$ independent external energies, while the $n$-th one is determined by the energy conservation, and are explicitly given as
\begin{equation}\label{eq:3sgFSax1ex}
 \begin{split}
  &A_3(x_2,x_3)\:=\:\frac{1}{y_{12}^2-(x_2+x_3)^2}\times\frac{1}{(y_{23}^3-x_3^2)},\\
  &A_2(x_2+y_{23})\:=\:\frac{1}{y_{12}^2-(x_2+y_{23})^2},\qquad A_2(x_3)\:=\:\frac{1}{y_{23}^2-x_3^2}
 \end{split}
\end{equation}
This representation can be also obtained via the energy space deformation $x_1\,\longrightarrow\,x_1+\zeta$ and looking at $\psi_3$ as a function of $\zeta$.

Let us now examine the residue of the pole $y_{12}+x_2+x_{23}\,=\,0$, which potentially can contain a different class of physical information. It turns out that it is not the case, and also the residue of this pole can be interpreted in terms of flat-space scattering amplitudes. First, it is straightforward to see that it can be written as product of two Lorentz-like propagators:

\begin{equation}\label{eq:3sgResc3}
 \begin{tikzpicture}[ball/.style = {circle, draw, align=center, anchor=north, inner sep=0}]
  \node[ball,text width=.18cm,fill,color=black, label=below:$\mbox{\tiny $x_1$}$] at (0,0) (x1) {};
  \node[ball,text width=.18cm,fill,color=black,right=1cm of x1.east, label=below:$\mbox{\tiny $x_2$}$] (x2) {};
  \node[ball,text width=.18cm,fill,color=black,right=1cm of x2.east, label=below:$\mbox{\tiny $x_3$}$] (x3) {};
  \draw[-,thick,color=black] (x1.east) -- (x2.west) node[midway, above] {\tiny $\displaystyle y_{12}$};
  \draw[-,thick,color=black] (x2.east) -- (x3.west) node[midway, above] {\tiny $\displaystyle y_{23}$};
  \draw[dashed, color=red, ultra thick] (x2) circle (.2cm);
  \node[right=.25cm of x3.east] (eq) {$\displaystyle =\:$};
  \node[ball,text width=.18cm,fill,color=black, right=.25cm of eq.east, label=below:$\mbox{\tiny $x_1$}$] (x1b) {};
  \node[ball,text width=.18cm,fill,color=black,right=1cm of x1b.east, label=below:$\mbox{\tiny $x_2+y_{23}$}$] (x2b) {};
  \draw[-,thick,color=black] (x1b.east) -- (x2b.west) node[midway, above] {\tiny $\displaystyle y_{12}$};
  \draw[dashed, color=red, ultra thick] (x2b) circle (.2cm);
  \node[right=.25cm of x2b.east] (ot) {$\displaystyle\otimes$};  
  \node[ball,text width=.18cm,fill,color=black, right=.25cm of ot.east, label=below:$\mbox{\tiny $y_{12}+x_2$}$] (x1c) {};
  \node[ball,text width=.18cm,fill,color=black,right=1cm of x1c.east, label=below:$\mbox{\tiny $x_3$}$] (x2c) {};
  \draw[-,thick,color=black] (x1c.east) -- (x2c.west) node[midway, above] {\tiny $\displaystyle y_{23}$};
  \draw[dashed, color=red, ultra thick] (x1c) circle (.2cm);
  \node[right=.25cm of x2c.east] (eq2) {$\displaystyle =\:$}; 
  \node[ball,text width=.18cm,fill,color=black, right=.25cm of eq2.east, label=below:$\mbox{\tiny $x_1$}$] (x1d) {};
  \node[ball,text width=.18cm,fill,color=black,right=1cm of x1d.east, label=below:$\mbox{\tiny $x_2+y_{23}$}$] (x2d) {};
  \draw[-,thick,color=black] (x1d.east) -- (x2d.west) node[midway, above] {\tiny $\displaystyle y_{12}$};
  \draw[dashed, color=red, ultra thick] ($(x1d)!0.5!(x2d)$) ellipse (.75cm and .4cm);
  \node[right=.25cm of x2d.east] (ot2) {$\displaystyle\otimes$};  
  \node[ball,text width=.18cm,fill,color=black, right=.25cm of ot2.east, label=below:$\mbox{\tiny $y_{12}+x_2$}$] (x1e) {};
  \node[ball,text width=.18cm,fill,color=black,right=1cm of x1e.east, label=below:$\mbox{\tiny $x_3$}$] (x2e) {};
  \draw[-,thick,color=black] (x1e.east) -- (x2e.west) node[midway, above] {\tiny $\displaystyle y_{23}$};
  \draw[dashed, color=red, ultra thick] ($(x1e)!0.5!(x2e)$) ellipse (.75cm and .4cm);
 \end{tikzpicture}
\end{equation}
Secondly, the very right-hand-side in \eqref{eq:3sgResc3} is actually the flat-space scattering amplitude $A_3(x_1,x_3)$ corresponding to the full three-site graph, exactly as it happens for the other poles. Hence, $\psi_3$ admits the following representation as well in terms of flat-space amplitudes
\begin{equation}\label{eq:3sgFSax2}
 \begin{split}
  &\Psi_3\:=\:A_3(x_1,x_3)\left[\frac{1}{x_1+x_2+x_3}-\frac{1}{x_1+x_2+y_{23}}-\frac{1}{y_{12}+x_2+x_3}+\frac{1}{y_{12}+x_2+y_{23}}\right],\\
  &A_3(x_1,x_3)\:=\:\frac{1}{y_{12}^2-x_1^2}\times\frac{1}{y_{23}^2-x_3^2}.
 \end{split}
\end{equation}
As in the previous case, it can be obtained via the energy space deformation $x_2\,\longrightarrow\,x_2+\zeta$ and looking at $\psi_3$ as a function of $\zeta$, and it is the very same one returned by the recursion relation discussed in Section 2.5 of \cite{Arkani-Hamed:2017fdk}.

\subsection{Loop factorisation}\label{subsubsc:LF}

The examples discussed so far concerns with tree-level wavefunctions only. A legitimate question is whether the wavefunction keeps being expressible in terms of the flat-space scattering amplitudes as {\it any} of its poles is approached. First, notice that the Cauchy theorem applied to the simplest possible loop graph -- the tadpole -- returns the following identity between its residues
\begin{equation}\label{eq:tpid}
 \begin{tikzpicture}[ball/.style = {circle, draw, align=center, anchor=north, inner sep=0}]
  \begin{scope}
    \node (eq) at (0,0) {$\displaystyle 0\:=$};
   \coordinate (cnt) at ($(eq)+(1.75,0)$);
   \draw[thick, color=black] (cnt) circle (.75cm);
   \node[ball, text width=.18cm,fill,color=black, label=left:$\mbox{\tiny $x$}$, left=.65cm of cnt] (x) {};
   \draw[dashed, color=red, ultra thick] (cnt) circle (.9cm);
  \end{scope}
  \begin{scope}[shift={(3.5,0)}]
    \node (eq) at (0,0) {$\displaystyle +$};
   \coordinate (cnt) at ($(eq)+(1.75,0)$);
   \draw[thick, color=black] (cnt) circle (.75cm);
   \node[ball, text width=.18cm,fill,color=black, label=left:$\mbox{\tiny $x$}$, left=.65cm of cnt] (x) {};
   \draw[dashed, color=red, ultra thick] (x) circle (.2cm);
  \end{scope}
 \end{tikzpicture}
\end{equation}
the first term on the right-hand-side being the flat-space integrand. Such an identity indeed allows for a representation of the tadpole wavefunction in terms of flat-space quantities only:
\begin{equation}\label{eq:fptd}
 \Psi_{1}\:=\:A_1(y)\left[\frac{1}{x}-\frac{1}{x+2y}\right],\qquad A_1(y)\:\equiv\:\frac{1}{2y}.
\end{equation}
A quick comment on the flat-space expression $A_1(y)$ above is in order. Contrarily to the tree-level case, this flat-space quantity does not have the form of a Lorentz propagator. One might wonder about the sense for which the above expression has anything to do with the flat-space tadpole integrand. It is important to remark that the wavefunction, upon taking the total energy conservation limit, returns a representation for the loop integrand with the integration over the time-like component $l_0$ of the loop momentum $l$ performed, living just $2y\,\equiv\,2|\overrightarrow{l}|$. Such a factor, together with the differential $d\overrightarrow{l}$, forms the Lorentz-invariant phase-space measure.

Going back to the singularity structure of the loop wavefunction, when any pole other than the total energy one is reached, the $L$-loop wavefunction factorises into lower-order flat-space amplitudes and lower-order wavefunctions computed at the location of the pole. As for the tree-level case, the identity \eqref{eq:tpid} as well as the Cauchy theorem applied to the wavefunction in question allow to express its residues at any poles in terms of flat-space objects. Taking the three-site one-loop graph as an example, its residues are given by
\begin{equation}\label{eq:3slpg}
 \begin{tikzpicture}[ball/.style = {circle, draw, align=center, anchor=north, inner sep=0}]
  \begin{scope}
   \coordinate (cnt) at (0,0);
   \def\r{.75}
   \def\rl{.9}
   \pgfmathsetmacro\ax{\r*cos(180)}
   \pgfmathsetmacro\ay{\r*sin(180)}
   \pgfmathsetmacro\bx{\r*cos(60)}
   \pgfmathsetmacro\by{\r*sin(60)}
   \pgfmathsetmacro\cx{\r*cos(-60)}
   \pgfmathsetmacro\cy{\r*sin(-60)}
   \pgfmathsetmacro\axl{\rl*cos(120)}
   \pgfmathsetmacro\ayl{\rl*sin(120)}
   \pgfmathsetmacro\bxl{\rl*cos(0)}
   \pgfmathsetmacro\byl{\rl*sin(0)}
   \pgfmathsetmacro\cxl{\rl*cos(240)}
   \pgfmathsetmacro\cyl{\rl*sin(240)}
   \coordinate[label=left:$\displaystyle \mbox{\tiny $x_1$}$] (x1) at (\ax,\ay);
   \coordinate[label=above:$\displaystyle \mbox{\tiny $x_2$}$] (x2) at (\bx,\by);   
   \coordinate[label=below:$\displaystyle \mbox{\tiny $x_3$}$] (x3) at (\cx,\cy);
   \coordinate[label=left:$\displaystyle \mbox{\tiny $y_{12}$}$] (y12) at (\axl,\ayl);
   \coordinate[label=right:$\displaystyle \mbox{\tiny $y_{23}$}$] (y23) at (\bxl,\byl);   
   \coordinate[label=below:$\displaystyle \mbox{\tiny $y_{31}$}$] (y31) at (\cxl,\cyl);    
   \draw[thick, color=black] (cnt) circle (.75cm);
   \draw[fill, color=black] (x1) circle (.1cm);
   \draw[fill, color=black] (x2) circle (.1cm);
   \draw[fill, color=black] (x3) circle (.1cm);
   \coordinate (x1a) at ($(x1)+(-.25,+.125)$);
   \coordinate (x1b) at ($(x1a)+(+.375,0)$);
   \coordinate (x13a) at ($(x1)!0.5!(x3)+(-.375,-.375)$);
   \coordinate (x13b) at ($(x13a)+(.25,.25)$);
   \coordinate (x3a) at ($(x3)+(+.125,-.125)$);
   \coordinate (x3b) at ($(x3)+(-.125,+.125)$);
   \coordinate (x32a) at ($(x3)!0.5!(x2)+(+.5,0)$);
   \coordinate (x32b) at ($(x32a)+(-.375,0)$);
   \coordinate (x2a) at ($(x2)+(+.125,+.25)$);
   \coordinate (x2b) at ($(x2a)+(-.375,-.25)$);
   \draw[dashed, color=red, ultra thick] plot [smooth cycle, tension=1] coordinates {(x1a) (x13a) (x3a) (x32a) (x2a) (x2b) (x32b) (x3b) (x13b) (x1b)};
   \node[right=1.5cm of cnt] (eq) {$\displaystyle =$};
  \end{scope}
  \begin{scope}[shift={(3.5,0)}]
   \coordinate (cnt) at (0,0);
   \def\r{.75}
   \pgfmathsetmacro\ax{\r*cos(0)}
   \pgfmathsetmacro\ay{\r*sin(0)}
   \pgfmathsetmacro\axl{\r*cos(180)}
   \pgfmathsetmacro\ayl{\r*sin(180)}
   \coordinate[label=right:$\displaystyle \mbox{\tiny $x_1+x_2$}$] (x12) at (\ax,\ay);
   \coordinate[label=left:$\displaystyle \mbox{\tiny $y_{12}$}$] (y12b) at (\axl, \ayl);
   \draw[thick, color=black] (cnt) circle (.75cm);
   \draw[fill, color=black] (x12) circle (.1cm);
   \draw[dashed, color=red, ultra thick] (x12) circle (.2cm);
   \node[right=1cm of x12] (ot) {$\displaystyle\otimes$};
   \coordinate[label=below:$\displaystyle\mbox{\tiny $x_1+y_{12}$}$, right=.5cm of ot] (n1) {};
   \coordinate[label=below:$\displaystyle\mbox{\tiny $x_3$}$, right=1.5cm of n1] (n2) {};
   \coordinate[label=below:$\displaystyle\mbox{\tiny $x_2+y_{12}$}$, right=1.5cm of n2] (n3) {};
   \draw[-, thick, color=black] (n1) edge node[label=above:$\displaystyle\mbox{\tiny $y_{31}$}$] {} (n2);
   \draw[-, thick, color=black] (n2) edge node[label=above:$\displaystyle\mbox{\tiny $y_{23}$}$] {} (n3);   
   \draw[fill, color=black] (n1) circle (.1cm);
   \draw[fill, color=black] (n2) circle (.1cm);
   \draw[fill, color=black] (n3) circle (.1cm);   
   \draw[dashed, color=red, ultra thick] (n2) ellipse (1.75cm and .3cm);
  \end{scope} 
  \begin{scope}[shift={(0,-2.5)}]
   \coordinate (cnt) at (0,0);
   \def\r{.75}
   \def\rl{.9}
   \pgfmathsetmacro\ax{\r*cos(180)}
   \pgfmathsetmacro\ay{\r*sin(180)}
   \pgfmathsetmacro\bx{\r*cos(60)}
   \pgfmathsetmacro\by{\r*sin(60)}
   \pgfmathsetmacro\cx{\r*cos(-60)}
   \pgfmathsetmacro\cy{\r*sin(-60)}
   \pgfmathsetmacro\axl{\rl*cos(120)}
   \pgfmathsetmacro\ayl{\rl*sin(120)}
   \pgfmathsetmacro\bxl{\rl*cos(0)}
   \pgfmathsetmacro\byl{\rl*sin(0)}
   \pgfmathsetmacro\cxl{\rl*cos(240)}
   \pgfmathsetmacro\cyl{\rl*sin(240)}
   \coordinate[label=left:$\displaystyle \mbox{\tiny $x_1$}$] (x1) at (\ax,\ay);
   \coordinate[label=above:$\displaystyle \mbox{\tiny $x_2$}$] (x2) at (\bx,\by);   
   \coordinate[label=below:$\displaystyle \mbox{\tiny $x_3$}$] (x3) at (\cx,\cy);
   \coordinate[label=left:$\displaystyle \mbox{\tiny $y_{12}$}$] (y12) at (\axl,\ayl);
   \coordinate[label=right:$\displaystyle \mbox{\tiny $y_{23}$}$] (y23) at (\bxl,\byl);   
   \coordinate[label=below:$\displaystyle \mbox{\tiny $y_{31}$}$] (y31) at (\cxl,\cyl);    
   \draw[thick, color=black] (cnt) circle (.75cm);
   \draw[fill, color=black] (x1) circle (.1cm);
   \draw[fill, color=black] (x2) circle (.1cm);
   \draw[fill, color=black] (x3) circle (.1cm);
   \coordinate (x1a) at ($(x1)+(-.25,+.125)$);
   \coordinate (x1b) at ($(x1a)+(+.375,0)$);
   \coordinate (x13a) at ($(x1)!0.5!(x3)+(-.375,-.375)$);
   \coordinate (x13b) at ($(x13a)+(.25,.25)$);
   \coordinate (x3a) at ($(x3)+(+.125,-.125)$);
   \coordinate (x3b) at ($(x3)+(-.125,+.125)$);
   \coordinate (x3c) at ($(x3)+(+.125,+.125)$);
   \draw[dashed, color=red, ultra thick] plot [smooth cycle, tension=1] coordinates {(x1a) (x13a) (x3a) (x3c) (x3b) (x13b) (x1b)};
   \node[right=1.5cm of cnt] (eq) {$\displaystyle =$};
  \end{scope}
  \begin{scope}[shift={(3.5,-2.5)}]
   \coordinate (cnt) at (0,0);
   \def\r{.75}
   \pgfmathsetmacro\ax{\r*cos(0)}
   \pgfmathsetmacro\ay{\r*sin(0)}
   \pgfmathsetmacro\axl{\r*cos(180)}
   \pgfmathsetmacro\ayl{\r*sin(180)}
   \pgfmathsetmacro\bxl{\r*cos(90)}
   \pgfmathsetmacro\byl{\r*sin(90)}
   \pgfmathsetmacro\cxl{\r*cos(-90)}
   \pgfmathsetmacro\cyl{\r*sin(-90)}
   \coordinate[label=right:$\displaystyle \mbox{\tiny $x_1+x_3$}$] (x13) at (\ax,\ay);
   \coordinate[label=left:$\displaystyle \mbox{\tiny $x_{2}$}$] (x2b) at (\axl, \ayl);
   \coordinate[label=above:$\displaystyle \mbox{\tiny $y_{12}$}$] (y12b) at (\bxl,\byl);
   \coordinate[label=below:$\displaystyle \mbox{\tiny $y_{23}$}$] (y23b) at (\cxl, \cyl);
   \draw[thick, color=black] (cnt) circle (.75cm);
   \draw[fill, color=black] (x13) circle (.1cm);
   \draw[fill, color=black] (x2b) circle (.1cm);
   \draw[dashed, color=red, ultra thick] (x13) circle (.2cm);
   \node[right=1cm of x13] (otb) {$\displaystyle\otimes$};
   \coordinate[label=below:$\displaystyle\mbox{\tiny $x_1+y_{12}$}$, right=.5cm of otb] (n1b) {};
   \coordinate[label=below:$\displaystyle\mbox{\tiny $x_3+y_{23}$}$, right=1.5cm of n1b] (n2b) {};
   \draw[-, thick, color=black] (n1b) edge node[label=above:$\displaystyle\mbox{\tiny $y_{31}$}$] {} (n2b);
   \draw[fill, color=black] (n1b) circle (.1cm);
   \draw[fill, color=black] (n2b) circle (.1cm);
   \draw[dashed, color=red, ultra thick] ($(n1b)!0.5!(n2b)$) ellipse (1cm and .3cm);
  \end{scope}
 \end{tikzpicture}
\end{equation}
where, in each line, the loop factors on the right-hand-side are (residues of) wavefunctions at the point in energy space where the total energy of the circled subgraph is conserved, and are related to flat-space amplitudes using the Cauchy theorem (which the identity \eqref{eq:tpid} is a manifestation for the simplest loop graph). The tree-level factors are, instead, flat-space amplitudes themselves. As far as the residues of the wavefunction related to the total energy of the single-node subgraph, they can be obtained in terms of the other ones (which we just got an interpretation in terms of flat-space quantities) via Cauchy theorem.

%%%%%%%%%%%%%%%%%%%%%%
%%%%%%%%%%%%%%%%%%%%%%
%%%%%%%%%%%%%%%%%%%%%%

\bibliographystyle{JHEP}
\bibliography{cprefs}

\end{document}